\documentclass[pre,aps,reprint,footinbib,longbibliography]{revtex4-1}
\usepackage{graphicx}
\usepackage{amssymb}
\usepackage{amsmath}
\usepackage{times}

\usepackage{xcolor} 

\begin{document}

\title{Phase reduction beyond the first order: The case of the mean-field complex Ginzburg-Landau equation}
\author{Iv\'an Le\'on}
\author{Diego Paz\'o}
\affiliation{Instituto de F\'{\i}sica de Cantabria (IFCA), CSIC-Universidad de 
Cantabria, 39005 Santander, Spain}

\date{\today}

\begin{abstract}
Phase reduction is a powerful technique that makes possible to describe the dynamics of a weakly perturbed
limit-cycle oscillator in terms of its phase. For ensembles of oscillators, 
a classical example of phase reduction is the derivation of 
the Kuramoto model from the mean-field complex Ginzburg-Landau equation (MF-CGLE).
Still, the Kuramoto model 
is a first-order phase approximation that displays either full synchronization or incoherence,
but none of the nontrivial dynamics of the MF-CGLE.
This fact calls for an expansion beyond the first order in the coupling constant.
We develop an isochron-based scheme to obtain the second-order phase approximation, 
which reproduces the weak coupling dynamics of the MF-CGLE.
The practicality of our method is evidenced by extending the calculation up to third order.
Each new term of the power series expansion contributes with additional higher-order multi-body (i.e.~non-pairwise) interactions. 
This points to intricate multi-body phase interactions as the source of pure collective chaos in the  MF-CGLE at moderate coupling.
\end{abstract}

  \maketitle
  
\section{Introduction}  
Networks of nonlinear elements with oscillatory behavior (`oscillators') are  found in a variety of disciplines, such as neuroscience or engineering \cite{Win80,HI97,Str03,PRK01}.
It is an empirical fact that some phenomena arising in these systems 
can be understood in terms of interacting phase oscillators.
This framework has proven to be useful modeling and engineering
experimental setups composed of many rhythmic elements, operating in a wide range of spatio-temporal scales,
and interacting through very different physical processes. We may cite small motors---cell phone vibrators---interacting through an elastic plate \cite{mertens11,*mertens11b},
networks of (electro-)chemical oscillators \cite{kiss07,totz15},
arrays of Josephson junctions \cite{wiesenfeld95,WCS96} and globally coupled electrical self-oscillators \cite{temirbayev12,english15}, 
or nanoelectromechanical oscillators in a ring \cite{matheny19}.  

Applying a phase reduction method \cite{Win80,Kur84,nakao16,monga19} is the rigorous way of describing a weakly perturbed oscillator 
solely in terms of its phase (the other degrees of freedom become enslaved). 
However, obtaining analytically the approximate `phase-only model' for a specific system is not an easy task.
% And, 
Moreover, phase reduction becomes inaccurate unless the disturbances are not sufficiently weak.   
While, according to common wisdom phase reduction of oscillator ensembles yields pairwise interacting phase oscillators \cite{Kur84},
 multi-body (i.e. non-pairwise) interactions may also be relevant in some contexts.
Apart from the idea of invoking hypothetical three-body interacting limit-cycle oscillators ~\cite{tanaka11},
multi-body 
% (i.e.~non-pairwise) 
phase interactions naturally arise if the coupling is nonlinear \cite{ashwin16}, see also \cite{RP07}. 
Instead, for linear pairwise coupling, three-body interactions are a distinctive element of second-order phase approximations,
as recently highlighted in \cite{matheny19}. Recognizing the ubiquity of multi-body interactions
may also be important for reconstructing phase interactions from data \cite{kralemann11,*kralemann14}.

Much of our knowledge on nonlinear dynamics relies on minimal models that capture the essential mechanisms behind complex phenomena.
For oscillatory dynamics, the conventional test bed is 
the normal form of the Hopf bifurcation above criticality: the so-called Stuart-Landau oscillator.
Concerning geometry, global coupling is a fruitful simplifying assumption \cite{Win80,Kur84,PikRos15}.
These two ingredients are combined in a standard model of collective dynamics:
the fully connected network of Stuart-Landau oscillators, or the mean-field version of the complex Ginzburg-Landau equation (MF-CGLE) 
\cite{HR92,NK93,NK94,NK95,chabanol97,banaji99,banaji02,DN06,DN07,TGC09,TC13,sethia14,KGO15,RP15,kemeth19,CP19}.
This system is particularly interesting for chaos theory since it
exhibits both microscopic (extensive) and macroscopic (collective) chaos, either 
% coexisting or separately 
{\color{black} combined or independently,}
depending on parameters
\cite{HR92,NK93,NK94,NK95,banaji99,TGC09,KGO15,CP19}.
Phase reduction of the MF-CGLE yields the Kuramoto-Sakaguchi model \cite{kur75,Kur84,NK93}, a first-order approximation that behaves in a 
pathological way (unless heterogeneities are present): it only displays full synchrony or incoherence.
Therefore, pure collective chaos and other phase dynamics of the MF-CGLE remain to be analytically described in terms of a phase model.
Such a phase reduction should provide additional insights into the nature of collective chaos (playing an
analogous role to the Kuramoto-Sivashinsky equation of phase turbulence).

The aim of this paper is two-fold: we introduce a phase-reduction method, 
and we investigate the phase model obtained from the MF-CGLE.
The paper is organized as follows. In Sec. II we reexamine the phase dynamics of
the MF-CGLE and the connection with its first-order phase reduction (the Kuramoto model). 
In section III, we present our systematic phase-reduction procedure,
based on the direct use of isochrons, which delivers a well-controlled
power expansion in the coupling strength parameter.
Section IV is devoted to investigating the weak coupling limit of the MF-CGLE
by means of the the second-order
phase reduction, which unfolds the degeneracies of the Kuramoto model; we address the cases of a large ensemble of oscillators,
as well as an small one of four oscillators. Section V presents the third-order contribution
to the phase reduction of the MF-CGLE. Finally, in Sec.~VI we discuss the implications of our work and some outlooks.

\section{mean-field complex Ginzburg-Landau equation}

The MF-CGLE
consists of $N$ diffusively coupled Stuart-Landau oscillators
governed by $N$ coupled (complex-valued) ordinary differential equations:
\begin{equation}
\dot{A}_j= A_j - ( 1 + i c_2) |A_j|^2 A_j
+\epsilon(1+ic_1) (\bar A- A_j) .
\label{CGLE}
\end{equation}
Here, $A_j=r_j e^{i\varphi_j}$ is a complex variable (index $j$ runs from 1 to $N$),
and the mean field is $\bar A=N^{-1} \sum_{k=1}^{N} A_k$. 
{\color{black} Apart from the population size $N$, there are three free parameters in Eq.~\eqref{CGLE}: $\epsilon$, $c_1$, and $c_2$.
Parameter $\epsilon$, controlling the coupling strength, is positive in order to preserve
the analogy with the (spatially extended) Ginzburg-Landau equation. 
Parameter $c_1$ introduces a cross-coupling between real and imaginary parts of the $A_j$'s. This non-dissipative coupling, so-called `reactive' \cite{PRK01},
generically appears from center manifold reduction \cite{Kur84}.
Finally, `nonisochronicity' (or `shear') parameter $c_2$
in Eq.~\eqref{CGLE} determines the dependence of the angular velocity of one oscillator on its radial coordinate. 
}
% Parameters $\epsilon$, $c_1$ and $c_2$ are real constants, controlling
% the coupling strength, the ``reactivity'' of the coupling, and the nonisochronicity
% of the oscillators, respectively.
% Diffusive coupling imposes $\epsilon>0$. And parameter $c_2$ is hereafter assumed to be nonnegative, since
% the case $c_2<0$ can be mapped to positive $c_2$ transforming 
% $c_1\to -c_1$, $A_j\to A_j^*$ (the asterisk stands for complex conjugation).
There are two important symmetries in system \eqref{CGLE}: invariance under a global phase shift $A_j\to A_j e^{i\phi}$,
and full permutation symmetry stemming from the mean-field coupling.

%%%%%%%%%%%%%%%%%%%%%%%%%%%%%%%%%%%%%%%%%%%%%%%%%%%%%%%%%%%%%%%%%%%%%
\begin{figure}
\begin{minipage}[t]{0.48\linewidth}
\includegraphics[width=\textwidth]{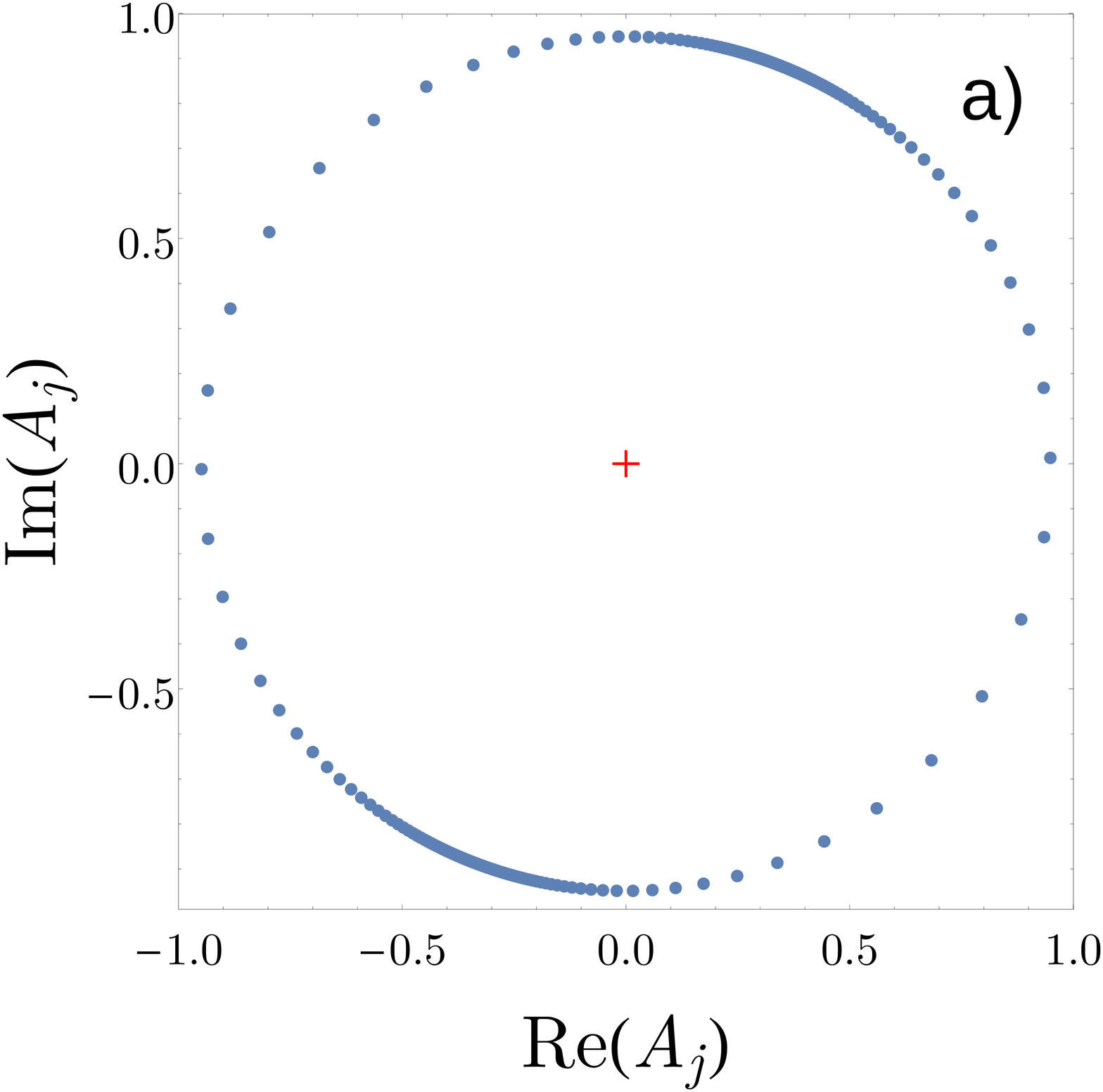}
\end{minipage}
\begin{minipage}[t]{0.48\linewidth}
\includegraphics[width=\textwidth]{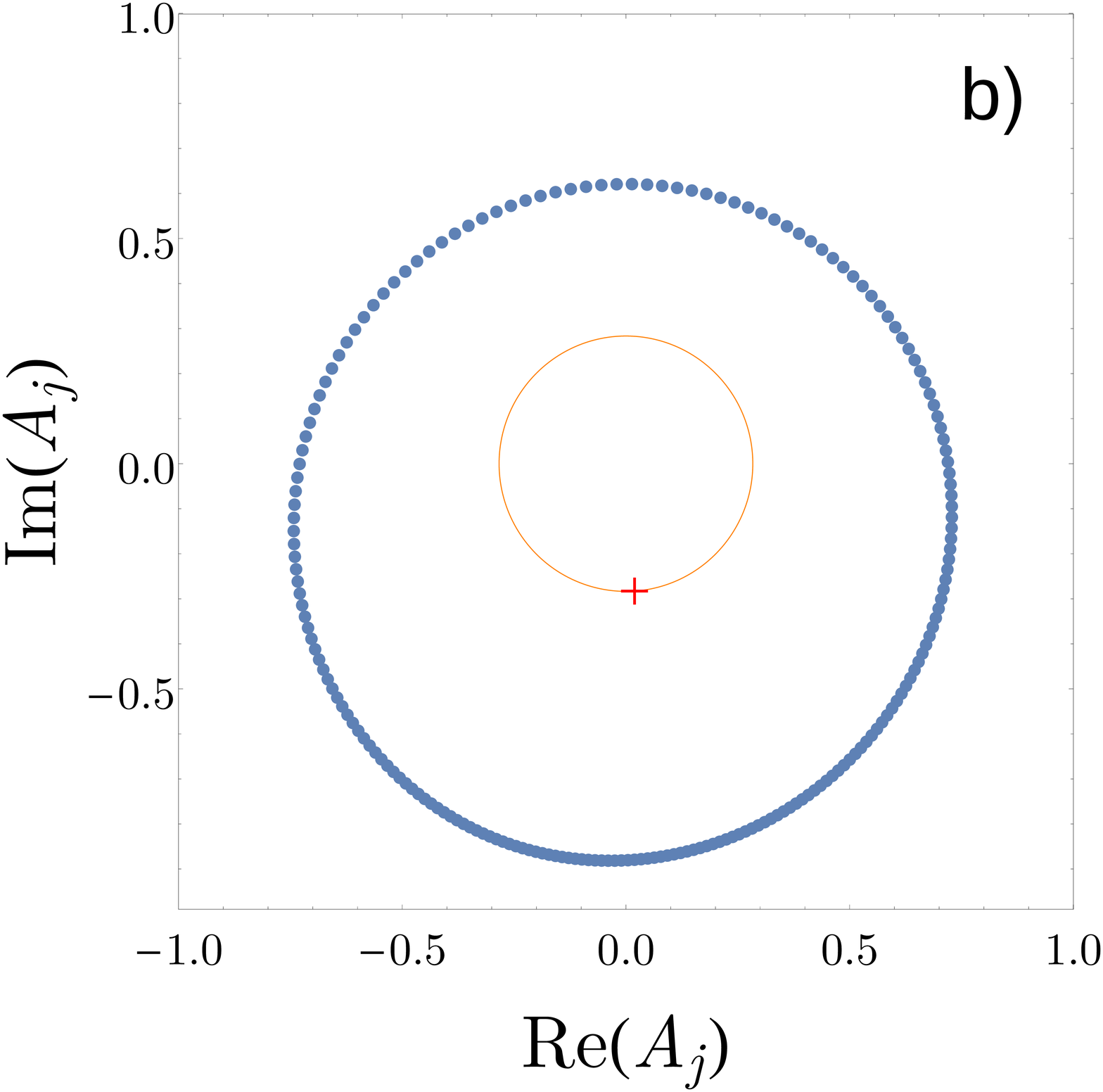}
\end{minipage}
\begin{minipage}[t]{0.48\linewidth}
\includegraphics[width=\textwidth]{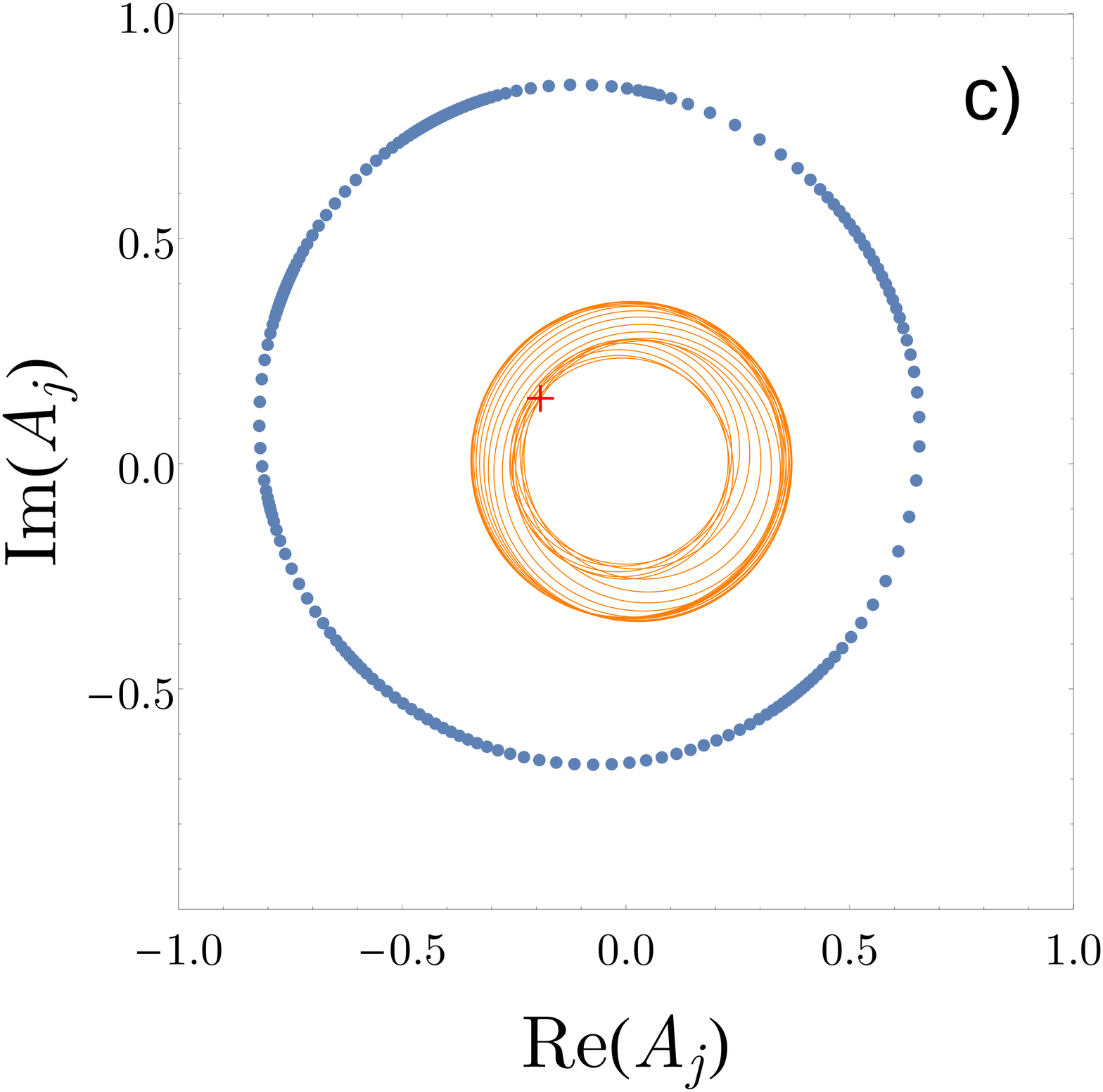}
\end{minipage}
\begin{minipage}[t]{0.48\linewidth}
\includegraphics[width=\textwidth]{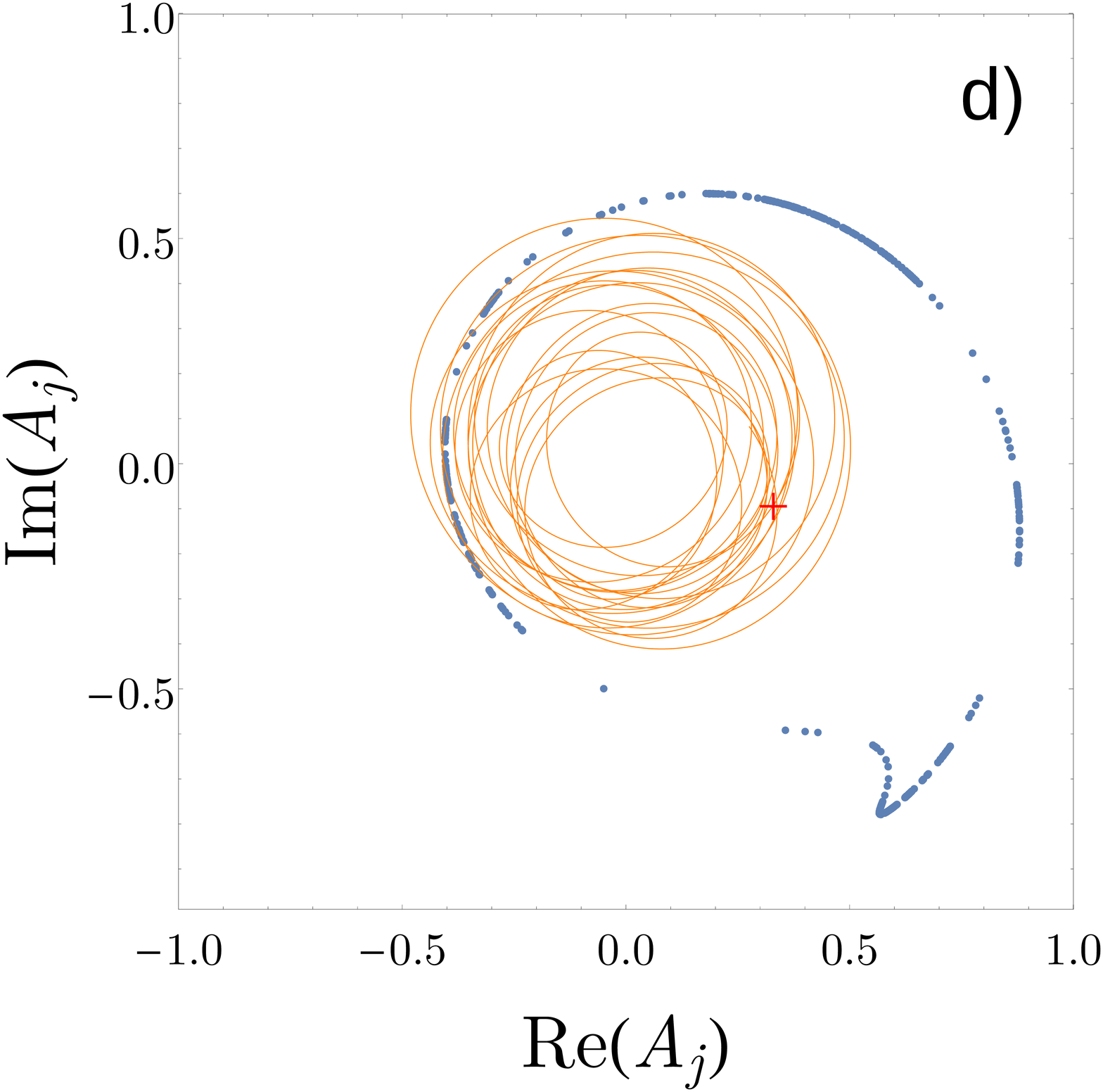}
\end{minipage}
\caption{Snapshot of the positions $A_j$ for a population of $N=200$ oscillators with $c_2=3$. The corresponding mean field $\bar A$ is marked by a red cross, and
a thin solid line is the trajectory of $\bar A(t)$ for an interval of 50 t.u. 
(a) NUIS state with $Q\approx 0.755$ ($c_1=-0.36$, $\epsilon=0.1$), 
(b) Quasiperiodic partial synchrony ($c_1=-2$, $\epsilon=0.4135$). 
(c) Pure collective chaos ($c_1=-2$, $\epsilon=0.4165$).
(d) Collective and microscopic chaos ($c_1=-2$, $\epsilon=0.47$) for $N=500$.}
\label{snapshot}
\end{figure}
%%%%%%%%%%%%%%%%%%%%%%%%%%%%%%%%%%%%%%%%%%%%%%%%%%%%%%%%%%%%%%%%%%

\subsection{Phenomenology}
For many parameter values, the global attractor of Eq.~\eqref{CGLE} is either full synchronization
(FS) $A_j=\bar A =e^{-ic_2 t}$ or one incoherent state with vanishing mean field  $\bar A=0$. In the latter case the oscillators
rotate freely, $A_j=\sqrt{1-\epsilon} \exp\{i[-c_2+\epsilon(c_2-c_1)]t+\phi_j\}$. Among all the states compatible with $\bar A=0$,
the most prominent one is the uniform incoherent state (UIS) in which the $\phi_j$ are uniformly distributed in the thermodynamic limit (for a finite
ensemble, the $\phi_j$ are evenly spaced, deserving the name of splay state or 
 ponies on a merry-go-round state). {\color{black} A continuum of} nonuniform incoherent states (NUISs)
coexists with UIS, but usually 
% any tiny amount of noise 
{\color{black} arbitrarily weak noise}
spreads the phases and UIS is eventually attained.
Nonetheless, for certain parameters values, such as those in Fig.~\ref{snapshot}(a), the UIS is unstable and one NUIS sets in spontaneously \cite{HR92,chabanol97}. 

In addition to FS, UIS, and NUIS, system \eqref{CGLE} exhibits a rich repertoire of collective states including
clustering \cite{NK93,banaji02,DN06,DN07,KGO15,kemeth19},
diffusion-induced inhomogeneity (or chimera) \cite{DN06,DN07,sethia14}, quasiperiodic partial synchronization (QPS) \cite{NK93,CP19},
as well as collective and microscopic chaos \cite{HR92,NK93,NK94,NK95,banaji99,TGC09,KGO15,CP19}.
{\color{black} In a QPS state, see e.g.~Fig.~\ref{snapshot}(b), the mean field $\bar A$ rotates uniformly,  while the individual oscillators behave quasiperiodically (since each
oscillators `feels' the periodic driving of the mean field).
Remarkably, increasing coupling QPS may undergo a couple of secondary Hopf bifurcations resulting in a state of pure collective chaos \cite{NK95,CP19}.
With this term we refer to a state in which the mean field behaves chaotically, while individual oscillators behave in seemingly chaotic-like fashion 
(neighboring oscillators remain close for ever due to the absence of microscopic chaos).
A shared feature of NUIS, QPS and pure collective chaos \cite{NK93,NK95,CP19} is that the
relative positions of the oscillators on top of a closed curve are preserved, see Figs.~\ref{snapshot}(a), \ref{snapshot}(b) and \ref{snapshot}(c). 
This fact suggests that a description in terms of oscillators' phases alone is possible. 
}
% Oscillator ordering on top of a closed curve is preserved in some states, allowing (in principle) a phase description. This is the case of NUIS 
% in Fig.~\ref{snapshot}(a), but also of QPS and pure collective chaos \cite{NK93,NK95,CP19}, 
% shown in Figs.~\ref{snapshot}(b) and \ref{snapshot}(c), respectively. 
% In QPS, the mean field rotates periodically and the individual oscillators behave quasiperiodically. For pure collective chaos,
% the mean field behaves chaotically, while individual oscillators behave in chaotic-like fashion (neighboring
% oscillators remain close for ever due to the absence of microscopic chaos).
In contrast to Fig.~\ref{snapshot}(c), Fig.~\ref{snapshot}(d)
shows a chaotic regime in which phase description breaks down, as it involves microscopic degrees of freedom and no phase ordering exists. 
Hence, our ultimate goal is to find a phase-reduced model of Eq.~\eqref{CGLE} that captures as much as possible of the phase-describable states
(NUIS, QPS, modulated QPS, pure collective chaos,etc.).

\subsection{Basic phase diagrams}

%%%%%%%%%%%%%%%%%%%%%%%%%%%%%%%%%%%%%%%%%%%%%%%%%%%%%%%%%%%%%%%%%%%%%
\begin{figure*}
\begin{minipage}[t]{0.3\linewidth}
\includegraphics[width=\textwidth]{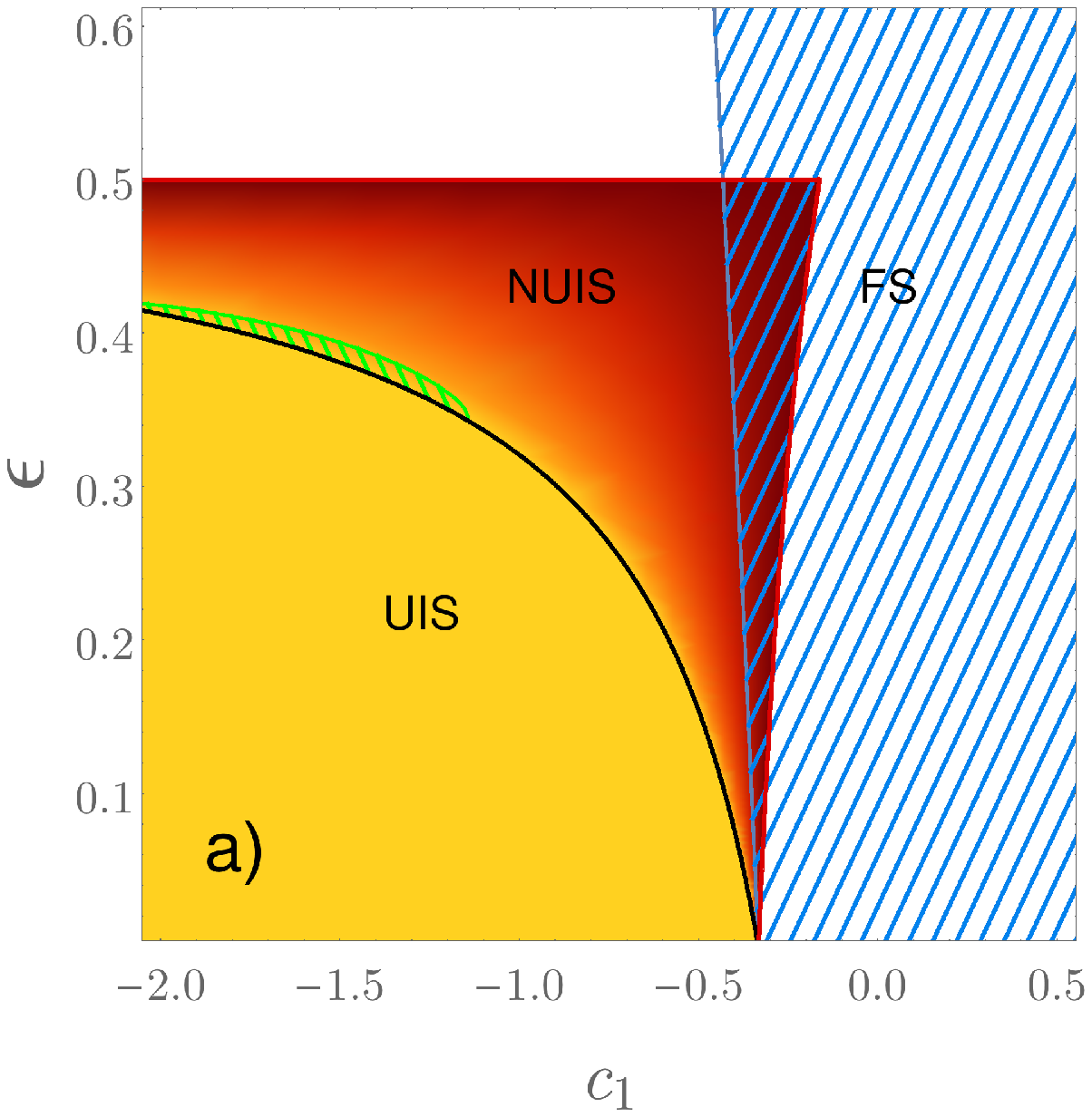}
\end{minipage}
\begin{minipage}[t]{0.3\linewidth}
\includegraphics[width=\textwidth]{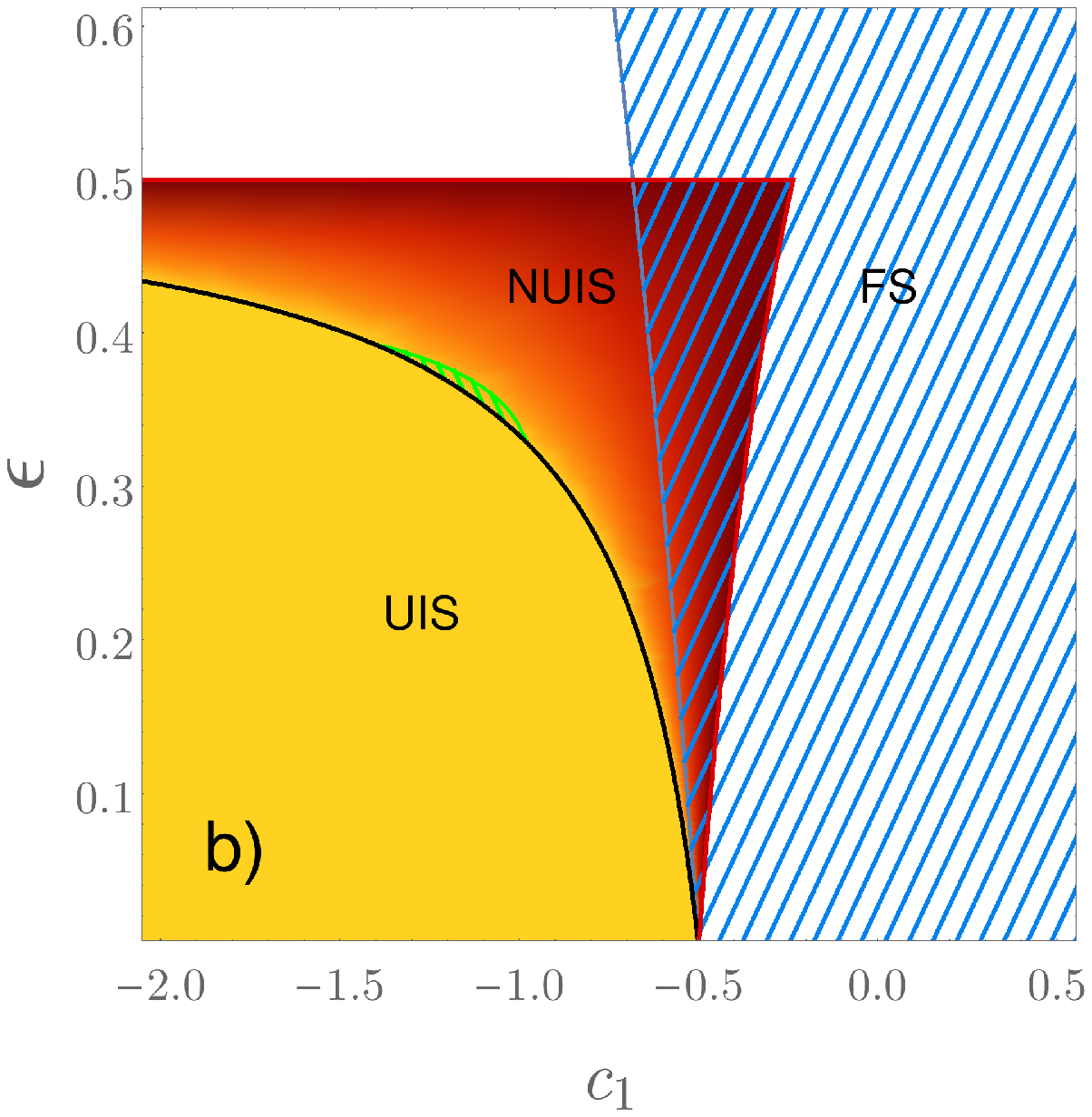}
\end{minipage}
\begin{minipage}[t]{0.3\linewidth}
\includegraphics[width=\textwidth]{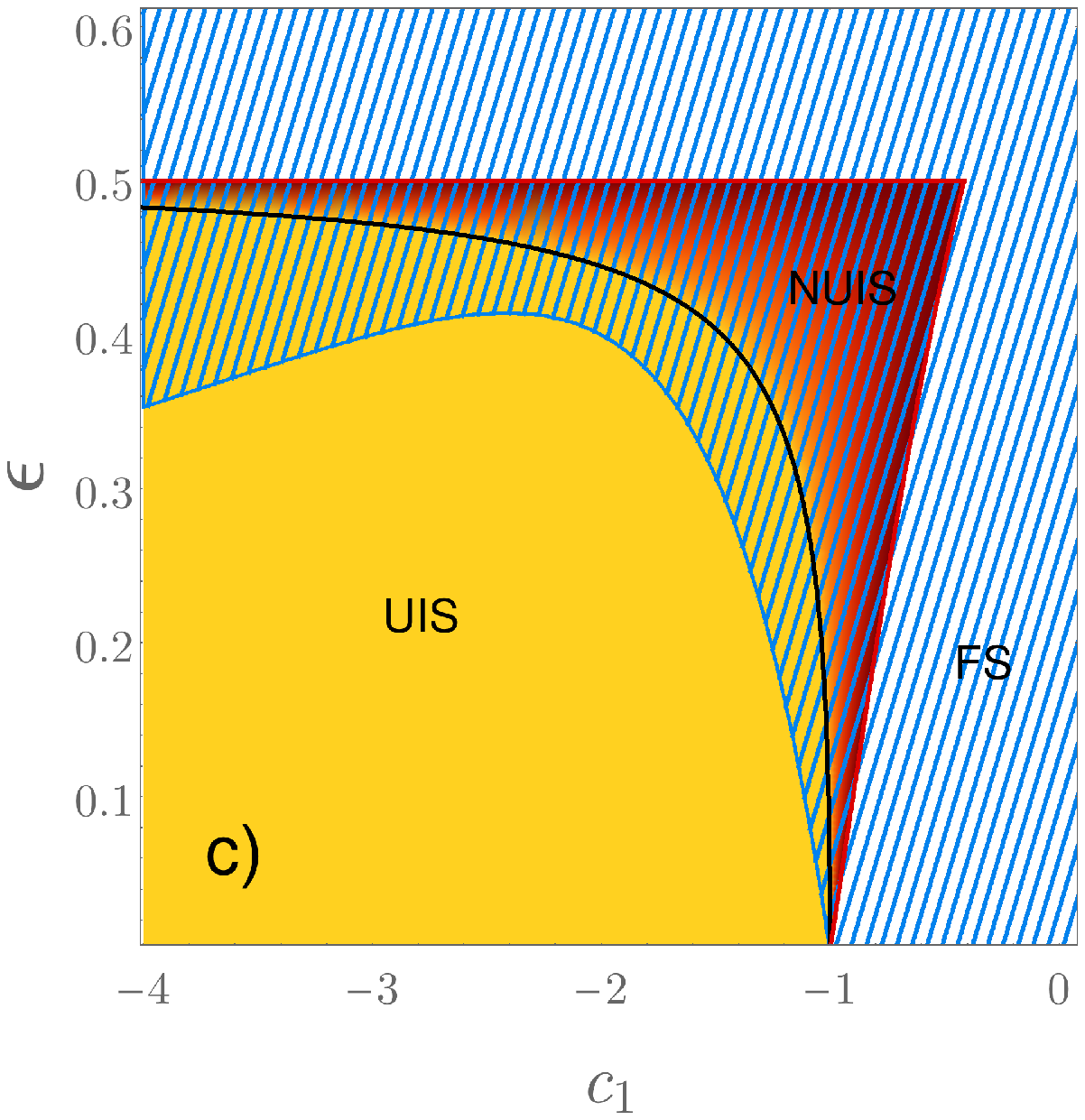}
\end{minipage}
\caption{Partial phase diagram of the MF-CGLE for $c_2=3$ (a), $2$ (b), and $1$ (c). In each panel, the region
with stable UIS is {\color{black} depicted in} yellow, and the region with 
%variable 
color {\color{black} gradation} corresponds to stable NUIS, with a color gradient
that indicates the actual $Q$ value (see text); it becomes darker as $Q\to1$. 
Stable FS is indicated by a blue hatched region. The stability boundaries of FS, UIS and ($Q=1$)-NUIS are depicted by blue, black and red lines,
respectively; following Eqs.~\eqref{FS}, \eqref{UIS},  and \eqref{NUIS} (setting $Q=1$).
In panels (a) and (b), there is  green-hatched region where 
other phase-describable states like the ones shown in Figs.~\ref{snapshot}(b) and \ref{snapshot}(c) are stable. 
}
\label{phase_diag}
\end{figure*}
%%%%%%%%%%%%%%%%%%%%%%%%%%%%%%%%%%%%%%%%%%%%%%%%%%%%%%%%%%%%%%%%%%

Before presenting our results it is convenient to review previous results on the MF-CGLE.
{\color{black} For fixed $c_1$ and $c_2$ values,} let us denote by $\epsilon_s$ and $\epsilon_0$, {\color{black} the $\epsilon$ values of marginal linear stability for FS and UIS.
Closed formulas for $\epsilon_s$ and $\epsilon_0$ are \cite{HR92,NK93}:}
% the stability boundaries of full synchrony  and UIS. 
% The analytic formulas are \cite{HR92,NK93}:
\begin{eqnarray}
\epsilon_s=-\frac{2(1+c_1c_2)}{1+c_1^2} ,\label{FS} \\
 \epsilon_0(2\epsilon_0-1)c_1^2+4(\epsilon_0-1)(2\epsilon_0-1)c_1 c_2 \nonumber \\- \epsilon_0(\epsilon_0-1)c_2^2+(3\epsilon_0-2)^2=0 . \label{UIS}
\end{eqnarray}
{\color{black} These formulas are also valid for finite ensembles, assuming $\epsilon_0$ refers to the splay state.}
% \$\epsilon_0$ is also valid for the splay state of a finite ensemble.
% To visualize these boundaries, in Fig.~\ref{phase_diag} we  plot the loci of $\epsilon_s$ and $\epsilon_0$ in the parameter plane $(c_1,\epsilon)$,
% for three different values of $c_2=1,2,3$.
{\color{black} To visualize the stability boundaries in Eqs.~\eqref{FS} and \eqref{UIS}, it is convenient to fix either $c_1$ or $c_2$. Following \cite{NK93} we choose to fix $c_2$,
and display the loci of $\epsilon_s$ and $\epsilon_0$ in the parameter plane $(c_1,\epsilon)$.
In the phase diagrams in Figs.~\ref{phase_diag}(a) \ref{phase_diag}(b) and \ref{phase_diag}(c) we selected $c_2=3$, $c_2=2$ and $c_2=1$, respectively.
This choice is motivated by the fact that most previous works on the MF-CGLE adopt either $c_2=2$ or $c_2=3$.}
One key observation is that, as $\epsilon_s$ and $\epsilon_0$ approach zero, the boundaries converge to the condition $1+c_1c_2=0$, which is
the well known Benjamin-Feir-Newell criterion for the stability of uniform oscillations in the 
%CGLE 
{\color{black} complex Ginzburg-Landau equation}
in arbitrary dimension \cite{Kur84,Walgraef,PRK01,AK02}.
There is a critical value $c_2=\sqrt{3}=1.732\ldots$ at which the boundaries $\epsilon_s$ and $\epsilon_0$ become tangent at $\epsilon=0$.
Accordingly, for $c_2=1$, see Fig.~\ref{phase_diag}(c), there is a region of bistability between UIS and synchrony,
in contrast {\color{black} e.g.}~to Fig.~\ref{phase_diag}(b).

The stability of a NUIS depends exclusively on the 
% real parameter 
{\color{black} mean field}
$Q=|N^{-1}\sum_j \exp(2i\varphi_j)|$. 
% The locus of each stability boundary $\epsilon_Q$ was obtained in Ref.~\cite{chabanol97}:
{\color{black} The coupling constant $\epsilon_Q$ at which one particular NUIS becomes unstable was obtained in Ref.~\cite{chabanol97}:}
\begin{eqnarray}
\left[\epsilon_Q(2\epsilon_Q-1)c_1^2+4(\epsilon_Q-1)(2\epsilon_Q-1)c_1c_2\right. \nonumber\\
-\epsilon_Q(\epsilon_Q-1)c_2^2 \left.+(3\epsilon_Q-2)^2\right][(2-3\epsilon_Q)^2+\epsilon_Q^2c_1^2] \label{NUIS}\\
=Q^2\epsilon_Q(1-\epsilon_Q)(3\epsilon_Q-2)^2 (c_1^2+1)(c_2^2+1)  .\nonumber
\end{eqnarray}
This formula is the generalization of \eqref{UIS} with the important qualitative information that 
the size of the stability region increases as $Q$ grows, reaching its maximum for $Q=1$. At $Q=1$ the NUIS collapses into a two-cluster state with equally 
populated groups.
The value of $Q$ is still far from breaking the degeneracy of a NUIS, provided $Q\ne1$, since the values of all 
% higher-order parameters
{\color{black} `higher-order' mean fields}
$f_n=|N^{-1}\sum_j \exp( n i\varphi_j)|$ ($n>2$)
are free. Nevertheless, the conclusion based on numerical simulation is that any small amount of noise causes $f_n$ to converge to zero, and $Q$ to take the smallest value among all
 non-unstable  {\color{black}(i.e.~neutrally stable})
NUISs. 
Therefore, it is assumed hereafter that the term NUIS is constrained to $f_n=0$ ($n>2$).

Figures~\ref{phase_diag} (a) and (b) include a green-hatched region, adjacent to the UIS region at moderate $\epsilon$ values,
where other phase-describable states are stable. These are QPS, modulated QPS and pure-collective chaos \cite{NK95,CP19}.
We determined the boundary through simulations with $N=200$ oscillators, but the result is
insensitive if a larger $N$ value is used.

\subsection{{\color{black} First-}order phase reduction: Kuramoto-Sakaguchi model}

At the lowest order, applying the classical averaging technique \cite{kur75,Kur84,PRK01} to Eq.~\eqref{CGLE}
yields the Kuramoto-Sakaguchi model \cite{SK86}.
In this model, each oscillator is described by a phase 
 $\theta_j$, and it is coupled to the other ones 
by pairwise interactions of the form $\sin(\theta_i-\theta_j+\alpha)$. In agreement with the mean-field character of the system, 
oscillators are coupled through the Kuramoto order parameter $Z_1\equiv R \, e^{i\Psi}= N^{-1} \sum_{k=1}^N e^{i\theta_k}$,
such that the ordinary differential equations governing the dynamics are:
\begin{equation}
\dot{\theta}_j=\Omega+
\epsilon \eta \,  R  \, \sin(\Psi-\theta_j+\alpha) ,
\label{kuramoto}
\end{equation}
with constants  $\Omega\equiv-c_2+\epsilon(c_2-c_1)$, $\eta\equiv\sqrt{(1+c_2^2)(1+c_1^2)}$, 
and phase lag 
\begin{equation}
\alpha=\arg[1+c_1 c_2+(c_1-c_2)i] .
\end{equation}
Equation~\eqref{kuramoto} is the disorder-free
version of the paradigmatic Kuramoto-Sakaguchi model \cite{Kur84,SK86} and related models \cite{MP11p,*MP11,*PM11}.
The dynamics of Eq.~\eqref{kuramoto} is determined
by the sign of $1+c_1 c_2$ (Benjamin-Feir-Newell criterion):
full synchrony ---corresponding to $R=1$--- is stable for $1+c_1 c_2>0$, and unstable for $1+c_1 c_2<0$. In the latter case,
among infinitely many oscillator densities with $R=0$, there is a convergence to
the UIS under {\color{black} an arbitrarily weak noise}
%a tiny amount of noise
\cite{NK93}.

As discussed above, the MF-CGLE has much richer dynamics than its first-order {\color{black}phase} reduction \eqref{kuramoto}, even arbitrarily close
to the $\epsilon=0$ limit. 
Therefore, it is mandatory to extend the phase reduction to order $O(\epsilon^2)$ if we wish to
avoid degeneracies in the phase approximation. This is what we do next.

\section{Systematic phase reduction}
\label{spr}
In spite of the relevance of Eq.~\eqref{CGLE} no phase reduction beyond the first order is currently available. Finding higher order terms 
in the phase reduction is necessary to unfold the singularity at $(c_1,\epsilon)=(-1/c_2,0)$, see Fig.~\ref{phase_diag}. 
This path of investigation should allow us to discern which are the true behaviors of the MF-CGLE in the small coupling limit $|\epsilon|\ll1$.
Moreover, it might serve to shed light on the mechanisms behind complex dynamics found (so far) for 
% not-so-small 
{\color{black} moderate}
$\epsilon$ values.

An isochron-based phase reduction approach is developed here. It allowed us to obtain the phase reduction of the MF-CGLE up to order $\epsilon^3$.
In this section we give the details of our phase reduction calculation. 
{\color{black} We anticipate that the results at second and third order in $\epsilon$ correspond to Eqs.~\eqref{o2} and \eqref{o3} below.}
% The reader not interested in this 
% is invited to jump to Eq.~\eqref{o2} and Sec.~\ref{sec:2nd}, where the
% contribution of order $\epsilon^2$ is analyzed in depth; in addition, the solution at order $\epsilon^3$ 
% can be found in Sec.~\ref{sec:3rd}.

\subsection{Isochrons}

The concept of isochron {\color{black} \cite{win74,guc75}} is the cornerstone of phase reduction methods \cite{Win80,Kur84}. Isochrons foliate the attraction basin of a stable limit cycle,  
each intersecting it at one point. The phase of that point is attributed to all points of the isochron,
motivated by their {\color{black} convergence} as time goes to infinity (the so-called `asymptotic phase' \cite{CL}). 
For the Stuart-Landau oscillator, polar coordinates $(r,\varphi)$ relate to the phase  $\theta$  according to \cite{Kur84,PRK01}:
\begin{equation}
\theta(r,\varphi)=\varphi-c_2 \ln r .
\label{iso}
\end{equation}
As mentioned above, on the limit cycle ($r=1$), $\theta=\varphi$. The term ``nonisochronicity'' or ``shear'' for parameter $c_2$ 
becomes clear in light of Eq.~\eqref{iso}, since $c_2$ controls how much the isochrons deviate from radial lines.

\subsection{Isochron-based phase reduction}

We continue the analysis writing Eq.~\eqref{CGLE} 
in polar coordinates:
\begin{eqnarray}
 \dot{r_j}=r_j(1-\epsilon-r_j^2) \nonumber\\+\frac{\epsilon}{N}\sum_{k=1}^{N}r_k\bigg[\cos(\varphi_k-\varphi_j)-c_1\sin(\varphi_k-\varphi_j)  \bigg] , \label{GLr} \\
\dot{\varphi_j}=-c_2 r_j^2-\epsilon c_1 \nonumber \\+\frac{\epsilon}{N r_j}\sum_{k=1}^{N}r_k\bigg[c_1 \cos(\varphi_k-\varphi_j)+\sin(\varphi_k-\varphi_j)  \bigg] .\label{GLtheta}
\end{eqnarray} 
After the change of variables $(r_j,\varphi_j)\to(r_j,\theta_j)$ {\color{black} through Eq.~\eqref{iso},} we get:
\begin{subequations}
	\label{eqsimples}
	\begin{eqnarray}
	\dot r_j=f(r_j)+\epsilon g_j(\mathbf{r},\boldsymbol{\theta}) ,\label{eqsimplea}\\
	\dot \theta_j=\epsilon h_j(\mathbf{r},\boldsymbol{\theta}) . \label{eqsimpleb}
	\end{eqnarray}
\end{subequations}
{\color{black} Here, we have also implemented the transformation $\theta_j\to \theta_j-c_2 t$
(by moving to a rotating frame with angular velocity $-c_2$). 
In this way, the time derivatives of the phases in \eqref{eqsimpleb} are proportional to $\epsilon$,
while the $r_j$ are fast variables that become enslaved to the dynamics of $\theta_j$.
}
In 
%this equation 
{\color{black} Eq.~\eqref{eqsimples}}
$f(r)=r(1-r^2)$, 
and functions $g_j$ and $h_j$ depend on the vectors $\mathbf{r}=(r_1,r_2,\ldots,r_N)^T$ and
$\boldsymbol{\theta}=(\theta_1,\theta_2,\ldots,\theta_N)^T$
{\color{black} as follows,}
%. The exact formulas of $g_j$ and $h_j$ are
\begin{widetext}
	\begin{subequations}
	\begin{eqnarray}
	g_j(\mathbf{r},\boldsymbol{\theta})&=&-r_j+\frac{1}{N}\sum_{k=1}^{N} \left\{r_k\bigg[\cos\big(\theta_k-\theta_j+c_2\ln\tfrac{r_k}{r_j}\big)-c_1\sin\big(\theta_k-\theta_j+c_2\ln\tfrac{r_k}{r_j}\big)  \bigg] \right\},\\
	h_j(\mathbf{r},\boldsymbol{\theta})&=&c_2-c_1
	+\frac{1}{N r_j}\sum_{k=1}^{N}\left\{r_k\bigg[(c_1-c_2) \cos\big(\theta_k-\theta_j+c_2\ln \tfrac{r_k}{r_j}\big)
	+(1+c_1c_2)\sin\big(\theta_k-\theta_j+c_2\ln\tfrac{r_k}{r_j}\big)  \bigg] \right\} .
	\end{eqnarray}
	\end{subequations}
\end{widetext}
% To obtain Eq.~\eqref{eqsimples} we have implemented the transformation $\theta_j\to \theta_j-c_2 t$
% (moving to a rotating frame with angular velocity $-c_2$); constant $-c_2$ is inserted back at the end of the calculation.
% In this way, the time derivatives of the phases in \eqref{eqsimpleb} are proportional to $\epsilon$,
% while the $r_j$ are fast variables that become enslaved to the dynamics of $\theta_j$.
{\color{black} The separation of time scales in Eq.~\eqref{eqsimples}}
% This 
suggests {\color{black}using} classical perturbation techniques like averaging, adiabatic approximation, or two-timing. 
However, the perturbation approach described next proved to be both conceptually simple and much less convoluted, permitting us to
obtain the phase reduction up to cubic order in $\epsilon$.
{\color{black} Based on the empirical observation that, at small $\epsilon$ values, the oscillators fall on a closed curve and preserve their phase ordering,}
 we 
%  simply 
 assume that the radii are completely determined by the phases $r_j=r_j(\theta_1,\theta_2,\ldots,\theta_N)$.
We also postulate  an expansion in powers of $\epsilon$ for the radii: $r=r_{j}^{(0)}+\epsilon r_{j}^{(1)}+\epsilon^2r_{j}^{(2)}+\cdots$;
or in vector notation $\mathbf{r}=\mathbf{r}^{(0)}+\epsilon \mathbf{r}^{(1)}+\epsilon^2 \mathbf{r}^{(2)}+\cdots$.
% The equation for each $\theta_j$ is:
{\color{black} Equation \eqref{eqsimpleb} for $\theta_j$ becomes:}
\begin{eqnarray}
\dot\theta_j=\epsilon h_j(\mathbf{r}^{(0)},\boldsymbol{\theta})+\epsilon^2\left(\boldsymbol{\nabla}_r h_j(\mathbf{r}^{(0)},\boldsymbol{\theta})\right)\cdot {\mathbf{r}}^{(1)} \nonumber\\ +
\epsilon^3\left[\left(\boldsymbol{\nabla}_r h_j(\mathbf{r}^{(0)},\boldsymbol{\theta})\right)\cdot {\mathbf{r}}^{(2)} +  (\mathbf{M r r})_j\right] + \cdots ,
\label{protom}
\end{eqnarray}
where  $\boldsymbol{\nabla}_r\equiv (\partial_{r_1},\partial_{r_2},\ldots,\partial_{r_N})$ and 
$(\mathbf{M r r})_j\equiv \tfrac{1}{2!} \sum_{k,l} \partial_{r_k} \partial_{r_l} h_j(\mathbf{r}^{(0)},\boldsymbol{\theta}) r_k^{(1)} r_l^{(1)}$.
Now, the explicit dependence on the radii in \eqref{protom} must be removed. This is accomplished equating both sides of \eqref{eqsimplea} at the same order.
The order $O(\epsilon^0)$ yields $r_j^{(0)}=1$, and \eqref{protom} becomes {\color{black} (at the lowest order) the} Kuramoto-Sakaguchi model \eqref{kuramoto}.
At order $\epsilon$:
\begin{equation}
\dot r_j^{(1)}=f'(r_j^{(0)}) r_j^{(1)}+  g_j(\mathbf{r}^{(0)},\boldsymbol{\theta}) .
\label{rd1}
\end{equation}
{\color{black} As $r_j$ depends exclusively on the phases, we can apply the chain rule:}
% $\dot r_j^{(1)}$ can be also obtained applying the chain rule, we have that 
$\dot 
r_j=(\boldsymbol{\nabla}_\theta r_j)\cdot \dot{\boldsymbol{\theta}}
$. 
% and 
At order $\epsilon${\color{black}, the time derivative vanishes}:
$$
\dot r_j^{(1)}=(\boldsymbol{\nabla}_\theta r_j^{(0)})\cdot {\mathbf{h}}=0 .
$$
Hence Eq.~\eqref{rd1} yields the result
\begin{equation}
r_j^{(1)}=-\frac{g_j(\mathbf{r}^{(0)},\boldsymbol{\theta})}{f'(r_j^{(0)})}=\frac{g_j(\mathbf{r}^{(0)},\boldsymbol{\theta})}{2} ,
\end{equation}
which can be inserted in \eqref{protom} to obtain the $\epsilon^2$ contribution.
%After some algebra 
{\color{black} Through elementary manipulations}
% we obtained 
the second-order phase reduction {\color{black} of Eq.~\eqref{CGLE}} {\color{black} can be condensed into this expression}:
\begin{widetext}
 \begin{eqnarray} \label{o2}
\dot{\theta}_j=\Omega+
\epsilon \eta \,  R  \, \sin(\Psi-\theta_j+\alpha) +
\frac{\epsilon^2 \eta^2}{4} 
\left\{R \, Q \sin(\Phi-\Psi-\theta_j) 
- \sum_{m=1}^2 (-R)^m \sin[m(\Psi-\theta_j)+\beta] \right\} + O(\epsilon^3) .
\end{eqnarray}
\end{widetext}
The $O(\epsilon^2)$ term depends
on $Z_1$ as well as on the second Kuramoto-Daido order parameter \cite{Dai96} $Z_2
\equiv Q \, e^{i\Phi}= N^{-1} \sum_{k=1}^N e^{2 i \theta_k}$. 
{\color{black} To enhance the clarity of Eq.~\eqref{o2}, we found it convenient to define a} %The 
phase lag 
%$\beta$ 
% in \eqref{o2} is independent of $c_2$: 
\begin{equation}
\beta= \arg(1-c_1^2+2 c_1 i)  ,
\end{equation}
{\color{black}  which turns out to be independent of $c_2$.} 
The other constants in Eq.~\eqref{o2} are the same as in 
Eq.~\eqref{kuramoto}{\color{black}; as the change to a rotating frame has been reversed, the
$O(\epsilon^0)$ term inside $\Omega$ is $-c_2$ (as before)}.

% \subsection{Relationship with other approaches}
% 
% In the next lines we argue why we believe our phase-reduction method is both conceptually simple and effective.
% 
% Applying the systematic averaging formulation in Chap.~4 of the book by Kuramoto \cite{Kur84},
% we were able to derive the same result than in \eqref{o2}.
% This calculation was, however, much more lengthy than the one schematized above. In fact 
% going to the next order in $\epsilon$ with the averaging approach \cite{Kur84}
% was a task of unsurmountable complexity while our method allowed us to succeed--- see the result in Sec.~\ref{sec:3rd} ---with symbolic software.
% 
% The same result to order $\epsilon^2$, can be also obtained assuming small variations of the radii, i.e.~setting $\dot r_j=0$. This procedure was
% followed in \cite{pikovsky06,matheny19}, with the difference that in those references the small quantity are deviations from the limit cycle,
% while here we pursued a bona-fide power expansion in terms of the coupling parameter.
% Finally, we mention that the two-timing approximation, such that the $r_j$ depend only on a slow time $\tau=\epsilon t$, can be also applied.

\section{Second-order phase reduction: three-body interactions}
\label{sec:2nd}

In this section we study in detail the phase model obtained from the second-order phase reduction of the MF-CGLE, i.e.~the 
% phase model 
system {\color{black} of phase oscillators governed} 
% defined 
by Eq.~\eqref{o2}.
Of the three $O(\epsilon^2)$ contributions to Eq.~\eqref{o2}, the first element of the sum ($m=1$)
% is simply 
{\color{black} entails}
a parameter shift to the $O(\epsilon)$ interaction, and it is therefore irrelevant in qualitative terms. 
% Remarkably, 
The other two terms in Eq.~\eqref{o2} correspond to three-body (i.e.~non-pairwise) interactions:
 \begin{equation}
R^2 \sin[2(\Psi-\theta_j)+\beta]=\frac{1}{N^2} % \sum_{k,l=1}^N
\sum_{k,l} \sin(\theta_k+\theta_l-2\theta_j+\beta) 
 \end{equation}
 \begin{equation}
R \, Q \sin(\Phi-\Psi-\theta_j)=\frac{1}{N^2} % \sum_{k,l=1}^N  
\sum_{k,l}  \sin(2\theta_k-\theta_l-\theta_j) 
 \end{equation}
The price of working only with the phases is that two-body interactions of the original MF-CGLE \eqref{CGLE}
become multi-body interactions, as higher orders of $\epsilon$ are considered.
In comparison to Eq.~\eqref{CGLE} our phase model 
can be much more efficiently analyzed, both analytically and numerically. We devote the remainder of this section
to analyze the phase model in Eq.~\eqref{o2}. We note that, as expected, the model
is invariant under global phase shift $\theta_j\to\theta_j+\phi$. For the sake of making the presentation
simpler we assume constant $\Omega=0$, since this can always be achieved by going to a rotating 
frame $\theta_j\to\theta_j+\Omega t$.

\subsection{Full synchronization}

\begin{figure}
 \begin{minipage}{0.48\linewidth}
  \includegraphics[width=\textwidth]{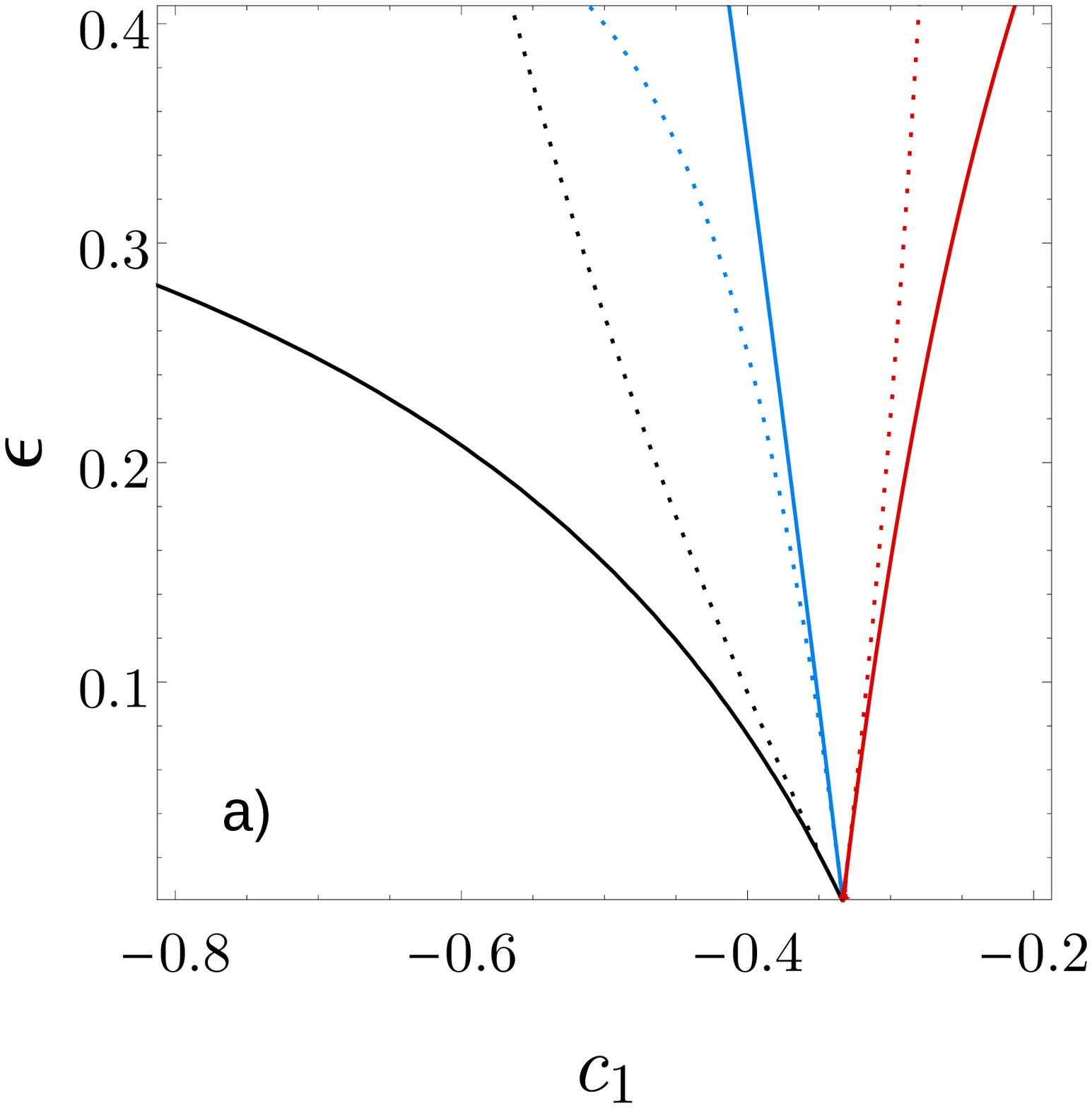}
  \end{minipage}
 \begin{minipage}{0.48\linewidth}
  \includegraphics[width=\textwidth]{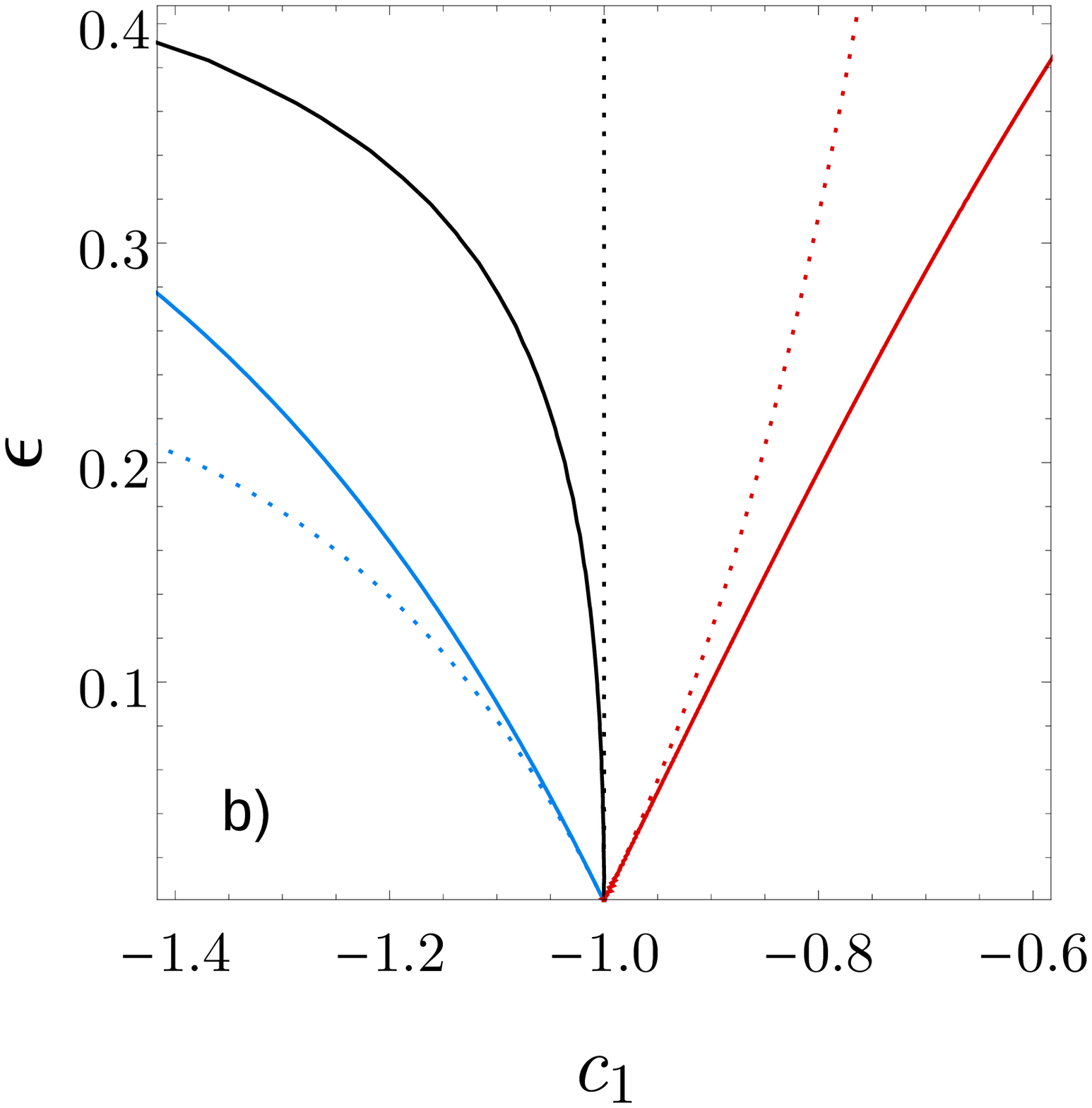}
  \end{minipage}
\caption {Comparison between {\color{black} the} bifurcation lines of FS, UIS and NUIS ($Q=1$)
for the MF-CGLE (solid lines) and for the second-order phase reduction (dashed lines). Line colors are the same as in Fig.~\ref{phase_diag}.
Panel (a) $c_2=3$, and panel (b) $c_2=1$.
}
\label{fig_accu}
\end{figure}

The stability boundary of FS ($\theta_j=\Psi=\Phi/2$) is easily calculated. 
In particular, for infinite $N$ it is almost immediate: we simply assume one oscillator is infinitesimally perturbed, say the first one, $\theta_1=\Psi+\delta\theta_1$. 
The evolution of the perturbation obeys the linear equation
$d\delta\theta_1/dt= \epsilon\eta \left[ \cos\alpha+ \frac{\epsilon\eta}{4}(1-\cos\beta)\right] \delta\theta_1$.
At threshold ($d\delta\theta_1/dt=0$) the coupling 
satisfies:
\begin{equation}
\epsilon_s=\frac{-2(1+c_1c_2)}{c_1^2 (1+c_2^2)}
\label{fs}
\end{equation}
where we have written $\cos\alpha$ and $\cos\beta$ in terms of $c_1$ and $c_2$.
% Equation~\eqref{fs} also holds for the splay state of a finite population
% (the proof is straightforward and we skip it).
{\color{black} For illustration, the curve defined by \eqref{fs} is represented by a blue dotted line in Figs.~\ref{fig_accu}(a) and (b) 
for $c_2=3$ and $c_1=1$, respectively. Equation \eqref{fs} is asymptotically exact as $\epsilon_s\to0$, and deviates progressively
from the FS boundary of the MF-CGLE (represented by a solid line) as $\epsilon_s$ increases.}

% %%%%%%%%%%%%%%%%%%%%%%%%%%%%%%%%%%%%%%%%%%%%%%%%%%%%%%%%%%%%%%%%%%%%%
% %%%%%%%%%%%%%%%%%%%%%%%%%%%%%%%%%%%%%%%%%%%%%%%%%%%%%%%%%%%%%%%%%%

\subsection{Incoherent states}

We adopt the thermodynamic limit and define a density $\rho$ such that
$\rho(\theta,t) d\theta$ is the fraction of oscillators with phases between $\theta$ and $\theta +d\theta$.
Now 
% $\theta$
{\color{black}$\theta\in[0,2\pi)$}
is a cyclic variable, i.e. $\rho(\theta+2\pi,t)=\rho(\theta,t)$, and we impose the normalization condition
$\int_{0}^{2\pi} \rho(\theta,t) d\theta =1$.
The oscillator density $\rho$ obeys the continuity equation because of the conservation of the number of oscillators:
\begin{equation}
\partial_t  \rho(\theta,t) + \partial_\theta[ v(\theta) \rho(\theta,t)] =0 .%D \partial_{\theta}^2 \rho(\theta,t)
\label{fp}
\end{equation}
Here $v=\dot\theta$ is the $\rho$-dependent velocity of an oscillator with phase $\theta$.
We define the Fourier modes of $\rho$:
\begin{equation}
\rho(\theta,t)=\frac{1}{2\pi} \sum_{n=-\infty}^\infty \rho_n e^{i n \theta} ,
\label{fourier}
\end{equation}
with $\rho_0=1$ and $\rho_n=\rho_{-n}^*$.
The mean fields {\color{black} $Z_n$} reduce to 
\[
Z_n=\int_{0}^{2\pi} \rho(\theta,t) \, e^{in\theta} \, d\theta =\rho_{-n} .
\]
Inserting the Fourier expansion \eqref{fourier} into the continuity equation \eqref{fp} allows us to rewrite our model in Fourier space:
\begin{widetext}
 \begin{equation}
\dot\rho_n =  
\frac{n}{2}\epsilon \eta\left\{e^{-i\alpha}\rho_1 \rho_{n-1}-e^{i\alpha}\rho_{1}^* \rho_{n+1}
+ \frac{\epsilon\eta}{4}\left[e^{-i\beta}\rho_1( \rho_{n-1}-\rho_1  \rho_{n-2})- 
e^{i\beta} \rho_{1}^*( \rho_{n+1}-\rho_{1}^*\rho_{n+2})-\rho_{2}^* \rho_1 \rho_{n+1}+\rho_2 \rho_{1}^*\rho_{n-1}\right]\right\}
\label{odes_modes}
\end{equation}
\end{widetext}

\subsubsection{Uniform incoherent state}
The stability boundary of the UIS ($\rho(\theta)=(2\pi)^{-1} \Leftrightarrow \rho_{n\ne0}=0$) is obtained
linearizing the previous equation. It {\color{black}is } easy to notice that only the first mode may destabilize.
We have for $|\rho_1|\ll1$:
\begin{equation}
\frac{d}{dt}\delta\rho_1 =  
\frac{\epsilon \eta }{2}\left[e^{-i\alpha}+\frac{\epsilon \eta e^{-i\beta}}{4} \right] \delta\rho_1  .
 %- D \, n^2   \rho_n
% &+&e^{i\beta} \rho_{-1}(\rho_{n+1}-\rho_{-1}\rho_{n+2})-e^{-i\beta} \rho_{1} (\rho_{n-1} - \rho_1  \rho_{n-2})] \nonumber
\label{ode_uis}
\end{equation}
Neglecting the trivial marginal case $\epsilon=0$, the stability boundary satisfies $\cos\alpha+(1/4)\epsilon_0 \eta\cos\beta=0$. Or in terms of $c_1$ and $c_2$:
\begin{equation}
\epsilon_0=\frac{4(1+c_1c_2)}{(c_1^2-1)(1+c_2^2)} .
\label{uis}
\end{equation}
% This expression differs from the FS boundary \eqref{fs}, evidencing the relevance of the second-order term in the power expansion. 
{\color{black} In Figs.~\ref{fig_accu}(a) and \ref{fig_accu}(b), we can contrast this formula with the exact one for the MF-CGLE, Eq.~\eqref{UIS},
for two $c_2$ values.}

\subsubsection{Nonuniform incoherent states}

According to \eqref{odes_modes}, in an incoherent state ($\rho_1=0$) higher-order modes are at rest:
$\dot\rho_n=0$ ($n>2$).
The linearization of \eqref{odes_modes} around $\rho_1=0$ and $\rho_n\ne0$ ($|n|\ge2$)
is (schematically) as follows:
\begin{equation}
%\frac{d}{dt} \begin{pmatrix}\delta \rho_1 \\ \delta\rho_2 \\ \delta\rho_3 \\ \delta\rho_4 \\  \vdots\end{pmatrix} =
% \left(
%\begin{array}{c|cccc}
%  \bullet & 0& 0& 0& \cdots  \\
% \hline
% \bullet & 0 & 0& 0& \cdots \\ 
% \bullet & 0 &0 & 0& \cdots \\
%  \bullet & 0 &0 & 0& \cdots \\
%  \vdots & \vdots &\vdots &\vdots & \ddots  \\
%  \end{array}
%  \right) ,
\frac{d}{dt} \begin{pmatrix}\delta \rho_1 \\ \delta\rho_1^* \\ \delta\rho_2 \\ \delta\rho_2^* \\  \vdots\end{pmatrix} =
\left(
\begin{array}{cc|ccc}
\bullet & \bullet& 0& 0& \cdots  \\
\bullet & \bullet & 0& 0& \cdots \\ 
\hline
\bullet & \bullet &0 & 0& \cdots \\
\bullet & \bullet &0 & 0& \cdots \\
\vdots & \vdots &\vdots &\vdots & \ddots  \\
\end{array}
\right) 
\begin{pmatrix} \delta \rho_1 \\ \delta\rho_1^* \\ \delta\rho_2 \\ \delta\rho_2^* \\\vdots\end{pmatrix}  ,
\end{equation}
where the $\bullet$ symbols denote {\color{black}nonzero} elements. Clearly, the structure of this equation yields an infinite set of vanishing eigenvalues
plus two eigenvalues coming from the first two rows. The equation for $\delta\rho_1$ is hence, the only relevant one. The linear terms in $\delta\rho_1$ yield:
% \begin{widetext}
 \begin{eqnarray}
\dot{\delta \rho_1} =  
\frac{\epsilon \,\eta }{2}\left\{\left[e^{-i\alpha}+\frac{\epsilon \eta }{4} \left(e^{-i\beta}-|\rho_2|^2\right)\right] \delta\rho_1 \right. \nonumber \\ \left.
-\left[e^{i\alpha}+\frac{\epsilon \eta }{4}(e^{i\beta}-1)\right]  \rho_{2}\, \delta\rho_{1}^*
 \right\}  .
\label{ode_rho1}
\end{eqnarray}
% \end{widetext}
All higher-order modes, save $\rho_2$, are absent in the equation.
As $\dot\rho_2=0$, we can choose the coordinate axes such that $\rho_2=Q\in \mathbb{R}$.
After some calculations we find that 
% each NUIS undergoes an oscillatory instability at the $Q$-dependent line:
{\color{black} NUIS with a specific $Q$ value is marginally stable at}:
\begin{equation}
\epsilon_{Q}=\frac{4(1+c_1c_2)}{(c_1^2-1)(1+c_2^2)+\eta^2 Q^2} .
\label{nuis}
\end{equation}
As occurs in the MF-CGLE the larger $Q$ is, the larger is the stability region of the NUIS.
Our empirical observation is that{\color{black}, for given $c_1$ and $c_2$, if $\epsilon$ is set at a certain} $\epsilon=\epsilon_{Q^*}$
the numerical integration 
of the system (either oscillators or Fourier modes), {\color{black} under a 
% tiny amount of 
very weak} noise, 
always converges to a NUIS with $\rho_{n\ge3}=0$;
and, $|\rho_2|=Q^*$.
{\color{black} In other words, the system adopts} 
% , i.e. 
the minimum value of $|\rho_2|$ among all allowed by Eq.~\eqref{nuis}.

The state $Q=1$ ($R=0$) ---the last NUIS to destabilize--- is singular, not only because it is just a two-cluster state with two equally populated
groups, but also because in contrast to the other NUIS, the instability is not oscillatory.
Eq.~\eqref{ode_rho1} takes the form 
% $\dot\rho_1\propto a\rho_1-a^* \rho_1^*$ 
$\dot{\delta\rho_1} \propto a\, \delta\rho_1-a^*\, \delta\rho_1^*$ 
what yields an additional zero eigenvalue corresponding to the direction $\mathrm{Im}(\delta \rho_1)=0$.

\subsection{Validity and accuracy}

{\color{black} From our previous results, we conclude that the phase reduction \eqref{o2} is free of degeneracies. The boundaries of 
FS, UIS and NUIS with different $Q$ values do not overlap.}
As a double-check of the correctness of our analysis, we verified that the boundaries \eqref{fs}, \eqref{uis}, and \eqref{nuis} obtained 
%by
{\color{black} through}
the phase reduction 
are tangent to the equivalent boundaries of the MF-CGLE, Eqs.~\eqref{FS}, \eqref{UIS} and \eqref{NUIS}, at $\epsilon=0$.
In Fig.~\ref{fig_accu} we depict together the boundaries of FS, UIS, and $(Q=1)$-NUIS of the MF-CGLE (solid lines) and phase reduction to second order (dashed lines)
for two 
% $c_2$ 
{\color{black} values of the nonisochronicity}: $c_2=3$ and $c_2=1$.
These plots permit us to identify the range of $\epsilon$ in which the second-order approximation is accurate.
For $c_2=1$ the approximate bifurcation lines are accurate up to $\epsilon\approx 0.05$, while this range
is certainly smaller for $c_2=3$.

For general $c_1$, $c_2$ values, the prefactor $(\epsilon\eta)^n$ appearing for first ($n=1$) and second ($n=2$) orders suggests
to extrapolate the {\color{black} relative} smallness of $\epsilon$ to other $c_1$, and $c_2$ values. Thus, if in Fig.~\ref{fig_accu}(b) 
accuracy is good up to $\epsilon\eta\approx 0.05\eta$, and $\eta\approx2$, we propose 
\begin{equation}
\epsilon\eta<0.1
\label{validity}
\end{equation}
as a conservative range of validity of the second-order approximation.
Nevertheless, Eq.~\eqref{validity} must be regarded with some caution, 
since the {\color{black} third-order} contribution 
% of third order 
to the phase reduction {\color{black} expansion} is not exactly proportional to $(\epsilon\eta)^3$, see Sec.~\ref{sec:3rd}.

\subsection{Quasiperiodic partial synchronization}

\begin{figure}%[b]
\centerline{\includegraphics[width=75mm]{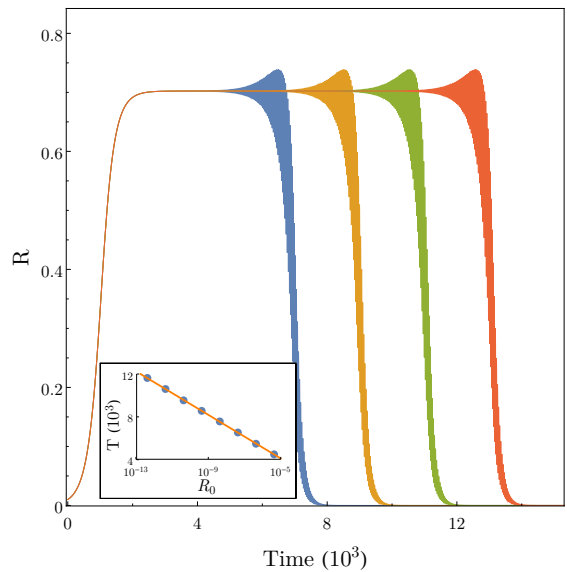}}
\caption{Evolution of $R(t)$ for $N=1000$ phase oscillators governed by \eqref{o2} initiated near the UIS state. For the selected parameter values ($c_2=3$, $c_1=-0.38$, $\epsilon=0.1$)
UIS is unstable but there are 
% non-unstable 
{\color{black} neutrally stable NUISs.}
% NUIS states.
After a transient in the neighborhood of QPS ($R(t)\approx \mathrm{const.}$), the system approaches 
a {\color{black} particular} NUIS 
% state 
($R=0$). From left to right the initial conditions are {\color{black} random perturbations of the}
%randomly perturbed 
UIS 
%states 
with $R_0=4.3 \times \{10^{-7}, 10^{-9}, 10^{-11}, 10^{-13}\}$.
The origin of times was shifted in all data sets to make the initial rise of $R$ coincident.
The inset shows the QPS transient time as a function of $R_0$. Note the 
logarithmic divergence of transient time $T\sim  \ln R_0$ (consistent with heteroclinicity).}
\label{fig_qps}
\end{figure}

{\color{black} The phenomenon of QPS was originally}
% QPS was initially 
reported in the MF-CGLE \cite{NK95} as 
{\color{black} a state emerging from the destabilization}
% an instability 
of the UIS, see Fig.~\ref{snapshot}(b), 
though its finding is usually attributed to a model of phase oscillators \cite{Vre96}. As mentioned above, in
a QPS state the mean-field rotates uniformly, but individual oscillators behave quasiperiodically. Each oscillator passes
periodically through a bottleneck located at the phase $\arg(\bar A)$. 
The onset of QPS looks like a Hopf bifurcation {\color{black} undergone by the UIS}, but 
% it is not 
{\color{black} this is not the case}
because of the infinitely many neutral directions pointing
to nearby NUISs. 
It is also important to emphasize that stable QPS does not settles any time that {\color{black} the} UIS becomes unstable. As can be
% distinguished 
{\color{black} appreciated}
in Fig.~\ref{phase_diag}(a) and (b), 
% this 
{\color{black} QPS is only observed}
% only occurs 
at moderate $\epsilon$ values when entering 
inside the green hatched region. 
Otherwise, what we observe in the MF-CGLE is that the QPS state born at the instability of {\color{black} the} UIS is a saddle. 
For parameter values with unstable UIS and FS ---outside the green-hatched regions--- initial conditions close to the UIS approach
QPS for long time, eventually converging to one NUIS. 
If any small amount of noise is present, the NUIS with the lowest $Q$ among the non-unstable ones is selected. 
The same behavior is displayed by the second-order phase reduction, Eq.~\eqref{o2};
see Fig.~\ref{fig_qps}. The logarithmic scaling 
of the residence times near QPS 
indicates a heteroclinic connection between UIS and QPS. 
The amplitude of {\color{black} the saddle} QPS {\color{black} depends on the particular parameter values.
The state of QPS} progressively grows {\color{black} as we move away from the UIS stability boundary, finally colliding} 
%colliding 
with FS ($|\rho_n|\to1$) at the point where FS becomes stable.

All in all, these results confirm the correctness of our expansion, but at the same time
prove the limitations of the second-order reduction, since the QPS attractor---found at moderate $\epsilon$ values, see Fig.~\ref{phase_diag}(a) and (b)---is not reproduced.

\subsection{Clustering}

Clustering is a much studied phenomenon in oscillator ensembles \cite{GHS+92}. In a clustered state there are several groups of oscillators, each group formed
by oscillators sharing the same phase. This kind of states are always possible in a mean-field model, so the relevant question
is the stability. Indeed, the MF-CGLE is known to exhibit stable cluster at certain parameter ranges \cite{HR92,NK93,banaji02,DN06,DN07,KGO15,kemeth19},
specifically for strong coupling ($\epsilon\approx1$). 

Are there stable clustered states at small coupling? Our phase model allows us to address this question 
in an analytical way.
Nonetheless, the general problem is intractable and we decided to restrict our study to states with two point clusters, 
where a fraction $p$ of the population is in the A-state $\theta^A$, and the {\color{black}remaining} $(1-p)$ fraction
is in the B-state $\theta^B\ne\theta^A$.
{\color{black} We now summarize the results; the corresponding calculations can be found
in Appendix I.}
% Even in this case the stability analysis is lengthy, and we moved most of the calculations (following \cite{KK01}) to Appendix I. 
% Here, in the main text, only the conclusions are outlined.

% %%%%%%%%%%%%%%%%%%%%%%%%%%%%%%%%%%%%%%%%%%%%%%%%%%%%%%%%%%%%%%%%%%%%%
 \begin{figure}
 \begin{minipage}{0.48\linewidth}
\includegraphics[width=\textwidth]{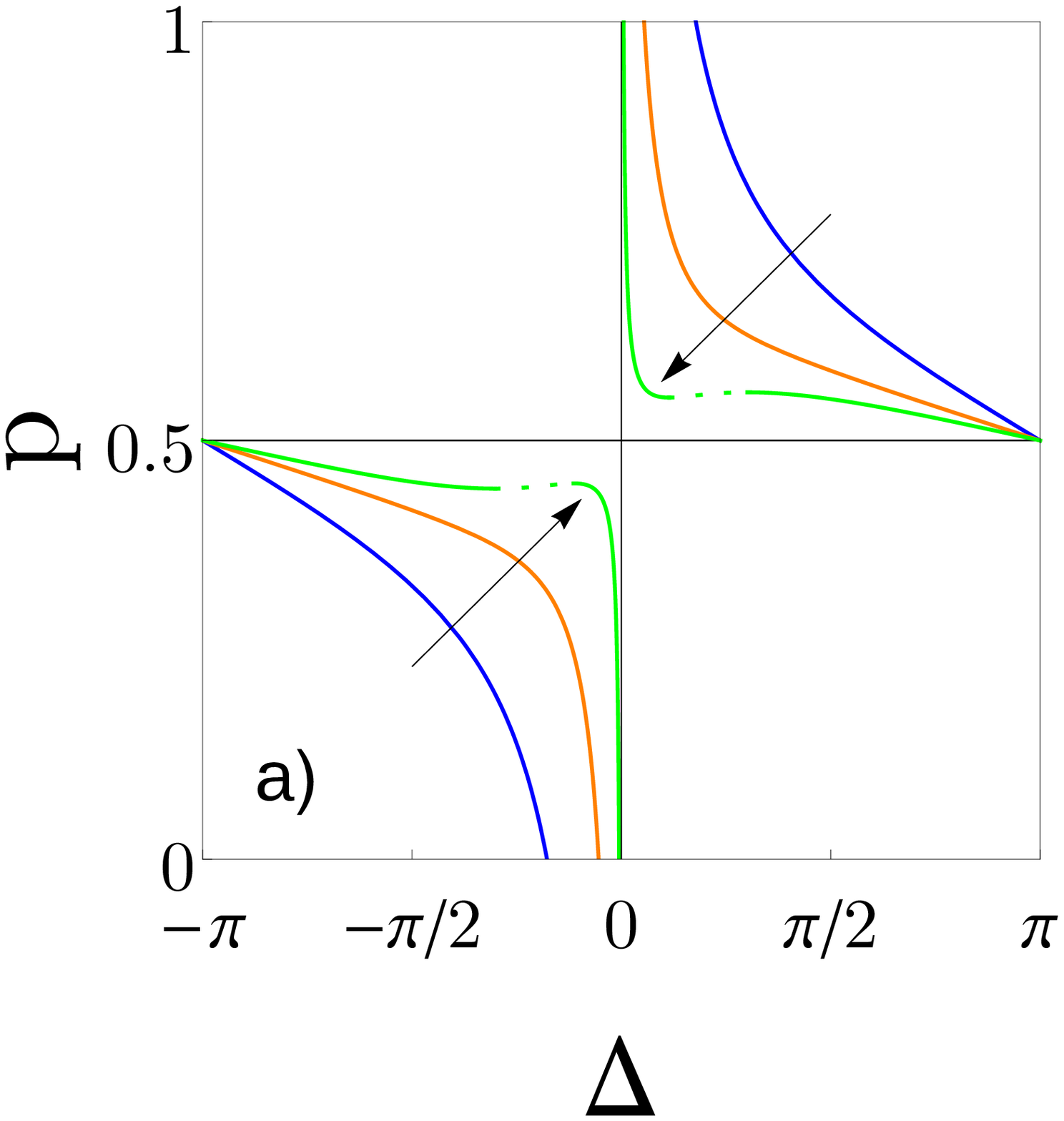}
 \end{minipage}
 \begin{minipage}{0.48\linewidth}
\includegraphics[width=\textwidth]{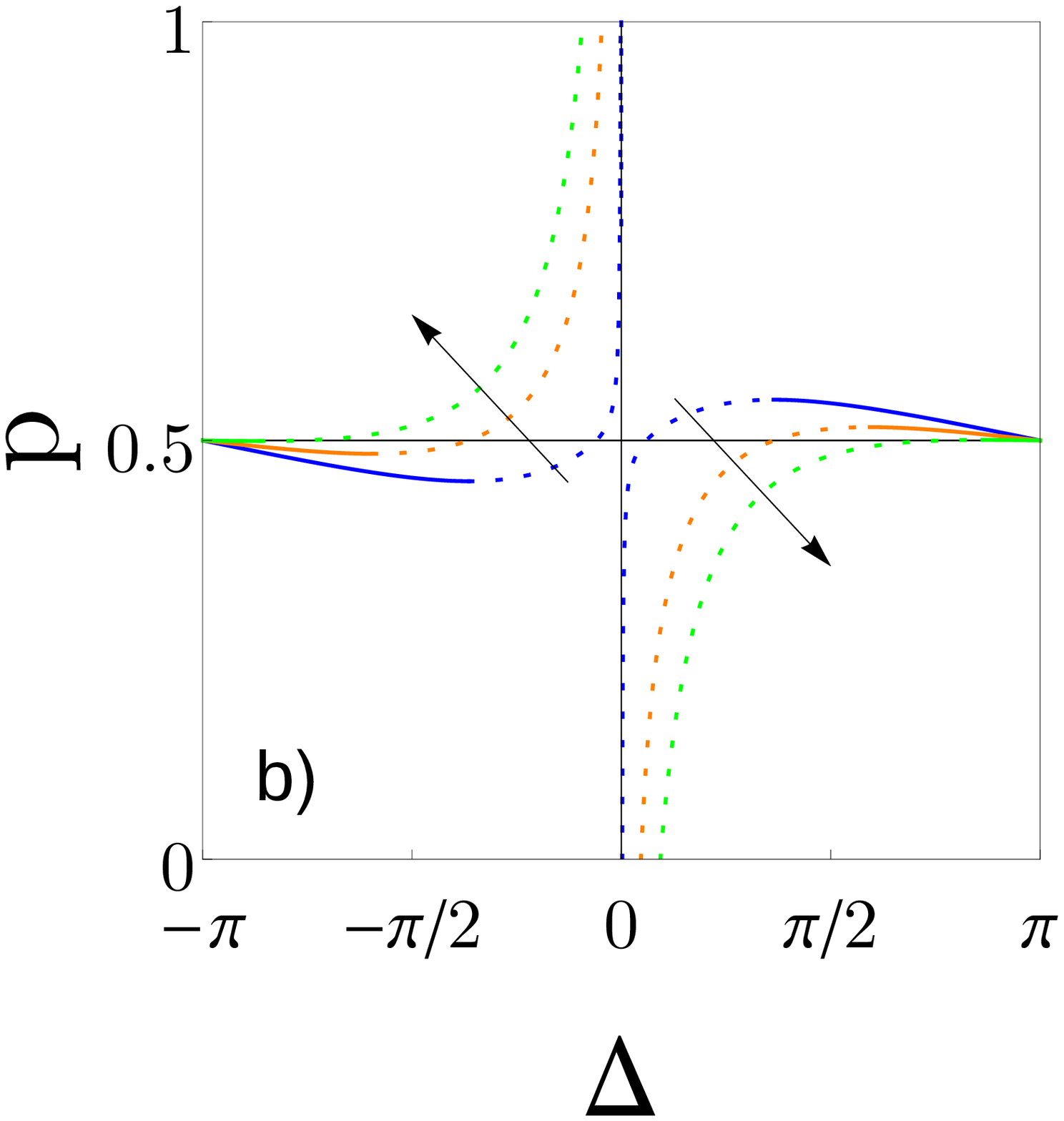}
 \end{minipage}
 \begin{minipage}{0.48\linewidth}
\includegraphics[width=\textwidth]{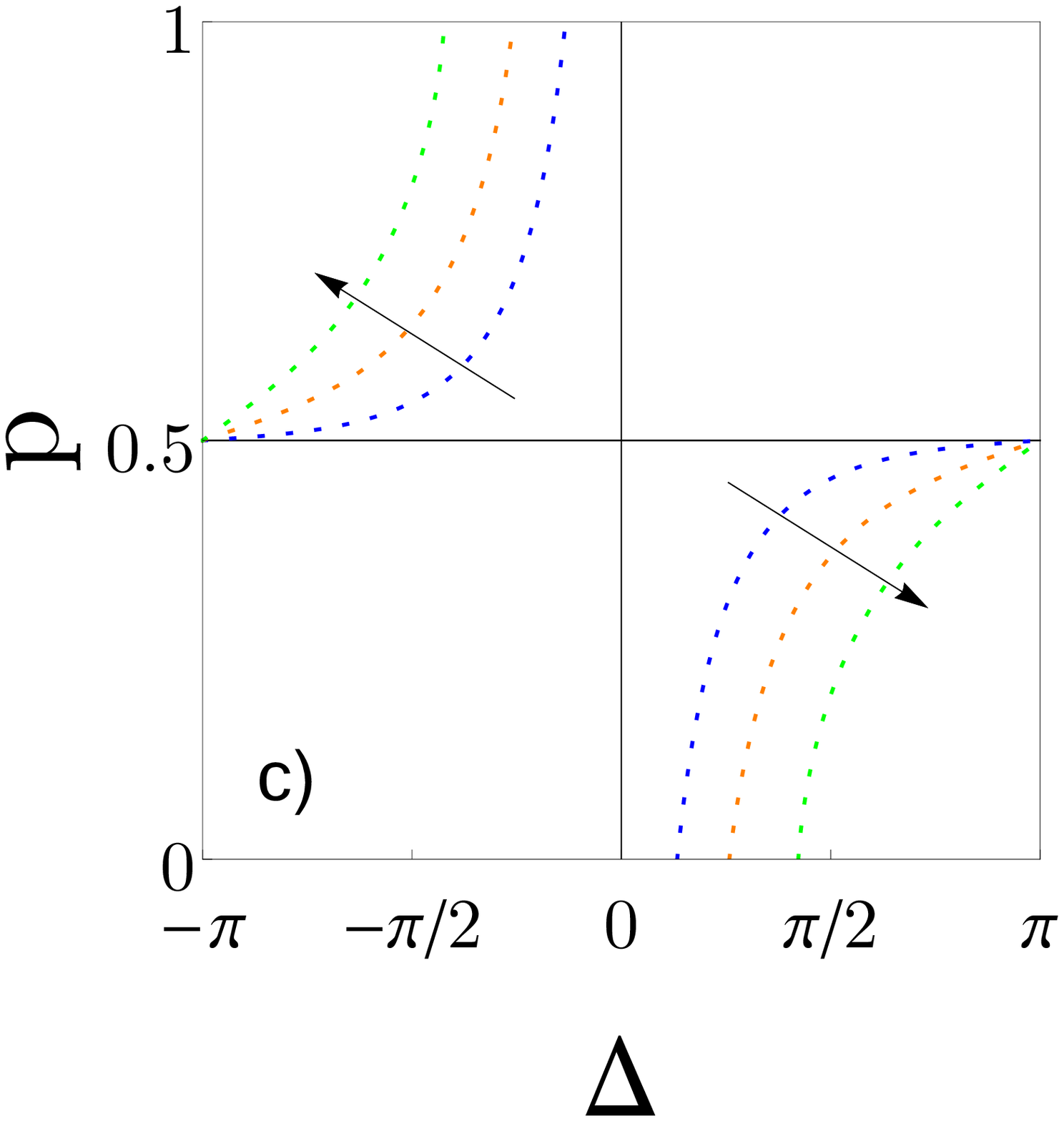}
 \end{minipage}
\caption{Two-cluster solutions of Eq.~\eqref{o2}. Each panel represents the fraction $p$ of oscillators in one cluster  
as a function of the phase lag between clusters $\Delta=\theta^A-\theta^B$. We fix $c_2=3$ and $\epsilon=0.3$ and 
select three values of $c_1$ in each panel. The arrow indicates the direction of increasing $c_1$. 
Solid (dashed) lines indicate stable (unstable) locking of the clusters.
(a) Unstable FS region, $c_1=-0.7,-0.5,-0.43$;
(b) stable FS  and not unstable ($Q=1$)-NUIS region, $c_1=-0.42,-0.36,-0.3$;
(c) stable FS and unstable ($Q=1$)-NUIS region,  $c_1=-0.25,-0.1,0.1$.}
\label{clust}
\end{figure}
% %%%%%%%%%%%%%%%%%%%%%%%%%%%%%%%%%%%%%%%%%%%%%%%%%%%%%%%%%%%%%%%%%%

As an {\color{black} illustrative} example, Fig.~\ref{clust} depicts the 
{\color{black} combinations of phase difference $\Delta=\theta^A-\theta^B$ and imbalance $p$ corresponding to actual} cluster solutions  
% for a set of 
{\color{black} for three different} 
$c_1$ values with fixed values of $c_2$ and $\epsilon$.
{\color{black} Each panel is a typical situation in a specific region of parameter space}. 
At the FS threshold, between panels (a) and (b), there is
an infinity of two-cluster solutions colliding with FS ($\Delta=0$). In consequence there is a reconnection of the two-cluster solutions. In Fig.~\ref{clust} solid lines 
represent stable locking of the clusters. However, these
solutions are fragile against disintegration of the largest cluster. Our conclusion after an extensive exploration of parameter space is that stable two-cluster states are not stable at small coupling.
To be more precise, what we observe in our second-order {\color{black} phase} reduction, {\color{black} Eq.~\eqref{o2}}, is that stable clustering is hardly found, and if so, 
it always requires moderate coupling strengths, violating \eqref{validity}.
And indeed, we could not replicate clustering in the MF-CGLE for the parameter values predicted by Eq.~\eqref{o2}.

The stability analysis of the two-cluster solutions also confirmed that slow switching \cite{hmm93} ---a stable heteroclinic connection between
two configurations of $\Delta=\theta^A-\theta^B$--- is not possible.

\subsection{Finite population, $N=4$}

This work focuses on the behavior of the MF-CGLE in the thermodynamic limit ($N\to\infty$). 
But the phase reduction \eqref{o2} is valid for an arbitrary population size. 
In this section we construct a bifurcation diagram for $N=4$ oscillators, 
{\color{black} one size previously considered in the MF-CGLE context \cite{kemeth18,kemeth19}. Here,}
this choice is motivated by the fact
that in globally coupled systems this is the smallest size with a continuum of states with $R=0$ \cite{ashwin92}, 
equivalent to the NUISs for $N=\infty$. 

{\color{black} In analogy with its thermodynamic limit, the finite-$N$ the  Kuramoto-Sakaguchi model 
has an exceptional transition between FS and the splay state at $1+c_1c_2=0$.
This degeneracy can be broken down, for instance, adding higher-order harmonics to the (pairwise) interactions \cite{ashwin08}. In our case,  
degeneracy is broken down by the three-body interactions of the second-order in the phase-reduction expansion.}

Working with a small number of oscillators has the advantage that we can track all the stationary solutions, 
in particular the clustered solutions.
As there are $3!$ orderings for the oscillators phases,
and phase ordering is preserved by the dynamics because of the mean-field interactions,
we choose the oscillators' labels  such that $\theta_j\le\theta_{j+1}$. 
{\color{black} (We assume here $\theta_j\in[0, 2\pi)$ to avoid artificial degeneracies.)}
The set of phases $\{\theta_1,\theta_2,\theta_3,\theta_4\}$,
may take several invariant configurations. Apart from the trivial FS state
$\{a,a,a,a\}$, there exists a continuum of
% ``finite-$N$ NUIS'' 
{\color{black} ``NUIS-like''
$\mathbf{Z}_2$-symmetric states   with  $\{a,a+b,a+\pi,a+b+\pi\}$, 
% with  $\{a,b,a+\pi,b+\pi\}$,
% where $b=a+c$ and 
where $b\in (0,\pi/2)\cup(\pi/2,\pi)$}. In the limit
$b\to\pi/2$ the NUIS becomes the splay state (the analogous of UIS).
In addition, in the limits {\color{black}  $b\to0$ 
%or
and
$b\to\pi$}
the NUIS collapses into a 2-cluster state with opposite phases.
Apart from this one, other 2-cluster solutions are possible. {\color{black} Namely, for some parameter values} there exist 
% one 
{\color{black} two symmetry-related}
2-2 configurations $\{a,a,b,b\}$ ($b\ne a+\pi$). {\color{black} Additionally,}
% and 
one 3-1 {\color{black} cluster exists: designated as} $\{a,a,a,b\}$ or $\{b,a,a,a\}$. Three-cluster solutions,
like $\{a,a,b,c\}$, do not exist in our 
{\color{black} phase reduction, in contrast to the MF-CGLE for strong coupling \cite{kemeth19}.}
%system.
Concerning the $N=4$ analogous of QPS, it is a periodic orbit, in which, due to the finiteness of the population, $R$ and $Q$ fluctuate
around 
their average values.

\begin{figure}
\centerline{\includegraphics[width=75mm,clip=true]{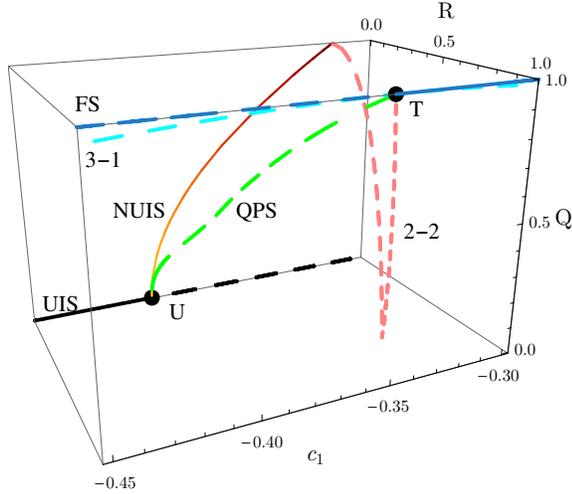}}
\caption{Bifurcation diagram for Eq.~\eqref{o2} with $N=4$ oscillators and $c_2=3$, $\epsilon=0.1$.
Solid (dashed) lines represent stable (unstable) solutions. In the case of UIS and NUIS the solution
depicted must be understood as the one observed under 
% a tiny amount of 
{\color{black}  arbitrarily weak}
noise (there is continuum of neutral solutions with 
$Q$ larger than the solution depicted). The saddle QPS orbit was continued by means of a Newton-Raphson algorithm, and the values 
of $R$ and $Q$ 
{\color{black} assigned in the diagram}
% depicted are 
{\color{black} correspond to their}
time averages.}
\label{fig_n4}
\end{figure}

We use $R$, 
% and 
$Q${\color{black}, and $c_1$} to 
% build
{\color{black} plot}
the bifurcation diagram
% , see 
{\color{black} in}
Fig.~\ref{fig_n4}. These 
%somewhat artificial 
coordinates
have the drawback of collapsing multiple equivalent states to a single point, hiding symmetries (e.g.~pitchfork bifurcations).
However, our choice intends to ease the comparison with the previous section, and with the same aim 
states are labeled borrowing the 
infinite-$N$ terminology; namely, we use the labels UIS, NUIS, and QPS instead of splay state, $\mathbf{Z}_2$-symmetric state, and limit cycle, respectively.

Due to permutation symmetry FS destabilizes {\color{black} at point T in Fig.~\ref{fig_n4}, as} three eigenvalues {\color{black} go through} zero {\color{black}simultaneously}. This comprises  
an equivariant transcritical bifurcation with the $3$-$1$ cluster, as well as a pitchfork bifurcation involving a $2$-$2$ cluster.
{\color{black} Moreover,} at 
%the same 
point {\color{black} T}, QPS collapses into a heteroclinic cycle. This coincidence of bifurcations is a known scenario in systems with full permutation symmetry \cite{ashwin92}.
Concerning  UIS, 
% it does not undergo a standard Hopf bifurcation {\color{black} (although there is a rotational instability)} {\color{black} at point U,} because of the neutral direction along the NUIS manifold {\color{black}.}
{\color{black} it undergoes an oscillatory instability at point U, but this is not a standard Hopf bifurcation because of the neutral direction along the NUIS manifold.}
QPS is a saddle, and not a stable limit cycle as {\color{black} it} might have been naively expected. 
In Fig.~\ref{fig_n4} we took $c_2=3$, and the QPS branch connects %UIS and FS 
{\color{black} points T and U}
in a simple way.
% If $c_2$ is decreased the QPS branch develops a fold, and finally for low enough $c_2$ the branch becomes reversed
% and a completely unstable QPS connects FS and UIS.
{\color{black} In contrast to Fig.~\ref{fig_n4}, for $c_2=1$ FS and UIS coexist, and points U and T switch their relative positions. In that case the QPS branch is 
completely reversed (not shown), and the QPS solution is fully unstable. 
Consistently, we found a range of $c_2$ values in between, $1<c_2<3$, where (depending on $\epsilon$) the QPS branch develops a fold.}

In summary, the bifurcation diagram for $N=4$ appears to capture the global picture of the transition from UIS to FS. Considering more oscillators
will increase the number of cluster solutions, see \cite{banaji02}, but no essential new features.

\section{Third-order phase reduction:  four-body interactions}
\label{sec:3rd}
Our reason to deal with the third-order term now is to illustrate the practicality of the phase reduction method, and 
get a glimpse of the power series expansion at higher orders.
% After tedious algebraic calculations we obtained the $O(\epsilon^3)$ term in Eq.~\eqref{o2}:
{\color{black} Evaluating the cubic term in Eq.~\eqref{protom} yields the $O(\epsilon^3)$ correction to Eq.~\eqref{o2}:
\begin{widetext}
	\begin{multline}
	\epsilon^3\frac{1+c_2^2}{16}\bigg\{C_1 R \sin (\Psi -\theta _j+\gamma_1)+C_2 R^2 \sin\left[2 (\Psi-\theta _j) +\gamma_2\right]+C_3 R  \,Q \sin (\Phi -\Psi -\theta _j+\gamma_3)
	+C_4 R  \, Q^2 \sin (\Psi -\theta _j+\gamma_4) \\+C_5 R^3\sin (\Psi -\theta _j+\gamma_5)
	+ C_6 R^2  Q \sin (\Phi -2 \theta _j+\gamma_6)
	+  C_7 R^3 \sin\left[3 (\Psi-\theta _j) +\gamma_7\right]
	+C_8 R^2 P \sin (\Xi -2 \Psi-\theta _j+\gamma_8 )\\+ C_9 R^2  Q \sin (\Phi -2 \Psi +\gamma_9)+D R^2
	\bigg\} .
	\label{o3}
	\end{multline} 
	\end{widetext}}% A new 
{\color{black}This expression depends on the third Kuramoto-Daido order parameter $Z_3\equiv P e^{i\Xi}=N^{-1}\sum_j e^{i 3\theta_j}$}. The 
% exact 
dependence of constants {\color{black}$\{C_j,\gamma_j\}_{j=1,\ldots,9}$ and $D$} on $c_1$ and $c_2$ 
% is convoluted and probably uninteresting, see the analytic expressions 
{\color{black} is tabulated}
in Appendix II. 
% Instead, t
The structure of Eq.~\eqref{o3} deserves some words here.
The terms proportional to  {\color{black} $C_j$} with indices $j=1,2,3$ are  
%  dull
{\color{black} higher-order} corrections {\color{black} to Eq.~\eqref{o2}, tantamount to a shift in parameter values.} 
% to 
% previous terms 
% of 
% {\color{black} in }
% the power series.
Four-body interactions appear in five different forms, corresponding to indices $j=4,\ldots,8$. For illustration we 
% write down explicitly 
{\color{black} expand}
a couple of these four-body contributions:
\begin{eqnarray}
R^3 \sin(\Psi-\theta_j)&=&\frac{1}{N^3} % \sum_{k,l=1}^N
\sum_{k,l,n} \sin(\theta_k+\theta_l-\theta_n-\theta_j) ,  \nonumber\\ 
R^2 P \sin(\Xi-2\Psi-\theta_j)&=&\frac{1}{N^3} % \sum_{k,l=1}^N
\sum_{k,l,n} \sin(3\theta_k-\theta_l-\theta_n-\theta_j) .\nonumber 
\end{eqnarray}
There are several qualitative  features in Eq.~\eqref{o3} that deserve to be pointed out:
\begin{enumerate}
    \item The overall $O(\epsilon^3)$ contribution is not proportional to $\eta^3$ ---though some terms indeed are--- in contrast to $O(\epsilon)$ 
  and $O(\epsilon^2)$, which  are
  proportional to $\eta$ and $\eta^2$, respectively. 
  
    \item From Eqs.~\eqref{o2} and \eqref{o3} we can expect that truncation of the power series to order $\epsilon^n$
  yields up to $(n+1)$-body interactions, but not higher-order non-pairwise couplings. We can also expect that only
  Kuramoto-Daido order parameters $Z_k$ with $k\le n$ appear.
  
    \item The last {\color{black} two} %three 
    terms in Eq.~\eqref{o3} are somewhat unexpected, {\color{black} (nonetheless see \cite{ashwin16})}, since
  they depend on the mean fields $Z_1$ and $Z_2$, but not on $\theta_j$ itself.
  They are hence irrelevant concerning synchronization boundaries.

    \item As occurs with the $O(\epsilon^2)$ term, FS and (N)UIS states are consistent with the MF-CGLE dynamics:
  (i) all terms in \eqref{o3} are proportional to $R$ ensuring that the contribution to the oscillators' frequencies
  vanishes in 
%   the
{\color{black} one}
  incoherent state;  (ii) in the FS state, the contribution also vanishes, as expected since the
  frequency of FS in the MF-CGLE varies linearly with $\epsilon$. Accordingly, it holds that {\color{black}$D+\sum_j C_j\sin\gamma_j=0$}, cf.~Appendix II.
 \end{enumerate}
Unfortunately, there is not a recognizable pattern in the new terms appearing in the power series expansion, 
so it is not 
% straightforward 
{\color{black} possible}
to extrapolate to higher orders in $\epsilon$.

From Eqs.\eqref{o2} and \eqref{o3} {\color{black} we can} 
% is possible to 
derive the stability boundary of FS, NUIS (for UIS just set $Q=0$) obtaining:
	\begin{equation}
	    2(1 + c_1 c_2) +  \epsilon_s c_1^2 (1 + c_2^2) + \epsilon_s^2 c_1^3 c_2 (1 + c_2^2)=0 ,
	    \label{fs3}
	\end{equation}
	\begin{eqnarray}
	 4(1+c_1 c_2)+\epsilon_Q(1+c_2^2)\bigg[(1-c_1^2)-Q^2(1+c_1^2)\bigg] \nonumber\\
	 +\frac{\epsilon_Q^2}{2}(1 + c_2^2)\bigg[ (2 - 2 c_1^2 - 3 c_1 c_2 + c_1^3 c_2)\nonumber \\- Q^2 (1 + c_1^2) (-2 + 3 c_1 c_2)\bigg]=0 . \label{nuis3}
	\end{eqnarray}
In Fig.~\ref{fig_ab} we depict (a) $\epsilon_0$, and (b) $\epsilon_s$ {\color{black} from the previous expressions}
and compare them with the 
result of the MF-CGLE, and with the second-order approximation. 
The slopes and the curvatures of the bifurcation lines of the 
third-order phase {\color{black} reduction} %model 
agree with those of the MF-CGLE at $\epsilon=0$.

 \section{Discussion}

\subsection{Alternative phase reduction(s)}

Our phase reduction is a genuine power series in the small parameter $\epsilon$. 
Another strategy to analyze \eqref{CGLE} is to absorb the $\epsilon A_j$ term prior to the phase reduction.
Specifically, setting $t'=(1-\epsilon)t$, and 
\begin{equation}
% \epsilon
\kappa=\frac{\epsilon}{1-\epsilon} ,
\end{equation}
we get 
\begin{equation}
 \frac{dB_j}{dt'}=B_j-(1+ic_2)|B_j|^2B_j+\kappa (1+ic_1)\bar B ,
\label{CGLEb}
 \end{equation}
where $B_j=A_j\exp(i \epsilon c_1  t)/\sqrt{1-\epsilon}$. Applying our phase-reduction method to \eqref{CGLEb} we obtain an alternative phase reduction
in powers of $\kappa$ (the result is not qualitatively different).

Is it worth transforming \eqref{CGLE} into \eqref{CGLEb}?
In other words, is the phase reduction of \eqref{CGLEb} up to order $\kappa^n$, 
% better or worse than 
{\color{black} superior to}
that of \eqref{CGLE} at order $\epsilon^n$? 
% First of all, the reader should 
% be 
{\color{black} Certainly, phase reductions at order $\epsilon^n$ and $\kappa^n$ are not equivalent}
% agree that both phase models 
% cannot be mapped one onto the other
since $\kappa=\epsilon+\epsilon^2+\epsilon^3+\cdots$. Any truncation at order $\kappa^n$ involves all powers of $\epsilon$.
% A good test  bed to confront the 
{\color{black} The relative accuracy of the}
phase reductions of \eqref{CGLEb} and \eqref{CGLE} at the same order
% is to compare 
{\color{black} can be assessed comparing}
the bifurcation loci. 
% obtained 
% at the same order. 
Instead of {\color{black} applying phase reduction to Eq.~\eqref{CGLEb}},
% working all out again explicitly, 
the quickest 
% form of proceeding 
{\color{black} strategy}
is to assume the existence of an exact phase reduction
involving all orders in $\epsilon$ such that
the exact critical value $\epsilon_*$ (the asterisk
denotes an arbitrary {\color{black} state}: UIS, FS, ...) 
% cancels out a power series:
{\color{black} satisfies:}
\begin{equation}
\sum_{n=1}^{\infty} a_n(c_1,c_2) \, \epsilon_*^n + 1+c_1c_2 =0 .
\label{power}
\end{equation}
The coefficients $a_n$ depend on the 
% bifurcation 
{\color{black} specific instability} we are considering.

{\color{black} Phase reduction of \eqref{CGLE} up to order $n$ results in a truncation of \eqref{power}
to order $n-1$.
 For instance, the second-order phase reduction of \eqref{CGLE} 
yields the linear relation [recall Eqs.~\eqref{fs} or \eqref{uis}]:
\begin{equation}
a_1 \epsilon_* + 1+c_1c_2 =0 .\label{o1a}
\end{equation}
At the same order, the phase reduction of \eqref{CGLEb} results in an analogous expression
\begin{equation}
a'_1 \kappa_* + 1+c_1c_2=0 .\label{o1b} 
\end{equation}}
%  For instance, 
% second-order phase reductions 
% of \eqref{CGLE} and \eqref{CGLEb} 
% yield, respectively, the linear relations [{\color{black}recall Eqs.~}%cf. 
% \eqref{fs} or \eqref{uis}]:
% \begin{subequations}
%  \begin{eqnarray}
% a_1 \epsilon_* + 1+c_1c_2 =0 ,\label{o1a}\\
% a'_1 \kappa_* + 1+c_1c_2=0 .\label{o1b} 
% \end{eqnarray}
% \end{subequations}
Given that $\kappa=\epsilon + O(\epsilon^2)$, consistency with \eqref{power} determines $a'_1=a_1$. Thus the bifurcation locus estimated
from the phase reduction of \eqref{CGLEb} satisfies (in coordinate $\epsilon$) $a_1 \epsilon_*/(1-\epsilon_*) + (1+c_1c_2)=0$, which is slightly different from \eqref{o1a}.
Analogous reasoning permits us to obtain the bifurcation lines for the third-order phase {\color{black} reduction}
of \eqref{CGLEb} 
from Eqs.~\eqref{fs3} and \eqref{nuis3}. %, without need of obtaining the phase model.

% %%%%%%%%%%%%%%%%%%%%%%%%%%%%%%%%%%%%%%%%%%%%%%%%%%%%%%%%%%%%%%%%%%%%%
 \begin{figure}
 \begin{minipage}{0.48\linewidth}
  \includegraphics[width=\textwidth]{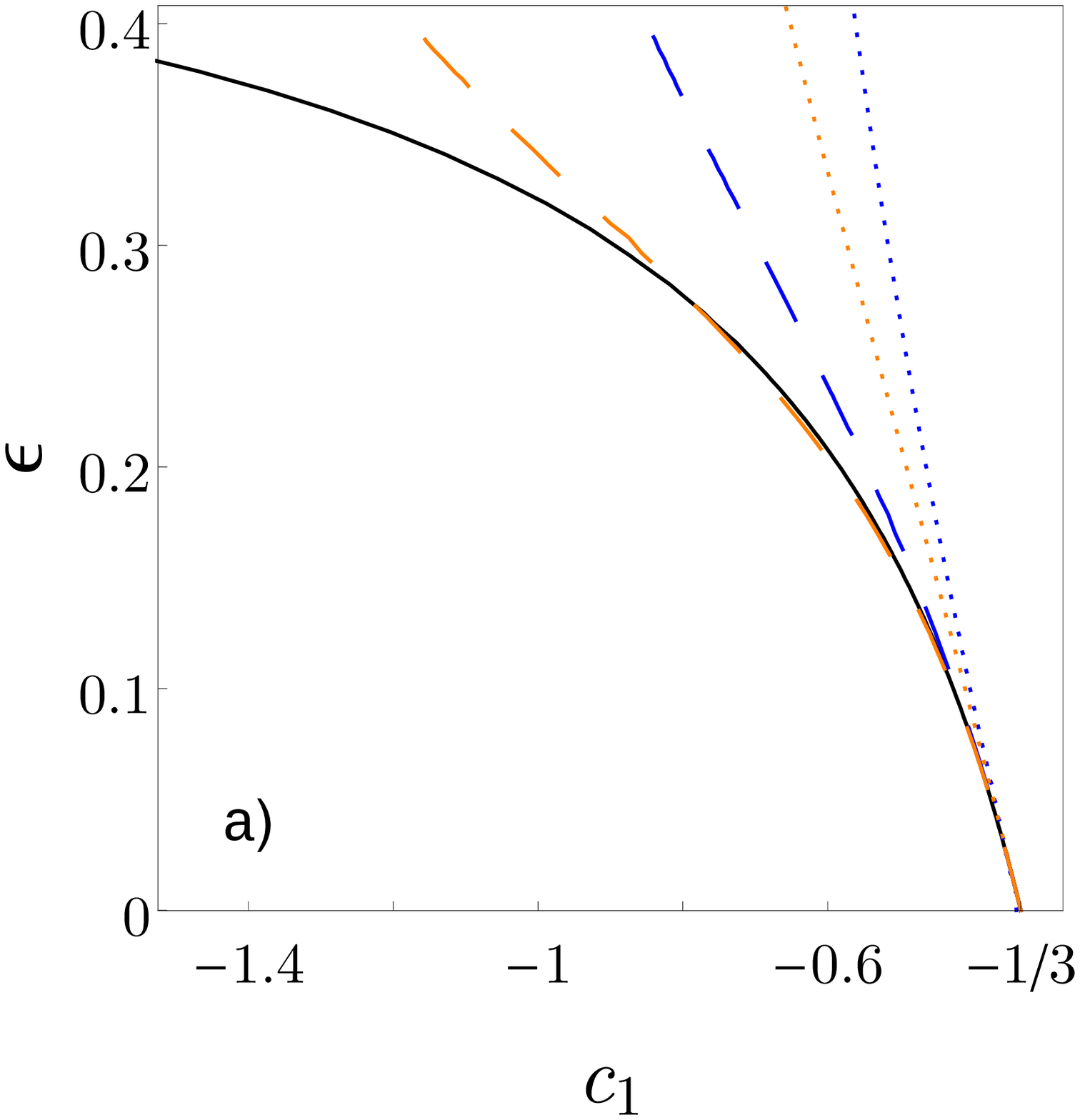}
  \end{minipage}
 \begin{minipage}{0.48\linewidth}
  \includegraphics[width=\textwidth]{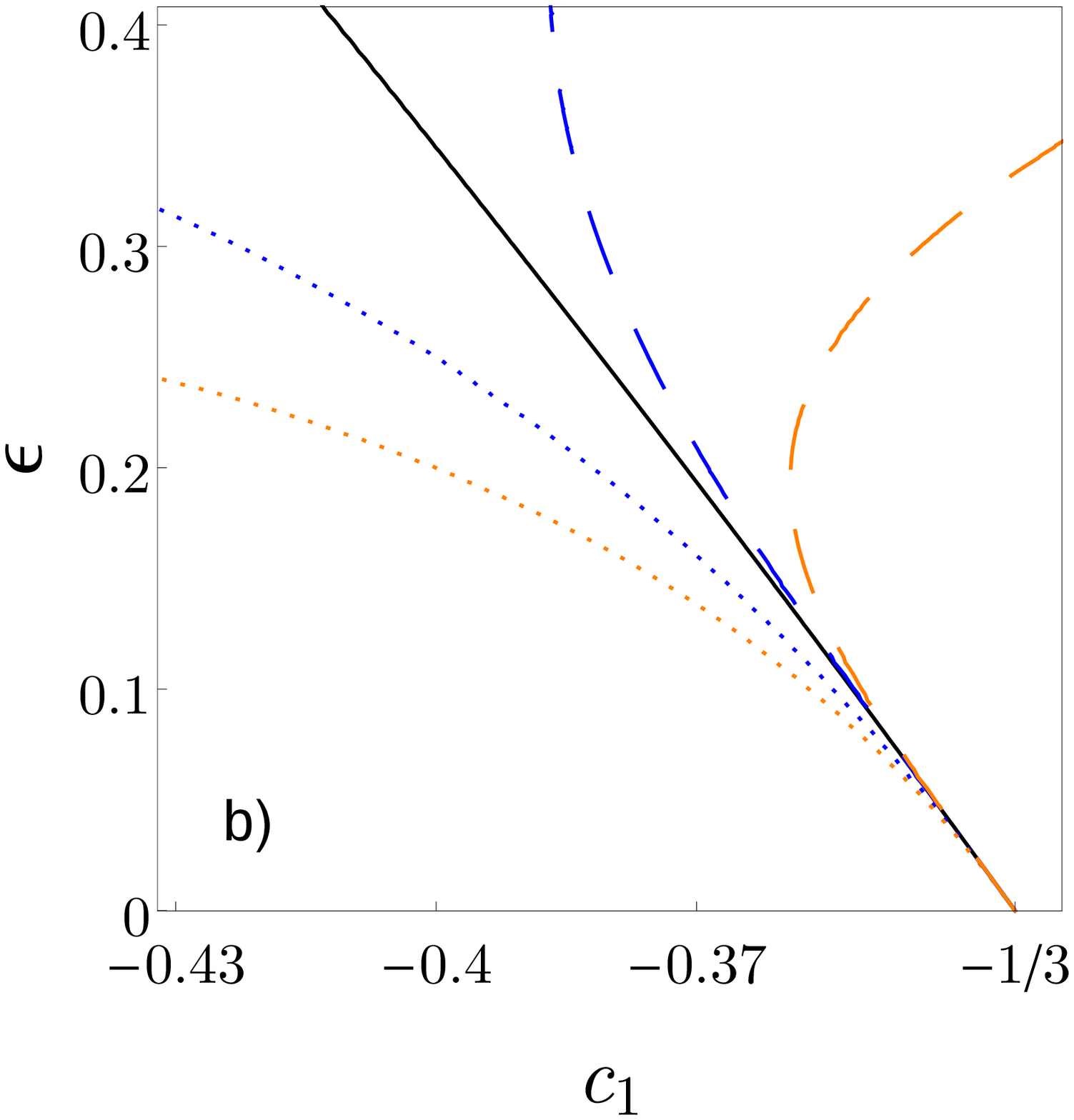}
  \end{minipage}
\caption{
Stability boundaries of (a) UIS and (b) FS obtained exactly and from phase approximations, for $c_2=3$.
The solid line correspond to the exact boundary of the MF-CGLE \eqref{CGLE}, while dotted and dashed
lines correspond to second- and third-order phase approximations{\color{black}, respectively}. Blue lines are
obtained from \eqref{o2} and \eqref{o3}. Orange lines are the results if prior to phase reduction
the MF-CGLE is transformed into \eqref{CGLEb}, performing an isochron-base phase reduction in powers of $\kappa$.
}
\label{fig_ab}
\end{figure}
% %%%%%%%%%%%%%%%%%%%%%%%%%%%%%%%%%%%%%%%%%%%%%%%%%%%%%%%%%%%%%%%%%%

A comparison of the bifurcations lines of UIS and FS is displayed in Figs.~\ref{fig_ab}(a) and \ref{fig_ab}(b) for $c_2=3$.
We see that the transformation of \eqref{CGLE} into \eqref{CGLEb} allows us to obtain a phase model
that captures better the stability boundary of UIS, but not of FS. It is easy to understand why. Each strategy
captures better the dynamics in which the quantities multiplying the coupling constant are small. Thus, Eq.~\eqref{CGLE}
is already a good starting point for states close to FS ($A_j\approx \bar A$), while \eqref{CGLEb} works
better close to incoherence ($\bar A \approx \bar B \approx0$). Finally, note that in addition to \eqref{CGLE} and \eqref{CGLEb},
there exists a continuum of alternative{\color{black}, intermediate} formulations, in which $\epsilon A_j$ is only partly absorbed by a coordinate transformation.

\subsection{Possible extensions of this work}

The phase-reduction procedure presented in this work can be easily implemented in other geometries, different from the fully connected network. 
In a networked architecture, phase reduction at first order in $\epsilon$ couples phases with the nearest-neighbor phases. 
% At second order 
{\color{black} At order $\epsilon^2$}, second nearest neighbors come also into play \cite{kralemann11,*kralemann14},
and progressively more distant nodes participate {\color{black} in the phase dynamics} at higher orders. 
Also, the case of non-locally coupled
Stuart-Landau oscillators \cite{Kur97} is analyzable with the phase reduction presented here.
Concerning the original complex Ginzburg-Landau equation, a partial differential equation of reaction-diffusion type, our phase reduction procedure is very simple and efficient
obtaining the coefficients of the second-order terms: $\nabla^4 \theta$, $(\nabla^2  \theta)^2$, $(\nabla  \theta)^2 \nabla^2  \theta$, $\nabla\theta \nabla^3\theta$ ~\cite{Kur84}.

Concerning the oscillator dynamics, the phase reduction carried out here can be easily applied to planar oscillators with polar symmetry ($\lambda-\omega$ systems). 
In the latter case, analogously to \eqref{iso}, the isochrons satisfy 
$\theta=\varphi+\chi(r)$ \cite{Win80}. Even if function $\chi$ does not have a closed form, 
it is still possible to obtain the phase model using the derivatives of the isochrons on the limit cycle.

{\color{black}
 \subsection{Relationship with other phase-reduction approaches}

In this subsection, we comment on the progress of our phase-reduction approach
with respect to previous works,
even if only directly applicable to $\lambda-\omega$ systems.

An alternative way of obtaining the second-order phase reduction of the MF-CGLE, Eq.~\eqref{o2}, 
is applying the systematic averaging formulation in Chap.~4 of the book by Kuramoto \cite{Kur84}.
This calculation is, however, much more lengthy than the one presented in Sec.~\ref{spr}. Not surprisingly, 
obtaining the order $\epsilon^3$ with the averaging approach \cite{Kur84}
is a totally impractical task, 
% a task of unsurmountable complexity 
while we succeeded with our method (with the assistance of symbolic software); see Eq.~\eqref{o3}.

Equation~\eqref{o2} can also be obtained assuming small variations of the radii, i.e.~setting $\dot r_j=0$. This procedure was
followed in \cite{pikovsky06,matheny19}, with the difference being that there the small quantities are deviations from the reference limit cycle.
Here, we pursued a bona-fide power expansion in terms of the coupling constant $\epsilon$, and the result differs from the one obtained following \cite{pikovsky06,matheny19}.
% Finally, we mention that t
In passing, we mention that instead of assuming $r_j=r_j(\theta_1,\theta_2,\ldots,\theta_N)$,
once  Eqs.~\eqref{eqsimples} are derived, the two-timing approximation, such that the $\theta_j$ depend only on a slow time $\tau=\epsilon t$, can also be applied.

In contrast to our work, Refs.~\cite{kori14,ashwin16} apply first-order phase reduction obtaining 
multi-body phase interactions.
The reason is that those works invoke amplitude equations for an ensemble close (but not asymptotically close) to a Hopf bifurcation.
The amplitude equation, which can be seen as a generalization of Eq.~\eqref{CGLE}, turns out to contain nonlinear interactions. The nonlinear coupling
among the $A_j$'s leads to multi-body interactions in the first-order phase reduction.
Applying second-order phase reduction, as described here, to the amplitude equations in \cite{ashwin16} or \cite{kori14}
may be interesting.
}

{\color{black}
\subsection{Towards a minimal phase model of pure collective chaos}
Pure collective chaos has been found in several phase models with heterogeneity \cite{bick18} or delay \cite{PM16,*devalle18}.
Collective chaos in the MF-CGLE, see Fig.~\ref{snapshot}(c), calls for a phase description in terms of identical phase oscillators
(without delays). The fact that we have not found evidence of
collective chaos in our numerical simulations of the second-order phase reduction \eqref{o2} ---nor in the third-order one--- can be reasonably attributed to
a too restrictive truncation of the power expansion.
We believe that a higher-order truncation will capture better the
behavior of the system at larger $\epsilon$ values, and eventually,
will exhibit collective chaos.  

As pairwise interactions through higher harmonics, like $Q\sin(\Phi-2\theta_j)=N^{-1}\sum_k\sin[2(\theta_k-\theta_j)]$,
do not show up in the phase reduction of the MF-CGLE \footnote{Otherwise, the frequency of (N)UIS would depend nonlinearly on $\epsilon$, in disagreement with the MF-CGLE.},
multi-body phase interactions appear to be the most promising ingredient to model collective chaos. 
In small ensembles of identical phase oscillators,
higher harmonics as well as multi-body interactions promote chaos alike, see \cite{bick11} and \cite{bick16}, respectively.
However, so far, collective chaos remains elusive in populations of higher-order pairwise interacting identical phase oscillators \cite{clusella16}.
We believe multi-body interactions could be the key element of collective chaos, instead.

In the MF-CGLE with parameter values close to those in Fig.~\ref{snapshot}(c),
we found chaos with a population sizes as small as $N=6$.
Does this say something about the order of the multi-body interactions needed in the phase reduction?
Is this chaos connected to collective chaos in the thermodynamic limit, 
as in \cite{bick18}?
}

\subsection{Conclusions}

Multi-body interactions are an unavoidable consequence of phase reduction, but
save for a few works \cite{tanaka11,bick16,cbick18,matheny19,SA19}, the role of multi-body phase interactions 
shaping exotic dynamics remains largely unexplored. 
In the weak-coupling regime of the MF-CGLE, multi-body phase interactions are essential
to describe {\em all} states apart from FS and UIS. 

In summary, in this work we achieve second- and third-order phase reductions of the MF-CGLE.
In our view, higher-order phase reductions promise to be crucial
for our understanding of collective chaos and other exotic phenomena \cite{matheny19}.
Moreover, analytic higher-order phase reductions may also serve as test beds for numerical phase reductions recently implemented \cite{RP19}. 
For these reasons, we 
{\color{black} regard}
% feel that 
phase reduction beyond the first order 
% is 
{\color{black} as}
an exciting battleground of nonlinear dynamics.

 \begin{acknowledgments}
We acknowledge support by MINECO (Spain) under Project
No.~FIS2016-74957-P. IL acknowledges support by Universidad de Cantabria and Government of Cantabria
under the Concepci\'on Arenal programme.
 \end{acknowledgments}

\section*{Appendix I: Clustering}

Our model \eqref{o2} in a more convenient form (recall that in the rotating frame $\Omega=0$) reads:
	\begin{eqnarray}\label{eqphasesorden2}
	\dot{\theta_i}=\frac{1}{N}\sum_{j}\Gamma(\theta_j-\theta_i) \nonumber\\
	+\frac{1}{N^2}\sum_{j,k}\left[G_1(\theta_j+\theta_k-2\theta_i)+G_2(2\theta_j-\theta_k-\theta_i)\right] ,
	\end{eqnarray}
with
\begin{eqnarray}
	\Gamma(x)&=&\epsilon\bigg[(c_1-c_2)\cos x+(1+c_1c_2)\sin x\bigg]-G_1(x) ,\\
	G_1(x)&=&-\epsilon^2(1+c_2^2)\bigg[\frac{c_1}{2}\cos x+\frac{(1-c_1^2)}{4}\sin x\bigg], \\
	G_2(x)&=&\frac{\epsilon^2(1+c_2^2)(1+c_1^2)}{4}\sin x .
\end{eqnarray}
Note that $G_2$ is an odd function.

Let us write first the evolution equation for cluster-$A$ phase $\theta^A$, defining
the phase difference $\Delta=\theta^A-\theta^B$:
\begin{widetext}
	\begin{eqnarray}
	\dot{\theta}^A=
	\left[p\Gamma(0)+(1-p)\Gamma(-\Delta)\right]+\left[p^2G_1(0)+2p(1-p)G_1(-\Delta)+(1-p)^2G_1(-2\Delta)\right] \nonumber\\
+	\left[-p(1-p)G_2(2\Delta)+(-2p^2+3p-1)G_2(\Delta)\right] ;
	\end{eqnarray}
the equivalent equation for the $B$-cluster is obtained with the substitution $\Delta\to-\Delta$ and $p \to (1-p)$.
The evolution of $\Delta(t)$ obeys
	\begin{eqnarray}\label{eqdesfase}
	\dot{\Delta}=(2p-1)\Gamma(0)+(1-p)\Gamma(-\Delta)-p\Gamma(\Delta) +\left[-2p(1-p)G_2(2\Delta)+(-4p^2+4p-1)G_2(\Delta)\right]\nonumber\\
	+\left\{(2p-1)G_1(0)+2p(1-p)\left[G_1(-\Delta)-G_1(\Delta)\right]+(1-p)^2G_1(-2\Delta)-p^2G_1(2\Delta)\right\}.
	\end{eqnarray} 
\end{widetext}
Setting $\dot\Delta=0$ we obtain a quadratic equation in $p$ that can be solved explicitly. 
We depict $p(\Delta)$ in Fig.~\ref{clust}
for selected parameter values. Note the symmetry of the curves because of the invariance under $(\Delta,p)\leftrightarrow(-\Delta,1-p)$.
There are $\Delta$ values for which $p$ is out of the range $(0,1)$,
indicating no two-cluster states with those particular $\Delta$ values exist. 
Conversely, different values of $\Delta$ may be consistent with the same $p$ value, indicating
the coexistence of multiple two-cluster solutions with the same sizes.

\subsubsection{Stability}
First of all, note that, one zero eigenvalue is always present due to the global phase shift invariance of 
the model, $\theta_j\to \theta_j+\mathrm{const.}$, and we ignore it hereafter.
For the analysis that follows it is simpler to assume the thermodynamic limit (eigenvalues are the same for finite $N$, but the calculation
is more convoluted.)
As already known from previous studies \cite{KK01}, perturbations on a two-cluster solution can be decomposed in three orthogonal modes.
Two of them are the disintegration of each respective cluster, and the third one is the unlocking of the two clusters.
We denote $\lambda_A$, $\lambda_B$ and $\lambda_L$ the corresponding eigenvalues.
For the stability of the $A$-cluster, we need to evaluate if one oscillator in the neighborhood of this cluster decays to it or departs (i.e.,~``evaporates'').
The eigenvalue $\lambda_A$ is simply obtained linearizing around the state. The result is:
\begin{eqnarray}
\lambda_A=-p\Gamma'(0)-(1-p)\Gamma'(-\Delta)-2p^2G_1'(0) \nonumber\\ 
-4p(1-p)G_1'(-\Delta) -2(1-p)^2G_1'(-2\Delta)\nonumber\\
-p^2 G_2'(0)-(1-p)G_2'(\Delta)-p(1-p)G_2'(2\Delta) .
\end{eqnarray}
The eigenvalue $\lambda_B$ is obtained from $\lambda_A$ after the substitution $p\to(1-p)$ and $\Delta\to-\Delta$, and viceversa.
Finally, the locking between the clusters is controlled by the eigenvalue obtained linearizing \eqref{eqdesfase}:
\begin{eqnarray}
\lambda_L&=&-(1-p)\Gamma'(-\Delta)-p\Gamma'(\Delta)-2(1-p)^2G_1'(-2\Delta)\nonumber\\
&&-2p(1-p)\big[G_1'(-\Delta)+G_1'(\Delta)\big]-2p^2G_1'(2\Delta)\nonumber\\
&&-(4p^2-4p+1)G_2'(\Delta)-4p(1-p)G_2'(2\Delta) .
\end{eqnarray}
Stability requires $\lambda_A,\lambda_B,\lambda_L <0$. 
For small $\epsilon$ we summarized our findings in the main text, 
distinguishing three different regions corresponding to the three panels of Fig.~\ref{clust}. 
As said in the main text we found stable clusters in the first region (FS unstable), e.g.~$\epsilon=0.1$, $c_1=-9$, $c_2=2$. However, 
for these parameters the condition \eqref{validity} does not hold, and in fact the cluster solution destabilized when we implemented it in the MF-CGLE.

\subsubsection{No slow switching}

With unstable two-cluster states, the system might still exhibit one nontrivial phenomenon called slow switching \cite{hmm93}. In this phenomenon,
the clusters switch between two different $\Delta$ values with identical $p$ value. The explanation
for this behavior is a stable heteroclinic connection between the pair of two-cluster states that causes the system to switch for ever between them with increasing
residence times \cite{hmm93,KK01}. In practice \cite{clusella16}, switching either terminates in one of the unstable two-cluster states (due to round-off errors), or 
it achieves a constant periodic switching (due to small noise).
According to \cite{KK01}, slow switching requires 
{\color{black} the coexistence of}
three 
%simultaneous 
two-cluster states 
{\color{black}$\Delta'$, $\Delta''$, and $\Delta'''$}
with an identical $p$ %and $\Delta$ 
value, such that $0<\Delta'<\Delta'''<\Delta''<2\pi${\color{black}, and}
% such, 
% that 
$\lambda_L<0$ for $\Delta'$ and $\Delta''$, while $\lambda_L>0$ for $\Delta'''$. 
{\color{black} As may be seen in Fig.~\ref{clust}, finding parameter values with three solutions for $\Delta$ at the same $p$ is aready difficult ---only for the green line in Fig.~\ref{clust}(a) do 
such $p$ values exist.
In addition, the condition for the eigenvalues is even more stringent: e.g., in Fig.~\ref{clust}(a) the three points share the stability of $\lambda_A$ and $\lambda_B$ 
making the heteroclinic connection impossible.}
% These conditions are virtually never met by our system.
% And even in the small parameter range where it does, see intermediate line in Fig.~\ref{clust}(a), the three points share the stability of $\lambda_A$ and $\lambda_B$ 
% making the heteroclinic connection impossible.

\section*{Appendix II: Constants $C_j$, $\gamma_j$ and $D$ in Eq.~(29)}

The dependences of constants $\{C_j,\gamma_j\}_{j=1,\dots,9}$ on $c_1$ and $c2$ is tabulated as follows:

% 	\begin{equation}
% 	\begin{aligned}
		\begin{eqnarray}
	{\color{black}C_j=\sqrt{A_j^2+B_j^2}} ,\\
	{\color{black} \gamma_j=\arg(A_j+i B_j)}, 
	\end{eqnarray}
% 	\end{aligned}
% 	\end{equation}
	where
% 	\begin{equation}
% 	\begin{aligned}
	\begin{eqnarray}
	A_1&=&2(c_2 c_1^3-2 c_1^2-3 c_2 c_1+2)\nonumber\\
	A_2&=&-(3 c_1^3 c_2-7 c_1^2-9 c_1 c_2+5)\nonumber\\
	A_3&=&-2(c_1^2+1) (2 c_1 c_2-3)\nonumber\\
	A_4&=&2(c_1^2+1)(c_1 c_2+1)\nonumber\\
	A_5&=&2(-c_2 c_1^3+c_1^2+2c_2 c_1\nonumber\\
	A_6&=&3(c_1^2+1) (c_1 c_2-1)\nonumber\\
	A_7&=&-\tfrac{1}{2}(-5 c_2 c_1^3+9 c_1^2+15 c_2 c_1-3)\nonumber\\
	A_8&=& \tfrac{1}{2}(1+c_1^2)\left(5 c_1 c_2-1\right)\nonumber\\
	A_9&=&(1+c_1^2)(1+c_1c_2) ,\nonumber
	\end{eqnarray}
% 		\end{aligned}
% 	\end{equation}
        and
        	\begin{eqnarray}
% 		\begin{equation}
% 	\begin{aligned}
	B_1&=&2(4c_1+c_2-3c_1^2c_2)\nonumber\\
	B_2&=&(c_1^3+9 c_1^2 c_2-11 c_1-3 c_2)\nonumber\\
	B_3&=&2 c_1( c_1^2 + 1)\nonumber\\
	B_4&=&2(c_1^2+1) (c_1-c_2) \nonumber\\
	B_5&=&(c_1^3+5 c_2 c_1^2-c_1-c_2)\nonumber\\
	B_6&=&-3(c_1^2+1) (c_1+c_2)\nonumber\\
	B_7&=&\tfrac{1}{2} \left(-3 c_1^3-15 c_2 c_1^2+9 c_1+5 c_2\right)\nonumber\\
	B_8&=&\tfrac{1}{2}\left(c_1^2+1\right) \left(c_1+5 c_2\right)\nonumber\\
	B_9&=&(1+c_1^2)(c_2-c_1) . \nonumber
	\end{eqnarray}
	% 	\end{aligned}
% 	\end{equation}
	
        Additionally,
        \begin{equation}
         	D=(c_1^2+1)(c_2-c_1)
        \end{equation}


%   \bibliography{bibliografia}

\begin{thebibliography}{70}%
\makeatletter
\providecommand \@ifxundefined [1]{%
 \@ifx{#1\undefined}
}%
\providecommand \@ifnum [1]{%
 \ifnum #1\expandafter \@firstoftwo
 \else \expandafter \@secondoftwo
 \fi
}%
\providecommand \@ifx [1]{%
 \ifx #1\expandafter \@firstoftwo
 \else \expandafter \@secondoftwo
 \fi
}%
\providecommand \natexlab [1]{#1}%
\providecommand \enquote  [1]{``#1''}%
\providecommand \bibnamefont  [1]{#1}%
\providecommand \bibfnamefont [1]{#1}%
\providecommand \citenamefont [1]{#1}%
\providecommand \href@noop [0]{\@secondoftwo}%
\providecommand \href [0]{\begingroup \@sanitize@url \@href}%
\providecommand \@href[1]{\@@startlink{#1}\@@href}%
\providecommand \@@href[1]{\endgroup#1\@@endlink}%
\providecommand \@sanitize@url [0]{\catcode `\\12\catcode `\$12\catcode
  `\&12\catcode `\#12\catcode `\^12\catcode `\_12\catcode `\%12\relax}%
\providecommand \@@startlink[1]{}%
\providecommand \@@endlink[0]{}%
\providecommand \url  [0]{\begingroup\@sanitize@url \@url }%
\providecommand \@url [1]{\endgroup\@href {#1}{\urlprefix }}%
\providecommand \urlprefix  [0]{URL }%
\providecommand \Eprint [0]{\href }%
\providecommand \doibase [0]{http://dx.doi.org/}%
\providecommand \selectlanguage [0]{\@gobble}%
\providecommand \bibinfo  [0]{\@secondoftwo}%
\providecommand \bibfield  [0]{\@secondoftwo}%
\providecommand \translation [1]{[#1]}%
\providecommand \BibitemOpen [0]{}%
\providecommand \bibitemStop [0]{}%
\providecommand \bibitemNoStop [0]{.\EOS\space}%
\providecommand \EOS [0]{\spacefactor3000\relax}%
\providecommand \BibitemShut  [1]{\csname bibitem#1\endcsname}%
\let\auto@bib@innerbib\@empty
%</preamble>
\bibitem [{\citenamefont {Winfree}(1980)}]{Win80}%
  \BibitemOpen
  \bibfield  {author} {\bibinfo {author} {\bibfnamefont {A.~T.}\ \bibnamefont
  {Winfree}},\ }\href@noop {} {\emph {\bibinfo {title} {The Geometry of
  Biological Time}}}\ (\bibinfo  {publisher} {Springer},\ \bibinfo {address}
  {New York},\ \bibinfo {year} {1980})\BibitemShut {NoStop}%
\bibitem [{\citenamefont {Hoppensteadt}\ and\ \citenamefont
  {Izhikevich}(1997)}]{HI97}%
  \BibitemOpen
  \bibfield  {author} {\bibinfo {author} {\bibfnamefont {F.~C.}\ \bibnamefont
  {Hoppensteadt}}\ and\ \bibinfo {author} {\bibfnamefont {E.~M.}\ \bibnamefont
  {Izhikevich}},\ }\href@noop {} {\emph {\bibinfo {title} {Weakly connected
  neural networks.}}}\ (\bibinfo  {publisher} {Spinger Verlag},\ \bibinfo
  {address} {N.Y.},\ \bibinfo {year} {1997})\BibitemShut {NoStop}%
\bibitem [{\citenamefont {Strogatz}(2003)}]{Str03}%
  \BibitemOpen
  \bibfield  {author} {\bibinfo {author} {\bibfnamefont {S.~H.}\ \bibnamefont
  {Strogatz}},\ }\href@noop {} {\emph {\bibinfo {title} {Sync: {T}he emerging
  science of spontaneous order.}}}\ (\bibinfo  {publisher} {Hyperion Press},\
  \bibinfo {address} {New York},\ \bibinfo {year} {2003})\BibitemShut {NoStop}%
\bibitem [{\citenamefont {Pikovsky}\ \emph {et~al.}(2001)\citenamefont
  {Pikovsky}, \citenamefont {Rosenblum},\ and\ \citenamefont {Kurths}}]{PRK01}%
  \BibitemOpen
  \bibfield  {author} {\bibinfo {author} {\bibfnamefont {A.~S.}\ \bibnamefont
  {Pikovsky}}, \bibinfo {author} {\bibfnamefont {M.~G.}\ \bibnamefont
  {Rosenblum}}, \ and\ \bibinfo {author} {\bibfnamefont {J.}~\bibnamefont
  {Kurths}},\ }\href@noop {} {\emph {\bibinfo {title} {Synchronization, a
  Universal Concept in Nonlinear Sciences}}}\ (\bibinfo  {publisher} {Cambridge
  University Press},\ \bibinfo {address} {Cambridge},\ \bibinfo {year}
  {2001})\BibitemShut {NoStop}%
\bibitem [{\citenamefont {Mertens}\ and\ \citenamefont
  {Weaver}(2011{\natexlab{a}})}]{mertens11}%
  \BibitemOpen
  \bibfield  {author} {\bibinfo {author} {\bibfnamefont {D.}~\bibnamefont
  {Mertens}}\ and\ \bibinfo {author} {\bibfnamefont {R.}~\bibnamefont
  {Weaver}},\ }\bibfield  {title} {\enquote {\bibinfo {title} {Synchronization
  and stimulated emission in an array of mechanical phase oscillators on a
  resonant support},}\ }\href {\doibase 10.1103/PhysRevE.83.046221} {\bibfield
  {journal} {\bibinfo  {journal} {Phys. Rev. E}\ }\textbf {\bibinfo {volume}
  {83}},\ \bibinfo {pages} {046221} (\bibinfo {year}
  {2011}{\natexlab{a}})}\BibitemShut {NoStop}%
\bibitem [{\citenamefont {Mertens}\ and\ \citenamefont
  {Weaver}(2011{\natexlab{b}})}]{mertens11b}%
  \BibitemOpen
  \bibfield  {author} {\bibinfo {author} {\bibfnamefont {D.}~\bibnamefont
  {Mertens}}\ and\ \bibinfo {author} {\bibfnamefont {R.}~\bibnamefont
  {Weaver}},\ }\bibfield  {title} {\enquote {\bibinfo {title} {Individual and
  collective behavior of vibrating motors interacting through a resonant
  plate},}\ }\href {\doibase 10.1002/cplx.20352} {\bibfield  {journal}
  {\bibinfo  {journal} {Complexity}\ }\textbf {\bibinfo {volume} {16}},\
  \bibinfo {pages} {45--53} (\bibinfo {year} {2011}{\natexlab{b}})}\BibitemShut
  {NoStop}%
\bibitem [{\citenamefont {Kiss}\ \emph {et~al.}(2007)\citenamefont {Kiss},
  \citenamefont {Rusin}, \citenamefont {Kori},\ and\ \citenamefont
  {Hudson}}]{kiss07}%
  \BibitemOpen
  \bibfield  {author} {\bibinfo {author} {\bibfnamefont {I.~Z.}\ \bibnamefont
  {Kiss}}, \bibinfo {author} {\bibfnamefont {C.~G.}\ \bibnamefont {Rusin}},
  \bibinfo {author} {\bibfnamefont {H.}~\bibnamefont {Kori}}, \ and\ \bibinfo
  {author} {\bibfnamefont {J.~L.}\ \bibnamefont {Hudson}},\ }\bibfield  {title}
  {\enquote {\bibinfo {title} {Engineering complex dynamical structures:
  Sequential patterns and desynchronization},}\ }\href {\doibase
  10.1126/science.1140858} {\bibfield  {journal} {\bibinfo  {journal}
  {Science}\ }\textbf {\bibinfo {volume} {316}},\ \bibinfo {pages} {1886--1889}
  (\bibinfo {year} {2007})}\BibitemShut {NoStop}%
\bibitem [{\citenamefont {Totz}\ \emph {et~al.}(2015)\citenamefont {Totz},
  \citenamefont {Snari}, \citenamefont {Yengi}, \citenamefont {Tinsley},
  \citenamefont {Engel},\ and\ \citenamefont {Showalter}}]{totz15}%
  \BibitemOpen
  \bibfield  {author} {\bibinfo {author} {\bibfnamefont {J.~F.}\ \bibnamefont
  {Totz}}, \bibinfo {author} {\bibfnamefont {R.}~\bibnamefont {Snari}},
  \bibinfo {author} {\bibfnamefont {D.}~\bibnamefont {Yengi}}, \bibinfo
  {author} {\bibfnamefont {M.~R.}\ \bibnamefont {Tinsley}}, \bibinfo {author}
  {\bibfnamefont {H.}~\bibnamefont {Engel}}, \ and\ \bibinfo {author}
  {\bibfnamefont {K.}~\bibnamefont {Showalter}},\ }\bibfield  {title} {\enquote
  {\bibinfo {title} {Phase-lag synchronization in networks of coupled chemical
  oscillators},}\ }\href {\doibase 10.1103/PhysRevE.92.022819} {\bibfield
  {journal} {\bibinfo  {journal} {Phys. Rev. E}\ }\textbf {\bibinfo {volume}
  {92}},\ \bibinfo {pages} {022819} (\bibinfo {year} {2015})}\BibitemShut
  {NoStop}%
\bibitem [{\citenamefont {Wiesenfeld}\ and\ \citenamefont
  {Swift}(1995)}]{wiesenfeld95}%
  \BibitemOpen
  \bibfield  {author} {\bibinfo {author} {\bibfnamefont {K.}~\bibnamefont
  {Wiesenfeld}}\ and\ \bibinfo {author} {\bibfnamefont {J.~W.}\ \bibnamefont
  {Swift}},\ }\bibfield  {title} {\enquote {\bibinfo {title} {Averaged
  equations for {J}osephson junction series arrays},}\ }\href {\doibase
  10.1103/PhysRevE.51.1020} {\bibfield  {journal} {\bibinfo  {journal} {Phys.
  Rev. E}\ }\textbf {\bibinfo {volume} {51}},\ \bibinfo {pages} {1020--1025}
  (\bibinfo {year} {1995})}\BibitemShut {NoStop}%
\bibitem [{\citenamefont {Wiesenfeld}\ \emph {et~al.}(1996)\citenamefont
  {Wiesenfeld}, \citenamefont {Colet},\ and\ \citenamefont {Strogatz}}]{WCS96}%
  \BibitemOpen
  \bibfield  {author} {\bibinfo {author} {\bibfnamefont {K.}~\bibnamefont
  {Wiesenfeld}}, \bibinfo {author} {\bibfnamefont {P.}~\bibnamefont {Colet}}, \
  and\ \bibinfo {author} {\bibfnamefont {S.~H.}\ \bibnamefont {Strogatz}},\
  }\bibfield  {title} {\enquote {\bibinfo {title} {Synchronization transitions
  in a disordered {J}osephson series array},}\ }\href {\doibase
  10.1103/PhysRevLett.76.404} {\bibfield  {journal} {\bibinfo  {journal} {Phys.
  Rev. Lett.}\ }\textbf {\bibinfo {volume} {76}},\ \bibinfo {pages} {404--407}
  (\bibinfo {year} {1996})}\BibitemShut {NoStop}%
\bibitem [{\citenamefont {Temirbayev}\ \emph {et~al.}(2012)\citenamefont
  {Temirbayev}, \citenamefont {Zhanabaev}, \citenamefont {Tarasov},
  \citenamefont {Ponomarenko},\ and\ \citenamefont {Rosenblum}}]{temirbayev12}%
  \BibitemOpen
  \bibfield  {author} {\bibinfo {author} {\bibfnamefont {A.~A.}\ \bibnamefont
  {Temirbayev}}, \bibinfo {author} {\bibfnamefont {Z.~Zh.}\ \bibnamefont
  {Zhanabaev}}, \bibinfo {author} {\bibfnamefont {S.~B.}\ \bibnamefont
  {Tarasov}}, \bibinfo {author} {\bibfnamefont {V.~I.}\ \bibnamefont
  {Ponomarenko}}, \ and\ \bibinfo {author} {\bibfnamefont {M.}~\bibnamefont
  {Rosenblum}},\ }\bibfield  {title} {\enquote {\bibinfo {title} {Experiments
  on oscillator ensembles with global nonlinear coupling},}\ }\href {\doibase
  10.1103/PhysRevE.85.015204} {\bibfield  {journal} {\bibinfo  {journal} {Phys.
  Rev. E}\ }\textbf {\bibinfo {volume} {85}},\ \bibinfo {pages} {015204(R)}
  (\bibinfo {year} {2012})}\BibitemShut {NoStop}%
\bibitem [{\citenamefont {English}\ \emph {et~al.}(2015)\citenamefont
  {English}, \citenamefont {Zeng},\ and\ \citenamefont {Mertens}}]{english15}%
  \BibitemOpen
  \bibfield  {author} {\bibinfo {author} {\bibfnamefont {L.~Q.}\ \bibnamefont
  {English}}, \bibinfo {author} {\bibfnamefont {Z.}~\bibnamefont {Zeng}}, \
  and\ \bibinfo {author} {\bibfnamefont {D.}~\bibnamefont {Mertens}},\
  }\bibfield  {title} {\enquote {\bibinfo {title} {Experimental study of
  synchronization of coupled electrical self-oscillators and comparison to the
  {S}akaguchi-{K}uramoto model},}\ }\href {\doibase 10.1103/PhysRevE.92.052912}
  {\bibfield  {journal} {\bibinfo  {journal} {Phys. Rev. E}\ }\textbf {\bibinfo
  {volume} {92}},\ \bibinfo {pages} {052912} (\bibinfo {year}
  {2015})}\BibitemShut {NoStop}%
\bibitem [{\citenamefont {Matheny}\ \emph {et~al.}(2019)\citenamefont
  {Matheny}, \citenamefont {Emenheiser}, \citenamefont {Fon}, \citenamefont
  {Chapman}, \citenamefont {Salova}, \citenamefont {Rohden}, \citenamefont
  {Li}, \citenamefont {Hudoba~de Badyn}, \citenamefont {P{\'o}sfai},
  \citenamefont {Duenas-Osorio}, \citenamefont {Mesbahi}, \citenamefont
  {Crutchfield}, \citenamefont {Cross}, \citenamefont
  {D{\textquoteright}Souza},\ and\ \citenamefont {Roukes}}]{matheny19}%
  \BibitemOpen
  \bibfield  {author} {\bibinfo {author} {\bibfnamefont {M.~H.}\ \bibnamefont
  {Matheny}}, \bibinfo {author} {\bibfnamefont {J.}~\bibnamefont {Emenheiser}},
  \bibinfo {author} {\bibfnamefont {W.}~\bibnamefont {Fon}}, \bibinfo {author}
  {\bibfnamefont {A.}~\bibnamefont {Chapman}}, \bibinfo {author} {\bibfnamefont
  {A.}~\bibnamefont {Salova}}, \bibinfo {author} {\bibfnamefont
  {M.}~\bibnamefont {Rohden}}, \bibinfo {author} {\bibfnamefont
  {J.}~\bibnamefont {Li}}, \bibinfo {author} {\bibfnamefont {M.}~\bibnamefont
  {Hudoba~de Badyn}}, \bibinfo {author} {\bibfnamefont {M.}~\bibnamefont
  {P{\'o}sfai}}, \bibinfo {author} {\bibfnamefont {L.}~\bibnamefont
  {Duenas-Osorio}}, \bibinfo {author} {\bibfnamefont {M.}~\bibnamefont
  {Mesbahi}}, \bibinfo {author} {\bibfnamefont {J.~P.}\ \bibnamefont
  {Crutchfield}}, \bibinfo {author} {\bibfnamefont {M.~C.}\ \bibnamefont
  {Cross}}, \bibinfo {author} {\bibfnamefont {R.~M.}\ \bibnamefont
  {D{\textquoteright}Souza}}, \ and\ \bibinfo {author} {\bibfnamefont {M.~L.}\
  \bibnamefont {Roukes}},\ }\bibfield  {title} {\enquote {\bibinfo {title}
  {Exotic states in a simple network of nanoelectromechanical oscillators},}\
  }\href {\doibase 10.1126/science.aav7932} {\bibfield  {journal} {\bibinfo
  {journal} {Science}\ }\textbf {\bibinfo {volume} {363}},\ \bibinfo {pages}
  {eaav7932} (\bibinfo {year} {2019})}\BibitemShut {NoStop}%
\bibitem [{\citenamefont {Kuramoto}(1984)}]{Kur84}%
  \BibitemOpen
  \bibfield  {author} {\bibinfo {author} {\bibfnamefont {Y.}~\bibnamefont
  {Kuramoto}},\ }\href@noop {} {\emph {\bibinfo {title} {Chemical Oscillations,
  Waves, and Turbulence}}}\ (\bibinfo  {publisher} {{S}pringer-{V}erlag},\
  \bibinfo {address} {Berlin},\ \bibinfo {year} {1984})\BibitemShut {NoStop}%
\bibitem [{\citenamefont {Nakao}(2016)}]{nakao16}%
  \BibitemOpen
  \bibfield  {author} {\bibinfo {author} {\bibfnamefont {H.}~\bibnamefont
  {Nakao}},\ }\bibfield  {title} {\enquote {\bibinfo {title} {Phase reduction
  approach to synchronisation of nonlinear oscillators},}\ }\href {\doibase
  10.1080/00107514.2015.1094987} {\bibfield  {journal} {\bibinfo  {journal}
  {Contemp. Phys.}\ }\textbf {\bibinfo {volume} {57}},\ \bibinfo {pages}
  {188--214} (\bibinfo {year} {2016})}\BibitemShut {NoStop}%
\bibitem [{\citenamefont {Monga}\ \emph {et~al.}(2019)\citenamefont {Monga},
  \citenamefont {Wilson}, \citenamefont {Matchen},\ and\ \citenamefont
  {Moehlis}}]{monga19}%
  \BibitemOpen
  \bibfield  {author} {\bibinfo {author} {\bibfnamefont {B.}~\bibnamefont
  {Monga}}, \bibinfo {author} {\bibfnamefont {D.}~\bibnamefont {Wilson}},
  \bibinfo {author} {\bibfnamefont {T.}~\bibnamefont {Matchen}}, \ and\
  \bibinfo {author} {\bibfnamefont {J.}~\bibnamefont {Moehlis}},\ }\bibfield
  {title} {\enquote {\bibinfo {title} {Phase reduction and phase-based optimal
  control for biological systems: a tutorial},}\ }\href {\doibase
  10.1007/s00422-018-0780-z} {\bibfield  {journal} {\bibinfo  {journal} {Biol.
  Cybern.}\ }\textbf {\bibinfo {volume} {113}},\ \bibinfo {pages} {11--46}
  (\bibinfo {year} {2019})}\BibitemShut {NoStop}%
\bibitem [{\citenamefont {Tanaka}\ and\ \citenamefont
  {Aoyagi}(2011)}]{tanaka11}%
  \BibitemOpen
  \bibfield  {author} {\bibinfo {author} {\bibfnamefont {T.}~\bibnamefont
  {Tanaka}}\ and\ \bibinfo {author} {\bibfnamefont {T.}~\bibnamefont
  {Aoyagi}},\ }\bibfield  {title} {\enquote {\bibinfo {title} {Multistable
  attractors in a network of phase oscillators with three-body interactions},}\
  }\href {\doibase 10.1103/PhysRevLett.106.224101} {\bibfield  {journal}
  {\bibinfo  {journal} {Phys. Rev. Lett.}\ }\textbf {\bibinfo {volume} {106}},\
  \bibinfo {pages} {224101} (\bibinfo {year} {2011})}\BibitemShut {NoStop}%
\bibitem [{\citenamefont {Ashwin}\ and\ \citenamefont
  {Rodrigues}(2016)}]{ashwin16}%
  \BibitemOpen
  \bibfield  {author} {\bibinfo {author} {\bibfnamefont {P.}~\bibnamefont
  {Ashwin}}\ and\ \bibinfo {author} {\bibfnamefont {A.}~\bibnamefont
  {Rodrigues}},\ }\bibfield  {title} {\enquote {\bibinfo {title} {Hopf normal
  form with sn symmetry and reduction to systems of nonlinearly coupled phase
  oscillators},}\ }\href {\doibase https://doi.org/10.1016/j.physd.2016.02.009}
  {\bibfield  {journal} {\bibinfo  {journal} {Physica D}\ }\textbf {\bibinfo
  {volume} {325}},\ \bibinfo {pages} {14 -- 24} (\bibinfo {year}
  {2016})}\BibitemShut {NoStop}%
\bibitem [{\citenamefont {Rosenblum}\ and\ \citenamefont
  {Pikovsky}(2007)}]{RP07}%
  \BibitemOpen
  \bibfield  {author} {\bibinfo {author} {\bibfnamefont {M.}~\bibnamefont
  {Rosenblum}}\ and\ \bibinfo {author} {\bibfnamefont {A.}~\bibnamefont
  {Pikovsky}},\ }\bibfield  {title} {\enquote {\bibinfo {title} {Self-organized
  quasiperiodicity in oscillator ensembles with global nonlinear coupling},}\
  }\href {\doibase 10.1103/PhysRevLett.98.064101} {\bibfield  {journal}
  {\bibinfo  {journal} {Phys. Rev. Lett.}\ }\textbf {\bibinfo {volume} {98}},\
  \bibinfo {pages} {064101} (\bibinfo {year} {2007})}\BibitemShut {NoStop}%
\bibitem [{\citenamefont {Kralemann}\ \emph {et~al.}(2011)\citenamefont
  {Kralemann}, \citenamefont {Pikovsky},\ and\ \citenamefont
  {Rosenblum}}]{kralemann11}%
  \BibitemOpen
  \bibfield  {author} {\bibinfo {author} {\bibfnamefont {B.}~\bibnamefont
  {Kralemann}}, \bibinfo {author} {\bibfnamefont {A.}~\bibnamefont {Pikovsky}},
  \ and\ \bibinfo {author} {\bibfnamefont {M.}~\bibnamefont {Rosenblum}},\
  }\bibfield  {title} {\enquote {\bibinfo {title} {Reconstructing phase
  dynamics of oscillator networks},}\ }\href {\doibase 10.1063/1.3597647}
  {\bibfield  {journal} {\bibinfo  {journal} {Chaos}\ }\textbf {\bibinfo
  {volume} {21}},\ \bibinfo {pages} {025104} (\bibinfo {year}
  {2011})}\BibitemShut {NoStop}%
\bibitem [{\citenamefont {Kralemann}\ \emph {et~al.}(2014)\citenamefont
  {Kralemann}, \citenamefont {Pikovsky},\ and\ \citenamefont
  {Rosenblum}}]{kralemann14}%
  \BibitemOpen
  \bibfield  {author} {\bibinfo {author} {\bibfnamefont {B.}~\bibnamefont
  {Kralemann}}, \bibinfo {author} {\bibfnamefont {A.}~\bibnamefont {Pikovsky}},
  \ and\ \bibinfo {author} {\bibfnamefont {M.}~\bibnamefont {Rosenblum}},\
  }\bibfield  {title} {\enquote {\bibinfo {title} {Reconstructing effective
  phase connectivity of oscillator networks from observations},}\ }\href
  {\doibase 10.1088/1367-2630/16/8/085013} {\bibfield  {journal} {\bibinfo
  {journal} {New J. Phys.}\ }\textbf {\bibinfo {volume} {16}},\ \bibinfo
  {pages} {085013} (\bibinfo {year} {2014})}\BibitemShut {NoStop}%
\bibitem [{\citenamefont {Pikovsky}\ and\ \citenamefont
  {Rosenblum}(2015)}]{PikRos15}%
  \BibitemOpen
  \bibfield  {author} {\bibinfo {author} {\bibfnamefont {A.}~\bibnamefont
  {Pikovsky}}\ and\ \bibinfo {author} {\bibfnamefont {M.}~\bibnamefont
  {Rosenblum}},\ }\bibfield  {title} {\enquote {\bibinfo {title} {Dynamics of
  globally coupled oscillators: Progress and perspectives},}\ }\href@noop {}
  {\bibfield  {journal} {\bibinfo  {journal} {Chaos}\ }\textbf {\bibinfo
  {volume} {25}},\ \bibinfo {eid} {097616} (\bibinfo {year}
  {2015})}\BibitemShut {NoStop}%
\bibitem [{\citenamefont {Hakim}\ and\ \citenamefont {Rappel}(1992)}]{HR92}%
  \BibitemOpen
  \bibfield  {author} {\bibinfo {author} {\bibfnamefont {V.}~\bibnamefont
  {Hakim}}\ and\ \bibinfo {author} {\bibfnamefont {W.~J.}\ \bibnamefont
  {Rappel}},\ }\bibfield  {title} {\enquote {\bibinfo {title} {Dynamics of the
  globally coupled complex {G}inzburg-{L}andau equation.}}\ }\href@noop {}
  {\bibfield  {journal} {\bibinfo  {journal} {Phys. Rev. A}\ }\textbf {\bibinfo
  {volume} {46}},\ \bibinfo {pages} {R7347--R7350} (\bibinfo {year}
  {1992})}\BibitemShut {NoStop}%
\bibitem [{\citenamefont {Nakagawa}\ and\ \citenamefont
  {Kuramoto}(1993)}]{NK93}%
  \BibitemOpen
  \bibfield  {author} {\bibinfo {author} {\bibfnamefont {N.}~\bibnamefont
  {Nakagawa}}\ and\ \bibinfo {author} {\bibfnamefont {Y.}~\bibnamefont
  {Kuramoto}},\ }\bibfield  {title} {\enquote {\bibinfo {title} {Collective
  chaos in a population of globally coupled oscillators},}\ }\href@noop {}
  {\bibfield  {journal} {\bibinfo  {journal} {Prog. Theor. Phys.}\ }\textbf
  {\bibinfo {volume} {89}},\ \bibinfo {pages} {313--323} (\bibinfo {year}
  {1993})}\BibitemShut {NoStop}%
\bibitem [{\citenamefont {Nakagawa}\ and\ \citenamefont
  {Kuramoto}(1994)}]{NK94}%
  \BibitemOpen
  \bibfield  {author} {\bibinfo {author} {\bibfnamefont {N.}~\bibnamefont
  {Nakagawa}}\ and\ \bibinfo {author} {\bibfnamefont {Y.}~\bibnamefont
  {Kuramoto}},\ }\bibfield  {title} {\enquote {\bibinfo {title} {From
  collective oscillations to collective chaos in a globally coupled oscillator
  system},}\ }\href@noop {} {\bibfield  {journal} {\bibinfo  {journal} {Physica
  D}\ }\textbf {\bibinfo {volume} {75}},\ \bibinfo {pages} {74--80} (\bibinfo
  {year} {1994})}\BibitemShut {NoStop}%
\bibitem [{\citenamefont {Nakagawa}\ and\ \citenamefont
  {Kuramoto}(1995)}]{NK95}%
  \BibitemOpen
  \bibfield  {author} {\bibinfo {author} {\bibfnamefont {N.}~\bibnamefont
  {Nakagawa}}\ and\ \bibinfo {author} {\bibfnamefont {Y.}~\bibnamefont
  {Kuramoto}},\ }\bibfield  {title} {\enquote {\bibinfo {title} {Anomalous
  lyapunov spectrum in globally coupled oscillators},}\ }\href@noop {}
  {\bibfield  {journal} {\bibinfo  {journal} {Physica D}\ }\textbf {\bibinfo
  {volume} {80}},\ \bibinfo {pages} {307--316} (\bibinfo {year}
  {1995})}\BibitemShut {NoStop}%
\bibitem [{\citenamefont {Chabanol}\ \emph {et~al.}(1997)\citenamefont
  {Chabanol}, \citenamefont {Hakim},\ and\ \citenamefont
  {Rappel}}]{chabanol97}%
  \BibitemOpen
  \bibfield  {author} {\bibinfo {author} {\bibfnamefont {M.-L.}\ \bibnamefont
  {Chabanol}}, \bibinfo {author} {\bibfnamefont {V.}~\bibnamefont {Hakim}}, \
  and\ \bibinfo {author} {\bibfnamefont {W.-J.}\ \bibnamefont {Rappel}},\
  }\bibfield  {title} {\enquote {\bibinfo {title} {Collective chaos and noise
  in the globally coupled complex {G}inzburg-{L}andau equation},}\ }\href
  {\doibase https://doi.org/10.1016/S0167-2789(96)00263-1} {\bibfield
  {journal} {\bibinfo  {journal} {Physica D}\ }\textbf {\bibinfo {volume}
  {103}},\ \bibinfo {pages} {273 -- 293} (\bibinfo {year} {1997})}\BibitemShut
  {NoStop}%
\bibitem [{\citenamefont {Banaji}\ and\ \citenamefont
  {Glendinning}(1999)}]{banaji99}%
  \BibitemOpen
  \bibfield  {author} {\bibinfo {author} {\bibfnamefont {M.}~\bibnamefont
  {Banaji}}\ and\ \bibinfo {author} {\bibfnamefont {P.}~\bibnamefont
  {Glendinning}},\ }\bibfield  {title} {\enquote {\bibinfo {title} {Towards a
  quasi-periodic mean field theory for globally coupled oscillators},}\ }\href
  {\doibase https://doi.org/10.1016/S0375-9601(98)00869-X} {\bibfield
  {journal} {\bibinfo  {journal} {Phys. Lett. A}\ }\textbf {\bibinfo {volume}
  {251}},\ \bibinfo {pages} {297 -- 302} (\bibinfo {year} {1999})}\BibitemShut
  {NoStop}%
\bibitem [{\citenamefont {Banaji}(2002)}]{banaji02}%
  \BibitemOpen
  \bibfield  {author} {\bibinfo {author} {\bibfnamefont {M.}~\bibnamefont
  {Banaji}},\ }\bibfield  {title} {\enquote {\bibinfo {title} {Clustering in
  globally coupled oscillators},}\ }\href {\doibase 10.1080/14689360210148485}
  {\bibfield  {journal} {\bibinfo  {journal} {Dynam. Syst.}\ }\textbf {\bibinfo
  {volume} {17}},\ \bibinfo {pages} {263--285} (\bibinfo {year}
  {2002})}\BibitemShut {NoStop}%
\bibitem [{\citenamefont {Daido}\ and\ \citenamefont {Nakanishi}(2006)}]{DN06}%
  \BibitemOpen
  \bibfield  {author} {\bibinfo {author} {\bibfnamefont {H.}~\bibnamefont
  {Daido}}\ and\ \bibinfo {author} {\bibfnamefont {K.}~\bibnamefont
  {Nakanishi}},\ }\bibfield  {title} {\enquote {\bibinfo {title}
  {Diffusion-induced inhomogeneity in globally coupled oscillators: Swing-by
  mechanism},}\ }\href@noop {} {\bibfield  {journal} {\bibinfo  {journal}
  {Phys. Rev. Lett.}\ }\textbf {\bibinfo {volume} {96}},\ \bibinfo {pages}
  {054101} (\bibinfo {year} {2006})}\BibitemShut {NoStop}%
\bibitem [{\citenamefont {Daido}\ and\ \citenamefont {Nakanishi}(2007)}]{DN07}%
  \BibitemOpen
  \bibfield  {author} {\bibinfo {author} {\bibfnamefont {H.}~\bibnamefont
  {Daido}}\ and\ \bibinfo {author} {\bibfnamefont {K.}~\bibnamefont
  {Nakanishi}},\ }\bibfield  {title} {\enquote {\bibinfo {title} {Aging and
  clustering in globally coupled oscillators},}\ }\href {\doibase
  10.1103/PhysRevE.75.056206} {\bibfield  {journal} {\bibinfo  {journal} {Phys.
  Rev. E}\ }\textbf {\bibinfo {volume} {75}},\ \bibinfo {pages} {056206}
  (\bibinfo {year} {2007})}\BibitemShut {NoStop}%
\bibitem [{\citenamefont {Takeuchi}\ \emph {et~al.}(2009)\citenamefont
  {Takeuchi}, \citenamefont {Ginelli},\ and\ \citenamefont {Chat\'e}}]{TGC09}%
  \BibitemOpen
  \bibfield  {author} {\bibinfo {author} {\bibfnamefont {K.~A.}\ \bibnamefont
  {Takeuchi}}, \bibinfo {author} {\bibfnamefont {F.}~\bibnamefont {Ginelli}}, \
  and\ \bibinfo {author} {\bibfnamefont {H.}~\bibnamefont {Chat\'e}},\
  }\bibfield  {title} {\enquote {\bibinfo {title} {Lyapunov analysis captures
  the collective dynamics of large chaotic systems},}\ }\href {\doibase
  10.1103/PhysRevLett.103.154103} {\bibfield  {journal} {\bibinfo  {journal}
  {Phys. Rev. Lett.}\ }\textbf {\bibinfo {volume} {103}},\ \bibinfo {pages}
  {154103} (\bibinfo {year} {2009})}\BibitemShut {NoStop}%
\bibitem [{\citenamefont {Takeuchi}\ and\ \citenamefont
  {Chat\'e}(2013)}]{TC13}%
  \BibitemOpen
  \bibfield  {author} {\bibinfo {author} {\bibfnamefont {K.~A.}\ \bibnamefont
  {Takeuchi}}\ and\ \bibinfo {author} {\bibfnamefont {H.}~\bibnamefont
  {Chat\'e}},\ }\bibfield  {title} {\enquote {\bibinfo {title} {Collective
  lyapunov modes},}\ }\href {http://stacks.iop.org/1751-8121/46/i=25/a=254007}
  {\bibfield  {journal} {\bibinfo  {journal} {J. Phys. A: Math. Theor.}\
  }\textbf {\bibinfo {volume} {46}},\ \bibinfo {pages} {254007} (\bibinfo
  {year} {2013})}\BibitemShut {NoStop}%
\bibitem [{\citenamefont {Sethia}\ and\ \citenamefont {Sen}(2014)}]{sethia14}%
  \BibitemOpen
  \bibfield  {author} {\bibinfo {author} {\bibfnamefont {G.~C.}\ \bibnamefont
  {Sethia}}\ and\ \bibinfo {author} {\bibfnamefont {A.}~\bibnamefont {Sen}},\
  }\bibfield  {title} {\enquote {\bibinfo {title} {Chimera states: The
  existence criteria revisited},}\ }\href {\doibase
  10.1103/PhysRevLett.112.144101} {\bibfield  {journal} {\bibinfo  {journal}
  {Phys. Rev. Lett.}\ }\textbf {\bibinfo {volume} {112}},\ \bibinfo {pages}
  {144101} (\bibinfo {year} {2014})}\BibitemShut {NoStop}%
\bibitem [{\citenamefont {Ku}\ \emph {et~al.}(2015)\citenamefont {Ku},
  \citenamefont {Girvan},\ and\ \citenamefont {Ott}}]{KGO15}%
  \BibitemOpen
  \bibfield  {author} {\bibinfo {author} {\bibfnamefont {Wai~Lim}\ \bibnamefont
  {Ku}}, \bibinfo {author} {\bibfnamefont {Michelle}\ \bibnamefont {Girvan}}, \
  and\ \bibinfo {author} {\bibfnamefont {Edward}\ \bibnamefont {Ott}},\
  }\bibfield  {title} {\enquote {\bibinfo {title} {Dynamical transitions in
  large systems of mean field-coupled {L}andau-{S}tuart oscillators: Extensive
  chaos and cluster states},}\ }\href@noop {} {\bibfield  {journal} {\bibinfo
  {journal} {Chaos}\ }\textbf {\bibinfo {volume} {25}},\ \bibinfo {pages}
  {123122} (\bibinfo {year} {2015})}\BibitemShut {NoStop}%
\bibitem [{\citenamefont {Rosenblum}\ and\ \citenamefont
  {Pikovsky}(2015)}]{RP15}%
  \BibitemOpen
  \bibfield  {author} {\bibinfo {author} {\bibfnamefont {M.}~\bibnamefont
  {Rosenblum}}\ and\ \bibinfo {author} {\bibfnamefont {A.}~\bibnamefont
  {Pikovsky}},\ }\bibfield  {title} {\enquote {\bibinfo {title} {Two types of
  quasiperiodic partial synchrony in oscillator ensembles},}\ }\href {\doibase
  10.1103/PhysRevE.92.012919} {\bibfield  {journal} {\bibinfo  {journal} {Phys.
  Rev. E}\ }\textbf {\bibinfo {volume} {92}},\ \bibinfo {pages} {012919}
  (\bibinfo {year} {2015})}\BibitemShut {NoStop}%
\bibitem [{\citenamefont {Kemeth}\ \emph {et~al.}(2019)\citenamefont {Kemeth},
  \citenamefont {Haugland},\ and\ \citenamefont {Krischer}}]{kemeth19}%
  \BibitemOpen
  \bibfield  {author} {\bibinfo {author} {\bibfnamefont {F.~P.}\ \bibnamefont
  {Kemeth}}, \bibinfo {author} {\bibfnamefont {S.~W.}\ \bibnamefont
  {Haugland}}, \ and\ \bibinfo {author} {\bibfnamefont {K.}~\bibnamefont
  {Krischer}},\ }\bibfield  {title} {\enquote {\bibinfo {title} {Cluster
  singularity: The unfolding of clustering behavior in globally coupled
  {S}tuart-{L}andau oscillators},}\ }\href {\doibase 10.1063/1.5055839}
  {\bibfield  {journal} {\bibinfo  {journal} {Chaos}\ }\textbf {\bibinfo
  {volume} {29}},\ \bibinfo {pages} {023107} (\bibinfo {year}
  {2019})}\BibitemShut {NoStop}%
\bibitem [{\citenamefont {Clusella}\ and\ \citenamefont {Politi}(2019)}]{CP19}%
  \BibitemOpen
  \bibfield  {author} {\bibinfo {author} {\bibfnamefont {P.}~\bibnamefont
  {Clusella}}\ and\ \bibinfo {author} {\bibfnamefont {A.}~\bibnamefont
  {Politi}},\ }\bibfield  {title} {\enquote {\bibinfo {title} {Between phase
  and amplitude oscillators},}\ }\href {\doibase 10.1103/PhysRevE.99.062201}
  {\bibfield  {journal} {\bibinfo  {journal} {Phys. Rev. E}\ }\textbf {\bibinfo
  {volume} {99}},\ \bibinfo {pages} {062201} (\bibinfo {year}
  {2019})}\BibitemShut {NoStop}%
\bibitem [{\citenamefont {Kuramoto}(1975)}]{kur75}%
  \BibitemOpen
  \bibfield  {author} {\bibinfo {author} {\bibfnamefont {Y.}~\bibnamefont
  {Kuramoto}},\ }\bibfield  {title} {\enquote {\bibinfo {title}
  {Self-entrainment of a population of coupled non-linear oscillators},}\ }in\
  \href@noop {} {\emph {\bibinfo {booktitle} {International Symposium on
  Mathematical Problems in Theoretical Physics}}},\ \bibinfo {series} {Lecture
  Notes in Physics}, Vol.~\bibinfo {volume} {39},\ \bibinfo {editor} {edited
  by\ \bibinfo {editor} {\bibfnamefont {Huzihiro}\ \bibnamefont {Araki}}}\
  (\bibinfo  {publisher} {Springer},\ \bibinfo {address} {Berlin},\ \bibinfo
  {year} {1975})\ pp.\ \bibinfo {pages} {420--422}\BibitemShut {NoStop}%
\bibitem [{\citenamefont {Walgraef}(1997)}]{Walgraef}%
  \BibitemOpen
  \bibfield  {author} {\bibinfo {author} {\bibfnamefont {D.}~\bibnamefont
  {Walgraef}},\ }\href@noop {} {\emph {\bibinfo {title} {Spatio-Temporal
  Pattern Formation}}}\ (\bibinfo  {publisher} {Springer-Verlag},\ \bibinfo
  {address} {New York},\ \bibinfo {year} {1997})\BibitemShut {NoStop}%
\bibitem [{\citenamefont {Aranson}\ and\ \citenamefont {Kramer}(2002)}]{AK02}%
  \BibitemOpen
  \bibfield  {author} {\bibinfo {author} {\bibfnamefont {I.~S.}\ \bibnamefont
  {Aranson}}\ and\ \bibinfo {author} {\bibfnamefont {L.}~\bibnamefont
  {Kramer}},\ }\bibfield  {title} {\enquote {\bibinfo {title} {The world of the
  complex {G}inzburg-{L}andau equation},}\ }\href {\doibase
  10.1103/RevModPhys.74.99} {\bibfield  {journal} {\bibinfo  {journal} {Rev.
  Mod. Phys.}\ }\textbf {\bibinfo {volume} {74}},\ \bibinfo {pages} {99}
  (\bibinfo {year} {2002})}\BibitemShut {NoStop}%
\bibitem [{\citenamefont {Sakaguchi}\ and\ \citenamefont
  {Kuramoto}(1986)}]{SK86}%
  \BibitemOpen
  \bibfield  {author} {\bibinfo {author} {\bibfnamefont {H.}~\bibnamefont
  {Sakaguchi}}\ and\ \bibinfo {author} {\bibfnamefont {Y.}~\bibnamefont
  {Kuramoto}},\ }\bibfield  {title} {\enquote {\bibinfo {title} {A soluble
  active rotator model showing phase transitions via mutual entrainment},}\
  }\href@noop {} {\bibfield  {journal} {\bibinfo  {journal}
  {Prog.~Theor.~Phys.}\ }\textbf {\bibinfo {volume} {76}},\ \bibinfo {pages}
  {576--581} (\bibinfo {year} {1986})}\BibitemShut {NoStop}%
\bibitem [{\citenamefont {Montbri\'o}\ and\ \citenamefont
  {Paz\'o}(2011{\natexlab{a}})}]{MP11p}%
  \BibitemOpen
  \bibfield  {author} {\bibinfo {author} {\bibfnamefont {E.}~\bibnamefont
  {Montbri\'o}}\ and\ \bibinfo {author} {\bibfnamefont {D.}~\bibnamefont
  {Paz\'o}},\ }\bibfield  {title} {\enquote {\bibinfo {title} {Collective
  synchronization in the presence of reactive coupling and shear diversity},}\
  }\href {\doibase 10.1103/PhysRevE.84.046206} {\bibfield  {journal} {\bibinfo
  {journal} {Phys. Rev. E}\ }\textbf {\bibinfo {volume} {84}},\ \bibinfo
  {pages} {046206} (\bibinfo {year} {2011}{\natexlab{a}})}\BibitemShut
  {NoStop}%
\bibitem [{\citenamefont {Montbri\'o}\ and\ \citenamefont
  {Paz\'o}(2011{\natexlab{b}})}]{MP11}%
  \BibitemOpen
  \bibfield  {author} {\bibinfo {author} {\bibfnamefont {E.}~\bibnamefont
  {Montbri\'o}}\ and\ \bibinfo {author} {\bibfnamefont {D.}~\bibnamefont
  {Paz\'o}},\ }\bibfield  {title} {\enquote {\bibinfo {title} {Shear diversity
  prevents collective synchronization},}\ }\href {\doibase
  10.1103/PhysRevLett.106.254101} {\bibfield  {journal} {\bibinfo  {journal}
  {Phys. Rev. Lett.}\ }\textbf {\bibinfo {volume} {106}},\ \bibinfo {pages}
  {254101} (\bibinfo {year} {2011}{\natexlab{b}})}\BibitemShut {NoStop}%
\bibitem [{\citenamefont {Paz{\'o}}\ and\ \citenamefont
  {Montbri{\'o}}(2011)}]{PM11}%
  \BibitemOpen
  \bibfield  {author} {\bibinfo {author} {\bibfnamefont {D.}~\bibnamefont
  {Paz{\'o}}}\ and\ \bibinfo {author} {\bibfnamefont {E.}~\bibnamefont
  {Montbri{\'o}}},\ }\bibfield  {title} {\enquote {\bibinfo {title} {The
  {K}uramoto model with distributed shear},}\ }\href {\doibase
  10.1209/0295-5075/95/60007} {\bibfield  {journal} {\bibinfo  {journal} {EPL
  (Europhys. Lett.)}\ }\textbf {\bibinfo {volume} {95}},\ \bibinfo {pages}
  {60007} (\bibinfo {year} {2011})}\BibitemShut {NoStop}%
\bibitem [{\citenamefont {Winfree}(1974)}]{win74}%
  \BibitemOpen
  \bibfield  {author} {\bibinfo {author} {\bibfnamefont {A.~T.}\ \bibnamefont
  {Winfree}},\ }\bibfield  {title} {\enquote {\bibinfo {title} {Patterns of
  phase compromise in biological cycles},}\ }\href {\doibase
  10.1007/BF02339491} {\bibfield  {journal} {\bibinfo  {journal} {J. Math.
  Biol.}\ }\textbf {\bibinfo {volume} {1}},\ \bibinfo {pages} {73--93}
  (\bibinfo {year} {1974})}\BibitemShut {NoStop}%
\bibitem [{\citenamefont {Guckenheimer}(1975)}]{guc75}%
  \BibitemOpen
  \bibfield  {author} {\bibinfo {author} {\bibfnamefont {J.}~\bibnamefont
  {Guckenheimer}},\ }\bibfield  {title} {\enquote {\bibinfo {title} {Isochrons
  and phaseless sets},}\ }\href {\doibase 10.1007/BF01273747} {\bibfield
  {journal} {\bibinfo  {journal} {J. Math. Biol.}\ }\textbf {\bibinfo {volume}
  {1}},\ \bibinfo {pages} {259--273} (\bibinfo {year} {1975})}\BibitemShut
  {NoStop}%
\bibitem [{\citenamefont {Coddington}\ and\ \citenamefont
  {Levinson}(1955)}]{CL}%
  \BibitemOpen
  \bibfield  {author} {\bibinfo {author} {\bibfnamefont {E.~A.}\ \bibnamefont
  {Coddington}}\ and\ \bibinfo {author} {\bibfnamefont {N.}~\bibnamefont
  {Levinson}},\ }\href@noop {} {\emph {\bibinfo {title} {Theory of Ordinary
  Differential Equations}}}\ (\bibinfo  {publisher} {McGraw-Hill},\ \bibinfo
  {address} {New York},\ \bibinfo {year} {1955})\ Chap.~\bibinfo {chapter}
  {13}\BibitemShut {NoStop}%
\bibitem [{\citenamefont {Daido}(1996)}]{Dai96}%
  \BibitemOpen
  \bibfield  {author} {\bibinfo {author} {\bibfnamefont {H.}~\bibnamefont
  {Daido}},\ }\bibfield  {title} {\enquote {\bibinfo {title} {Onset of
  cooperative entrainment in limit-cycle oscillators with uniform all-to-all
  interactions: bifurcation of the order function},}\ }\href@noop {} {\bibfield
   {journal} {\bibinfo  {journal} {Physica D}\ }\textbf {\bibinfo {volume}
  {91}},\ \bibinfo {pages} {24--66} (\bibinfo {year} {1996})}\BibitemShut
  {NoStop}%
\bibitem [{\citenamefont {{van Vreeswijk}}(1996)}]{Vre96}%
  \BibitemOpen
  \bibfield  {author} {\bibinfo {author} {\bibfnamefont {C.}~\bibnamefont {{van
  Vreeswijk}}},\ }\bibfield  {title} {\enquote {\bibinfo {title} {Partial
  synchronization in populations of pulse-coupled oscillators},}\ }\href
  {\doibase 10.1103/PhysRevE.54.5522} {\bibfield  {journal} {\bibinfo
  {journal} {Phys. Rev. E}\ }\textbf {\bibinfo {volume} {54}},\ \bibinfo
  {pages} {5522--5537} (\bibinfo {year} {1996})}\BibitemShut {NoStop}%
\bibitem [{\citenamefont {Golomb}\ \emph {et~al.}(1992)\citenamefont {Golomb},
  \citenamefont {Hansel}, \citenamefont {Shraiman},\ and\ \citenamefont
  {Sompolinsky}}]{GHS+92}%
  \BibitemOpen
  \bibfield  {author} {\bibinfo {author} {\bibfnamefont {D.}~\bibnamefont
  {Golomb}}, \bibinfo {author} {\bibfnamefont {D.}~\bibnamefont {Hansel}},
  \bibinfo {author} {\bibfnamefont {B.}~\bibnamefont {Shraiman}}, \ and\
  \bibinfo {author} {\bibfnamefont {H.}~\bibnamefont {Sompolinsky}},\
  }\bibfield  {title} {\enquote {\bibinfo {title} {Clustering in globally
  coupled phase oscillators},}\ }\href {\doibase 10.1103/PhysRevA.45.3516}
  {\bibfield  {journal} {\bibinfo  {journal} {Phys. Rev. A}\ }\textbf {\bibinfo
  {volume} {45}},\ \bibinfo {pages} {3516--3530} (\bibinfo {year}
  {1992})}\BibitemShut {NoStop}%
\bibitem [{\citenamefont {Hansel}\ \emph {et~al.}(1993)\citenamefont {Hansel},
  \citenamefont {Mato},\ and\ \citenamefont {Meunier}}]{hmm93}%
  \BibitemOpen
  \bibfield  {author} {\bibinfo {author} {\bibfnamefont {D.}~\bibnamefont
  {Hansel}}, \bibinfo {author} {\bibfnamefont {G.}~\bibnamefont {Mato}}, \ and\
  \bibinfo {author} {\bibfnamefont {C.}~\bibnamefont {Meunier}},\ }\bibfield
  {title} {\enquote {\bibinfo {title} {Clustering and slow switching in
  globally coupled phase oscillators},}\ }\href {\doibase
  10.1103/PhysRevE.48.3470} {\bibfield  {journal} {\bibinfo  {journal} {Phys.
  Rev. E}\ }\textbf {\bibinfo {volume} {48}},\ \bibinfo {pages} {3470--3477}
  (\bibinfo {year} {1993})}\BibitemShut {NoStop}%
\bibitem [{\citenamefont {Kemeth}\ \emph {et~al.}(2018)\citenamefont {Kemeth},
  \citenamefont {Haugland},\ and\ \citenamefont {Krischer}}]{kemeth18}%
  \BibitemOpen
  \bibfield  {author} {\bibinfo {author} {\bibfnamefont {F.~P.}\ \bibnamefont
  {Kemeth}}, \bibinfo {author} {\bibfnamefont {S.~W.}\ \bibnamefont
  {Haugland}}, \ and\ \bibinfo {author} {\bibfnamefont {K.}~\bibnamefont
  {Krischer}},\ }\bibfield  {title} {\enquote {\bibinfo {title} {Symmetries of
  chimera states},}\ }\href {\doibase 10.1103/PhysRevLett.120.214101}
  {\bibfield  {journal} {\bibinfo  {journal} {Phys. Rev. Lett.}\ }\textbf
  {\bibinfo {volume} {120}},\ \bibinfo {pages} {214101} (\bibinfo {year}
  {2018})}\BibitemShut {NoStop}%
\bibitem [{\citenamefont {Ashwin}\ and\ \citenamefont
  {Swift}(1992)}]{ashwin92}%
  \BibitemOpen
  \bibfield  {author} {\bibinfo {author} {\bibfnamefont {P.}~\bibnamefont
  {Ashwin}}\ and\ \bibinfo {author} {\bibfnamefont {J.~W.}\ \bibnamefont
  {Swift}},\ }\bibfield  {title} {\enquote {\bibinfo {title} {The dynamics ofn
  weakly coupled identical oscillators},}\ }\href {\doibase 10.1007/BF02429852}
  {\bibfield  {journal} {\bibinfo  {journal} {J. Nonlin. Sci.}\ }\textbf
  {\bibinfo {volume} {2}},\ \bibinfo {pages} {69--108} (\bibinfo {year}
  {1992})}\BibitemShut {NoStop}%
\bibitem [{\citenamefont {Ashwin}\ \emph {et~al.}(2008)\citenamefont {Ashwin},
  \citenamefont {Burylko},\ and\ \citenamefont {Maistrenko}}]{ashwin08}%
  \BibitemOpen
  \bibfield  {author} {\bibinfo {author} {\bibfnamefont {P.}~\bibnamefont
  {Ashwin}}, \bibinfo {author} {\bibfnamefont {O.}~\bibnamefont {Burylko}}, \
  and\ \bibinfo {author} {\bibfnamefont {Y.}~\bibnamefont {Maistrenko}},\
  }\bibfield  {title} {\enquote {\bibinfo {title} {Bifurcation to heteroclinic
  cycles and sensitivity in three and four coupled phase oscillators},}\ }\href
  {\doibase https://doi.org/10.1016/j.physd.2007.09.015} {\bibfield  {journal}
  {\bibinfo  {journal} {Physica D}\ }\textbf {\bibinfo {volume} {237}},\
  \bibinfo {pages} {454 -- 466} (\bibinfo {year} {2008})}\BibitemShut {NoStop}%
\bibitem [{\citenamefont {Kuramoto}(1997)}]{Kur97}%
  \BibitemOpen
  \bibfield  {author} {\bibinfo {author} {\bibfnamefont {Y.}~\bibnamefont
  {Kuramoto}},\ }\bibfield  {title} {\enquote {\bibinfo {title} {Phase- and
  center-manifold reductions for large populations of coupled oscillators with
  application to non-locally coupled systems},}\ }\href {\doibase
  10.1142/S0218127497000595} {\bibfield  {journal} {\bibinfo  {journal} {Int.
  J. Bifurcat. Chaos}\ }\textbf {\bibinfo {volume} {7}},\ \bibinfo {pages}
  {789--805} (\bibinfo {year} {1997})}\BibitemShut {NoStop}%
\bibitem [{\citenamefont {Pikovsky}\ and\ \citenamefont
  {Rosenau}(2006)}]{pikovsky06}%
  \BibitemOpen
  \bibfield  {author} {\bibinfo {author} {\bibfnamefont {A.}~\bibnamefont
  {Pikovsky}}\ and\ \bibinfo {author} {\bibfnamefont {P.}~\bibnamefont
  {Rosenau}},\ }\bibfield  {title} {\enquote {\bibinfo {title} {Phase
  compactons},}\ }\href {\doibase https://doi.org/10.1016/j.physd.2006.04.015}
  {\bibfield  {journal} {\bibinfo  {journal} {Physica D}\ }\textbf {\bibinfo
  {volume} {218}},\ \bibinfo {pages} {56 -- 69} (\bibinfo {year}
  {2006})}\BibitemShut {NoStop}%
\bibitem [{\citenamefont {Kori}\ \emph {et~al.}(2014)\citenamefont {Kori},
  \citenamefont {Kuramoto}, \citenamefont {Jain}, \citenamefont {Kiss},\ and\
  \citenamefont {Hudson}}]{kori14}%
  \BibitemOpen
  \bibfield  {author} {\bibinfo {author} {\bibfnamefont {H.}~\bibnamefont
  {Kori}}, \bibinfo {author} {\bibfnamefont {Y.}~\bibnamefont {Kuramoto}},
  \bibinfo {author} {\bibfnamefont {S.}~\bibnamefont {Jain}}, \bibinfo {author}
  {\bibfnamefont {I.~Z.}\ \bibnamefont {Kiss}}, \ and\ \bibinfo {author}
  {\bibfnamefont {J.~L.}\ \bibnamefont {Hudson}},\ }\bibfield  {title}
  {\enquote {\bibinfo {title} {Clustering in globally coupled oscillators near
  a hopf bifurcation: Theory and experiments},}\ }\href {\doibase
  10.1103/PhysRevE.89.062906} {\bibfield  {journal} {\bibinfo  {journal} {Phys.
  Rev. E}\ }\textbf {\bibinfo {volume} {89}},\ \bibinfo {pages} {062906}
  (\bibinfo {year} {2014})}\BibitemShut {NoStop}%
\bibitem [{\citenamefont {Paz\'o}\ and\ \citenamefont
  {Montbri\'o}(2016)}]{PM16}%
  \BibitemOpen
  \bibfield  {author} {\bibinfo {author} {\bibfnamefont {D.}~\bibnamefont
  {Paz\'o}}\ and\ \bibinfo {author} {\bibfnamefont {E.}~\bibnamefont
  {Montbri\'o}},\ }\bibfield  {title} {\enquote {\bibinfo {title} {From
  quasiperiodic partial synchronization to collective chaos in populations of
  inhibitory neurons with delay},}\ }\href {\doibase
  10.1103/PhysRevLett.116.238101} {\bibfield  {journal} {\bibinfo  {journal}
  {Phys. Rev. Lett.}\ }\textbf {\bibinfo {volume} {116}},\ \bibinfo {pages}
  {238101} (\bibinfo {year} {2016})}\BibitemShut {NoStop}%
\bibitem [{\citenamefont {Devalle}\ \emph {et~al.}(2018)\citenamefont
  {Devalle}, \citenamefont {Montbri\'o},\ and\ \citenamefont
  {Paz\'o}}]{devalle18}%
  \BibitemOpen
  \bibfield  {author} {\bibinfo {author} {\bibfnamefont {F.}~\bibnamefont
  {Devalle}}, \bibinfo {author} {\bibfnamefont {E.}~\bibnamefont {Montbri\'o}},
  \ and\ \bibinfo {author} {\bibfnamefont {D.}~\bibnamefont {Paz\'o}},\
  }\bibfield  {title} {\enquote {\bibinfo {title} {Dynamics of a large system
  of spiking neurons with synaptic delay},}\ }\href {\doibase
  10.1103/PhysRevE.98.042214} {\bibfield  {journal} {\bibinfo  {journal} {Phys.
  Rev. E}\ }\textbf {\bibinfo {volume} {98}},\ \bibinfo {pages} {042214}
  (\bibinfo {year} {2018})}\BibitemShut {NoStop}%
\bibitem [{Note1()}]{Note1}%
  \BibitemOpen
  \bibinfo {note} {Otherwise, the frequency of (N)UIS would depend nonlinearly
  on $\epsilon $, in disagreement with the MF-CGLE.}\BibitemShut {Stop}%
\bibitem [{\citenamefont {Bick}\ \emph {et~al.}(2011)\citenamefont {Bick},
  \citenamefont {Timme}, \citenamefont {Paulikat}, \citenamefont {Rathlev},\
  and\ \citenamefont {Ashwin}}]{bick11}%
  \BibitemOpen
  \bibfield  {author} {\bibinfo {author} {\bibfnamefont {C.}~\bibnamefont
  {Bick}}, \bibinfo {author} {\bibfnamefont {M.}~\bibnamefont {Timme}},
  \bibinfo {author} {\bibfnamefont {D.}~\bibnamefont {Paulikat}}, \bibinfo
  {author} {\bibfnamefont {D.}~\bibnamefont {Rathlev}}, \ and\ \bibinfo
  {author} {\bibfnamefont {P.}~\bibnamefont {Ashwin}},\ }\bibfield  {title}
  {\enquote {\bibinfo {title} {Chaos in symmetric phase oscillator networks},}\
  }\href {\doibase 10.1103/PhysRevLett.107.244101} {\bibfield  {journal}
  {\bibinfo  {journal} {Phys. Rev. Lett.}\ }\textbf {\bibinfo {volume} {107}},\
  \bibinfo {pages} {244101} (\bibinfo {year} {2011})}\BibitemShut {NoStop}%
\bibitem [{\citenamefont {Bick}\ \emph {et~al.}(2016)\citenamefont {Bick},
  \citenamefont {Ashwin},\ and\ \citenamefont {Rodrigues}}]{bick16}%
  \BibitemOpen
  \bibfield  {author} {\bibinfo {author} {\bibfnamefont {C.}~\bibnamefont
  {Bick}}, \bibinfo {author} {\bibfnamefont {P.}~\bibnamefont {Ashwin}}, \ and\
  \bibinfo {author} {\bibfnamefont {A.}~\bibnamefont {Rodrigues}},\ }\bibfield
  {title} {\enquote {\bibinfo {title} {Chaos in generically coupled phase
  oscillator networks with nonpairwise interactions},}\ }\href {\doibase
  10.1063/1.4958928} {\bibfield  {journal} {\bibinfo  {journal} {Chaos}\
  }\textbf {\bibinfo {volume} {26}},\ \bibinfo {pages} {094814} (\bibinfo
  {year} {2016})}\BibitemShut {NoStop}%
\bibitem [{\citenamefont {Clusella}\ \emph {et~al.}(2016)\citenamefont
  {Clusella}, \citenamefont {Politi},\ and\ \citenamefont
  {Rosenblum}}]{clusella16}%
  \BibitemOpen
  \bibfield  {author} {\bibinfo {author} {\bibfnamefont {P.}~\bibnamefont
  {Clusella}}, \bibinfo {author} {\bibfnamefont {A.}~\bibnamefont {Politi}}, \
  and\ \bibinfo {author} {\bibfnamefont {M.}~\bibnamefont {Rosenblum}},\
  }\bibfield  {title} {\enquote {\bibinfo {title} {A minimal model of
  self-consistent partial synchrony},}\ }\href {\doibase
  10.1088/1367-2630/18/9/093037} {\bibfield  {journal} {\bibinfo  {journal}
  {New J. Phys.}\ }\textbf {\bibinfo {volume} {18}},\ \bibinfo {pages} {093037}
  (\bibinfo {year} {2016})}\BibitemShut {NoStop}%
\bibitem [{\citenamefont {Bick}\ \emph {et~al.}(2018)\citenamefont {Bick},
  \citenamefont {Panaggio},\ and\ \citenamefont {Martens}}]{bick18}%
  \BibitemOpen
  \bibfield  {author} {\bibinfo {author} {\bibfnamefont {C.}~\bibnamefont
  {Bick}}, \bibinfo {author} {\bibfnamefont {M.~J.}\ \bibnamefont {Panaggio}},
  \ and\ \bibinfo {author} {\bibfnamefont {E.~A.}\ \bibnamefont {Martens}},\
  }\bibfield  {title} {\enquote {\bibinfo {title} {Chaos in {K}uramoto
  oscillator networks},}\ }\href {\doibase 10.1063/1.5041444} {\bibfield
  {journal} {\bibinfo  {journal} {Chaos}\ }\textbf {\bibinfo {volume} {28}},\
  \bibinfo {pages} {071102} (\bibinfo {year} {2018})}\BibitemShut {NoStop}%
\bibitem [{\citenamefont {Bick}(2018)}]{cbick18}%
  \BibitemOpen
  \bibfield  {author} {\bibinfo {author} {\bibfnamefont {C.}~\bibnamefont
  {Bick}},\ }\bibfield  {title} {\enquote {\bibinfo {title} {Heteroclinic
  switching between chimeras},}\ }\href {\doibase 10.1103/PhysRevE.97.050201}
  {\bibfield  {journal} {\bibinfo  {journal} {Phys. Rev. E}\ }\textbf {\bibinfo
  {volume} {97}},\ \bibinfo {pages} {050201(R)} (\bibinfo {year}
  {2018})}\BibitemShut {NoStop}%
\bibitem [{\citenamefont {Skardal}\ and\ \citenamefont {Arenas}()}]{SA19}%
  \BibitemOpen
  \bibfield  {author} {\bibinfo {author} {\bibfnamefont {P.~S.}\ \bibnamefont
  {Skardal}}\ and\ \bibinfo {author} {\bibfnamefont {A.}~\bibnamefont
  {Arenas}},\ }\href@noop {} {\enquote {\bibinfo {title} {Abrupt
  desynchronization and extensive multistability in globally coupled oscillator
  simplices},}\ }\href {\doibase 10.1103/PhysRevLett.122.248301} {\bibfield  {journal}
  {\bibinfo  {journal} {Phys. Rev. Lett.}\ }\textbf {\bibinfo {volume} {122}},\
  \bibinfo {pages} {248301} (\bibinfo {year} {2019})}\BibitemShut {NoStop}%
\bibitem [{\citenamefont {Rosenblum}\ and\ \citenamefont
  {Pikovsky}(2019)}]{RP19}%
  \BibitemOpen
  \bibfield  {author} {\bibinfo {author} {\bibfnamefont {M.}~\bibnamefont
  {Rosenblum}}\ and\ \bibinfo {author} {\bibfnamefont {A.}~\bibnamefont
  {Pikovsky}},\ }\bibfield  {title} {\enquote {\bibinfo {title} {Numerical
  phase reduction beyond the first order approximation},}\ }\href {\doibase
  10.1063/1.5079617} {\bibfield  {journal} {\bibinfo  {journal} {Chaos}\
  }\textbf {\bibinfo {volume} {29}},\ \bibinfo {pages} {011105} (\bibinfo
  {year} {2019})}\BibitemShut {NoStop}%
\bibitem [{\citenamefont {Kori}\ and\ \citenamefont {Kuramoto}(2001)}]{KK01}%
  \BibitemOpen
  \bibfield  {author} {\bibinfo {author} {\bibfnamefont {H.}~\bibnamefont
  {Kori}}\ and\ \bibinfo {author} {\bibfnamefont {Y.}~\bibnamefont
  {Kuramoto}},\ }\bibfield  {title} {\enquote {\bibinfo {title} {Slow switching
  in globally coupled oscillators: robustness and occurrence through delayed
  coupling},}\ }\href {\doibase 10.1103/PhysRevE.63.046214} {\bibfield
  {journal} {\bibinfo  {journal} {Phys. Rev. E}\ }\textbf {\bibinfo {volume}
  {63}},\ \bibinfo {pages} {046214} (\bibinfo {year} {2001})}\BibitemShut
  {NoStop}%
\end{thebibliography}
	
\end{document}